 \documentclass[10pt]{jfm1}

\usepackage{graphicx}

\usepackage{natbib}            
\usepackage{epic,eepic,units}
\usepackage{psfrag}
\usepackage{latexsym}
\usepackage{longtable}
\usepackage{mathrsfs}
\usepackage{bigstrut}
\usepackage{pifont}
\usepackage{amsmath}
\usepackage{setspace}
\usepackage{subfigure}
\usepackage{mathtools}
\usepackage{mathcomp}
\usepackage{mathdots}
\usepackage{rotating}
\usepackage{array}
\usepackage{multirow}
\usepackage{multicol}
\usepackage{color}
\usepackage{url}
\usepackage{algorithm}
\usepackage{algorithmic}

\usepackage[dvipsnames]{xcolor}

\usepackage{euscript,yfonts,dsfont,bbm,wasysym,pdfsync,arydshln,framed}
\usepackage{amstext}
\usepackage{amssymb}

\usepackage{tikz}
\usetikzlibrary{arrows,shapes,trees,calc,positioning}
\usetikzlibrary{matrix}

\newcommand{\bbC}{\mathbb{C}}

\newcommand{\ds}{\displaystyle}

\newcommand{\mrd}{\mathrm{d}}
\newcommand{\mre}{\mathrm{e}}
\newcommand{\mri}{\mathrm{i}}

\newcommand{\bd}{{\bf d}}

\newcommand{\bA}{\mathbf{A}}
\newcommand{\bB}{\mathbf{B}}
\newcommand{\bC}{\mathbf{C}}

\newcommand{\bI}{{\bf I}}

\newcommand{\bu}{{\bf u}}

\newcommand{\bQ}{{\bf Q}}

\newcommand{\bphi}{\mbox{\boldmath$\phi$}}
\newcommand{\bvarphi}{\mbox{\boldmath$\varphi$}}
\newcommand{\bpsi}{\mbox{\boldmath$\psi$}}
\newcommand{\bchi}{\mbox{\boldmath$\chi$}}

\newcommand{\tc}{\textcolor}

\newcommand{\DefinedAs}[0]{\mathrel{\mathop:}=}
\newcommand{\DefinedAsLtoR}[0]{=\mathrel{\mathop:}}

\DeclareMathOperator*{\logdet}{log\,det}

\DeclareMathOperator*{\minimize}{minimize}

\DeclareMathOperator*{\subject}{subject~to}
\DeclareMathOperator*{\rank}{rank}

\definecolor{bgblue}{rgb}{0.04,0.19,0.53}
\definecolor{dblue1}{rgb}{0,0.3,0.7}
\definecolor{dred}{rgb}{0.4,0.2,0}

\newtheorem{remark}{Remark}

\newcommand{\enma}[1]   {\ensuremath{#1}}

\newcommand{\beq}{\begin{equation}}
\newcommand{\eeq}{\end{equation}}
\newcommand{\bseq}{\begin{subequations}}
\newcommand{\eseq}{\end{subequations}}
\newcommand{\beqn}{\begin{eqnarray}}
\newcommand{\eeqn}{\end{eqnarray}}
\newcommand{\ba}{\begin{array}}
\newcommand{\ea}{\end{array}}
\newcommand{\bct}{\begin{center}}
\newcommand{\ect}{\end{center}}
\newcommand{\btmz}{\begin{itemize}}
\newcommand{\etmz}{\end{itemize}}
\newcommand{\benum}{\begin{enumerate}}
\newcommand{\eenum}{\end{enumerate}}

\newcommand{\norm}[1]{\| #1 \|}                 

\newcommand{\diag}      {\enma{\mathrm{diag}}}

\newcommand{\trace}     {\enma{\mathrm{trace}}}

\newcommand{\bv}{{\bf v}}

\newcommand{\matbegin}{
        \left[
}
\newcommand{\matend}{
        \right]
}

\newcommand{\tbo}[2]{
  \matbegin \begin{array}{c}
       #1 \\ #2
       \end{array} \matend }

\newcommand{\obt}[2]{
  \matbegin \begin{array}{cc}
       #1 & #2
       \end{array} \matend }

\newcommand{\tbt}[4]{
  \matbegin \begin{array}{cc}
       #1 & #2 \\ #3 & #4
       \end{array} \matend }

\newcommand{\be}{\begin{equation}}
\newcommand{\ee}{\end{equation}}

\newcommand{\cplxs}{ C\kern -.35em \rule{0.03 em}{.7 ex}~   }

\def\complex{\hbox{C\kern -.45em \rule{0.03 em}{1.5 ex}}~}

\newcommand{\bi}{\begin{itemize}}
\newcommand{\ei}{\end{itemize}}

\newcommand{\bk}{{\bf{k}}}
\newcommand{\bw}{{\bf{w}}}

\title{Color of turbulence}

\shorttitle{Color of turbulence}

\author{Armin Zare, Mihailo R.\ Jovanovi\'c, and Tryphon T.\ Georgiou}

\affiliation{Department of Electrical and Computer Engineering,
	\\
University of Minnesota, Minneapolis, MN 55455, USA}

\shortauthor{A.\ Zare, M.\ R.\ Jovanovi\'c, and T.\ T.\ Georgiou}

\begin{document}

    \maketitle

\begin{abstract}
In this paper, we address the problem of how to account for second-order statistics of turbulent flows using low-complexity stochastic dynamical models based on the linearized Navier-Stokes equations. The complexity is quantified by the number of degrees of freedom in the linearized evolution model that are directly influenced by stochastic excitation sources. For the case where only a subset of velocity correlations are known, we develop a framework to complete unavailable second-order statistics in a way that is consistent with linearization around turbulent mean velocity. In general, white-in-time stochastic forcing is not sufficient to explain turbulent flow statistics. We develop models for colored-in-time forcing using a maximum entropy formulation together with a regularization that serves as a proxy for rank minimization. We show that colored-in-time excitation of the Navier-Stokes equations can also be interpreted as a low-rank modification to the generator of the linearized dynamics. Our method provides a data-driven refinement of models that originate from first principles and captures complex dynamics of turbulent flows in a way that is tractable for analysis, optimization, and control design.
\end{abstract}

	\vspace*{-4ex}
\section{Introduction}
    \label{sec.intro}

The advent of advanced measurement techniques and the availability of parallel computing have played a pivotal role in improving our understanding of turbulent flows. Experimentally and numerically generated data sets are becoming increasingly available for a wide range of flow configurations and Reynolds numbers. An accurate statistical description of turbulent flows may provide insights into flow physics and will be instrumental in model-based control design for suppressing or promoting turbulence. Thus, it is increasingly important to understand how structural and statistical features of turbulent flows can be embedded in models of low-complexity that are suitable for analysis, optimization, and control design. 

Nonlinear dynamical models of wall-bounded shear flows that are based on the Navier-Stokes (NS) equations typically have a large number of degrees of freedom. This makes them unsuitable for analysis and control synthesis. The existence of coherent structures in turbulent wall-bounded shear flows~\citep*{rob91,adr07,smimckmar11} has inspired the development of data-driven techniques for reduced-order modeling of the NS equations. However, control actuation and sensing may significantly alter the identified modes in nonlinear reduced-order modeling schemes. This introduces nontrivial challenges for model-based control design~\citep*{noamortad11,tadnoa11}. In contrast, linearization of the NS equations around mean-velocity gives rise to models that are well-suited for analysis and synthesis using tools of modern robust control. 
Further, such linearized models, subject to white-in-time stochastic excitation, have been shown to qualitatively replicate structural features of transitional~\citep{farioa93,bamdah01,jovbamJFM05} and turbulent~\citep{hwacosJFM10a,hwacosJFM10b,moajovJFM12} wall-bounded shear flows. However, it has also been recognized that white-in-time stochastic excitation is insufficient to accurately reproduce statistics of the fluctuating velocity field~\citep{jovbamCDC01,jovgeoAPS10}.

In this paper, we introduce {\em colored-in-time\/} stochastic excitation to the linearized NS equations {and} develop an optimization framework to identify low-complexity models for such excitation sources. We show that these models are suitable to replicate available second-order statistics of wall-bounded shear flows. Our models contain the same number of degrees of freedom as the finite-dimensional approximation of the NS equations and, moreover, they can be interpreted as {\em low-rank\/} perturbations of the linearized dynamics.

	\vspace*{-2ex}
\subsection{Linear analysis of transitional and turbulent shear flows}
	\label{sec.linear}

The linearized NS equations have been effectively used to capture the early stages of transition in wall-bounded shear flows and to identify key mechanisms for subcritical transition. It has been demonstrated that velocity fluctuations around {a} laminar base flow exhibit high sensitivity to different sources of perturbations. This has provided reconciliation with experimental observations~\citep{kletidsar62,kle71,kli92,wesboiklikozalf94,matalf01} that, even in the absence of modal instability, bypass transition can be triggered by large transient growth~\citep{gus91,butfar92,redhen93,henred94,schhen94} or large amplification of deterministic and stochastic disturbances~\citep{tretrereddri93,farioa93b,bamdah01,mj-phd04,jovbamJFM05}. The non-normality of the linearized dynamical generator introduces interactions of exponentially decaying normal modes~\citep{tretrereddri93,sch07}, which in turn result in high flow sensitivity. In the presence of mean shear and spanwise-varying fluctuations, vortex tilting induces high sensitivity of the laminar flow and provides a mechanism for the appearance of streamwise streaks and oblique modes~\citep*{lan75}.

Linear mechanisms also play an important role in the formation and maintenance of streamwise streaks in turbulent shear flows. \cite*{leekimmoi90} used numerical simulations to show that, even in the absence of a solid boundary, streaky structures appear in homogeneous turbulence subject to large mean shear. The formation of such structures has been attributed to the linear amplification of eddies that interact with background shear. 
{These authors also demonstrated the ability of linear rapid distortion theory~\citep{pop00} to predict the long time anisotropic behavior as well as the qualitative features of the instantaneous velocity field in homogeneous turbulence.
}
 \cite{kimlim00} highlighted the importance of linear mechanisms in maintaining near-wall streamwise vortices in wall-bounded shear flows. Furthermore, \cite{chebai05} used the linearized NS equations to predict the spacing of near-wall streaks and relate their formation to a combination of lift-up {due to} the mean profile, mean shear, and viscous dissipation. The linearized NS equations also reveal large transient growth of fluctuations around turbulent mean velocity~\citep{butfar93,farioa93a} and a high amplification of stochastic disturbances~\citep{farioa98}.~\cite*{schhus02,hoebrahen05} further identified a secondary growth (of the streaks) which may produce much larger transient responses than a secondary instability. All of these studies support the relevance of linear mechanisms in the self-sustaining regeneration cycle~\citep{hamkimwal95,wal97} and motivate {low-complexity} dynamical modeling of turbulent shear flows.  

Other classes of linear models have also been utilized to study {the} spatial structure of {the} most energetic fluctuations in turbulent flows. In particular, augmentation of molecular viscosity with turbulent eddy-viscosity yields the turbulent mean flow as the exact steady-state solution of the modified NS equations~\citep{reytie67,reyhus72-3}. The analysis of the resulting eddy-viscosity-enhanced linearized model reliably predicts the length scales of near-wall structures in turbulent wall-bounded shear flows~\citep{deljimjfm06,cospujdep09,pujgarcosdep09}. This model was also used to study the optimal response to initial conditions and body forcing fluctuations in turbulent channel~\citep{hwacosJFM10b} and Couette flows~\citep{hwacosJFM10a}, and {served as the basis for} model-based control design in turbulent channel flow~\citep{moajovJFM12}.

Recently, a gain-based decomposition of fluctuations around turbulent mean velocity has been used to characterize energetic structures in terms of their wavelengths and convection speeds~\citep{mcksha10,mckshajac13,shamck13,moashatromck13}. For turbulent pipe flow,~\cite{mcksha10} used resolvent analysis to explain the extraction of energy from the mean flow. Resolvent analysis provides further insight into linear amplification mechanisms associated with critical-layers. \cite{moashatromck13} extended this approach to turbulent channel flow and studied the Reynolds number scaling and geometric self-similarity of the dominant resolvent modes. In addition, they showed that decomposition of the resolvent operator can be used to provide a low-order description of the energy intensity of streamwise velocity fluctuations. Finally,~\cite{moajovtroshamckPOF14} used a weighted sum of a few resolvent modes to approximate the velocity spectra in turbulent channel flow.

	\vspace*{-2ex}
\subsection{Stochastic forcing and flow statistics}
	\label{sec.sffs}
	
The nonlinear terms in the NS equations are conservative and, as such, they do not contribute to the transfer of energy between the mean flow and velocity fluctuations; they only transfer energy between different {Fourier} modes~\citep{mcc91,durpet00}. This observation has inspired researchers to {\em model the effect of nonlinearity\/} via an {\em additive stochastic forcing\/} to the equations that govern the dynamics of fluctuations. Early efforts focused on homogeneous isotropic turbulence~\citep{kra59,kra71,ors70,monyag75}. In these studies, the conservative nature of the equations was maintained via a balanced combination of dynamical damping and stochastic forcing terms. However, imposing similar dynamical constraints in anisotropic and inhomogeneous flows is challenging and requires significant increase in \mbox{computational complexity.}  

The NS equations linearized around the mean velocity capture the interactions between the background flow and velocity fluctuations. In the absence of body forcing {and neutrally stable modes}, linearized models predict either asymptotic decay or unbounded growth of fluctuations. Thus, without a stochastic source of excitation linearized models {around stationary mean profiles} cannot generate the statistical responses that are ubiquitous in turbulent flows. For quasi-geostrophic turbulence, linearization around the time-averaged mean profile was used to model heat and momentum fluxes as well as spatio-temporal spectra~\citep{farioa93c,farioa94a,farioa95,delfar95,delfar96}. In these studies, the linearized model was driven with white-in-time stochastic forcing and the dynamical generator was augmented with a source of constant dissipation.~\cite{del96} examined the ability of Markov models (of different orders) subject to white forcing to explain time-lagged covariances of quasi-geostrophic turbulence.~\cite*{majtimeij99,majtimeij01} employed singular perturbation methods in an attempt to justify the use of stochastic models for climate prediction. Their analysis suggests that more sophisticated models, which involve not only additive but also multiplicative noise sources, may be required. All of these studies demonstrate encouraging agreement {between predictions resulting from stochastically driven linearized models and} available data and highlight {the} challenges that arise in modeling dissipation and the statistics of forcing~{\citep{del00,del04}}.
	
	\cite{farioa98} examined {the} statistics of the NS equations linearized around the Reynolds-Tiederman velocity profile subject to white stochastic forcing. It was demonstrated that velocity correlations over a finite interval determined by the eddy turnover time qualitatively {agree with second-order statistics} of turbulent channel flow.~\cite{jovbamCDC01} studied {the NS equations linearized around turbulent mean velocity and examined the influence of second-order spatial statistics of white-in-time stochastic disturbances on velocity correlations}. It was shown that portions of one-point correlations in turbulent channel flow can be approximated by the appropriate choice of forcing covariance. This was done in an {\em ad hoc\/} fashion by computing the steady-state velocity statistics for a variety of spatial forcing correlations. This line of work has inspired the development of optimization algorithms for approximation of full covariance matrices using stochastically-forced linearized NS equations~\citep{hoe05,linjovTAC09}. Moreover,~\cite{moajovJFM12} demonstrated that the energy spectrum of turbulent channel flow can be exactly reproduced using the linearized NS equations driven by white-in-time stochastic forcing with variance proportional to the turbulent energy spectrum. This choice was motivated by the observation that the second-order statistics of homogeneous isotropic turbulence can be exactly matched by such forcing spectra \citep{jovgeoAPS10,rashad-phd12}. 

{Stochastically forced models were also utilized in the context of stochastic structural stability theory to study jet formation and equilibration in barotropic beta-plane turbulence~\citep{farioa03,farioa07,bakioa11,confarioa14,bakioa14}. Recently, it was} demonstrated that a feedback interconnection of the streamwise-constant NS equations with {a} stochastically-driven streamwise-varying linearized model can generate self-sustained turbulence in Couette and Poiseuille flows~\citep{farioa12,tholiejovfarioagayPOF14,conloznikfarioajim14}. {Turbulence was triggered by the stochastic forcing and was maintained even after the forcing had been turned off. Even in the absence of stochastic forcing, certain measures of turbulence, e.g., the correct mean velocity profile, are maintained through interactions between the mean flow and a small subset of streamwise varying modes. Even though turbulence can be triggered with white-in-time stochastic forcing, correct statistics cannot be obtained without accounting for the dynamics of the streamwise averaged mean flow or without manipulation of the underlying \mbox{dynamical modes~\citep{bremengay15,thofarioagay15}.}}

	
	\vspace*{-2ex}
\subsection{Preview of modeling and optimization framework}
	\label{sec.preview}
	
As {already noted}, the linearized NS equations with white-in-time stochastic forcing have been used to predict coherent structures in transitional and turbulent shear flows and to yield statistics that are in qualitative agreement with experiments and simulations. For homogeneous isotropic turbulence, this model can completely recover second-order statistics of the velocity field~\citep{jovgeoAPS10,rashad-phd12}. For turbulent channel flow, however, we demonstrate that the linearized NS equations with white-in-time stochastic excitation {\em cannot reproduce\/} second-order statistics that originate from direct numerical simulations (DNS). This {observation} exposes {the} limitations of the white-in-time forcing model.

In the present paper, we show that {\em colored-in-time\/} stochastic forcing provides sufficient flexibility to account for statistical signatures of turbulent channel flow. We develop a systematic method for identifying the spectral content of colored-in-time forcing to the linearized NS equations that allows us to capture second-order statistics of fully-developed turbulence. Most of our discussion focuses on channel flow, yet the methodology and theoretical framework are applicable to more complex flow configurations.

We are interested in completing partially available second-order statistics of velocity fluctuations in a way that is consistent with the known dynamics.
The statistics of forcing to the linearized equations around turbulent mean velocity are unknown and sought to match the available velocity correlations and to complete any missing data. Our approach utilizes an algebraic relation that characterizes steady-state covariances of linear systems subject to colored-in-time excitation sources~\citep{geo02b,geo02a}. This relation extends the standard algebraic Lyapunov equation, which maps white-in-time forcing correlations to state statistics, and it imposes a structural constraint on admissible covariances. We follow a maximum entropy formalism to obtain positive definite velocity covariance matrices and use suitable regularization to identify forcing correlation structures of low rank. This restricts the number of degrees of freedom that are directly influenced by the stochastic forcing and, thus, the complexity of the colored-in-time forcing model~\citep{chejovgeoCDC13,zarchejovgeoTAC16}.

Minimizing the rank, in general, leads to difficult non-convex optimization problems. Thus, instead, we employ nuclear norm regularization as a surrogate for rank minimization~\citep*{faz02,canrec09,canpla10,recfazpar10}. The nuclear norm of a matrix is determined by the sum of its singular values and it provides a means for controlling {the} complexity of the model for stochastic forcing to the linearized NS equations. The covariance completion problem that we formulate is convex and its globally optimal solution can be efficiently computed using customized algorithms that we recently developed~\citep{zarjovgeoACC15,zarchejovgeoTAC16}.

\begin{figure}
	\begin{center}
%
%
%
%
%
%
\input{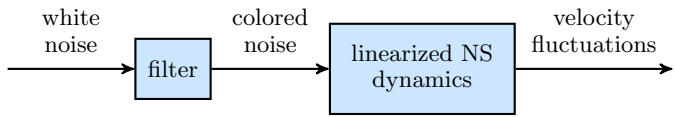}
%
%
\noindent
\begin{tikzpicture}[scale=1, auto, >=stealth']
  
    \small

    
     \node[block, minimum height = 0.8cm, top color=RoyalBlue!20, bottom color=RoyalBlue!20] (sys1) {filter};
     
     \node[block, minimum height = 1.2cm, top color=RoyalBlue!20, bottom color=RoyalBlue!20] (sys2) at ($(sys1.east) + (2.8cm,0)$) {$\ba{c} \mbox{linearized NS} \\ \mbox{dynamics}\ea$};
     
     \node[] (output-node) at ($(sys2.east) + (2.2cm,0)$) {};
     
     \node[] (input-node) at ($(sys1.west) - (1.8cm,0)$) {}; 
      
%
%
%
%
%
%
%


    \draw [connector] (input-node) -- node [midway, above] {$\ba{c} \mbox{white} \\ \mbox{noise} \ea$} (sys1.west);
    \draw [connector] (sys2.east) -- node [midway, above] {$\ba{c} \mbox{velocity} \\ \mbox{fluctuations} \ea$} (output-node);
    \draw [connector] (sys1.east) -- node [midway, above] {$\ba{c} \mbox{colored} \\ \mbox{noise} \ea$} (sys2.west);
\end{tikzpicture}
	\end{center}
	\caption{A spatio-temporal filter is designed to provide colored stochastic input to the linearized NS equations in order to reproduce partially available second-order statistics of turbulent channel flow.}
	\label{fig.filter_sys}
\end{figure}

We use the solution to the covariance completion problem to develop a dynamical model for colored-in-time stochastic forcing to the linearized NS equations {(see figure~\ref{fig.filter_sys}) and} provide a state-space realization for spatio-temporal filters that generate the appropriate forcing. These filters are generically minimal in the sense that their state dimension coincides with the number of degrees of freedom in the finite-dimensional approximation of the NS equations. We also show that colored-in-time stochastic forcing can be equivalently interpreted as a {\em low-rank modification\/} to the dynamics of the NS equations linearized around turbulent mean velocity. This dynamical perturbation provides a {\em data-driven refinement\/} of a {\em physics-based model\/} (i.e., the linearized NS equations) and it guarantees statistical consistency with fully-developed turbulent channel flow. This should be compared and contrasted to alternative modifications proposed in the literature, e.g., the eddy-viscosity-enhanced linearization~\citep{reyhus72-3,deljimjfm06,cospujdep09,pujgarcosdep09,hwacosJFM10a,hwacosJFM10b} or the addition of a dissipation operator~\citep{del04}; see \S~\ref{sec.role-color} for additional details.
	
We consider the mean velocity profile and one-point velocity correlations in the wall-normal direction at various wavenumbers as available data for our optimization problem. These are obtained using DNS of turbulent channel flow~\citep*{kimmoimos87,moskimman99,deljim03,deljimzanmos04,hoyjim06}. We show that stochastically-forced linearized NS equations can be used to exactly reproduce all one-point correlations (including one-dimensional energy spectra) and to provide good completion of unavailable two-point correlations of the turbulent velocity field. The resulting modified dynamics have the same number of degrees of freedom as the finite-dimensional approximation of the linearized NS equations. Thus, they are convenient for conducting linear stochastic simulations.  The ability of our model to account for the statistical signatures of turbulent channel flow is verified using these simulations. We also demonstrate that our approach captures velocity correlations at higher Reynolds numbers. We close the paper by employing tools from linear systems theory to analyze the spatio-temporal features of our model in the presence of stochastic and deterministic excitation sources.

\subsection{Paper outline}
	\label{sec.outline}
	
The rest of our presentation is organized as follows. In \S~\ref{sec.LNS-stats}, we introduce the stochastically-forced linearized NS equations and describe the algebraic relation that linear dynamics impose on admissible state and forcing correlations. In \S~\ref{sec.mcp}, we formulate the covariance completion problem, provide {a} state-space realization for spatio-temporal filters, and show that the linearized NS equations with colored-in-time forcing can be equivalently represented as a low-rank modification to the original linearized dynamics. In \S~\ref{sec.application_turbchannel}, we apply our framework to turbulent channel flow and verify our results using linear stochastic simulations. In \S~\ref{sec.frequency-anal}, we examine spatio-temporal frequency responses of the identified model, visualize dominant flow structures, and compute two-point temporal correlations. In \S~\ref{sec.discussion}, we discuss features of our framework and offer perspective on future research directions. We conclude with a summary of our contributions in  \S~\ref{sec.remarks}.

	\vspace*{-2ex}
\section{Linearized Navier-Stokes equations and flow statistics}
    \label{sec.LNS-stats}

In this section, we present background material on stochastically-forced linearized NS equations and second-order statistics of velocity fluctuations. Specifically, we provide an algebraic relation that is dictated by the linearized dynamics and that connects the steady-state covariance of the state in the linearized evolution model to the spectral content of the forcing. We focus on colored-in-time forcing inputs and extend the standard algebraic Lyapunov equation, which maps white-in-time disturbances to state statistics, to this more general case. Even though most of our discussion focuses on turbulent channel flow, the methodology and theoretical framework presented herein {are} applicable to other flow configurations.

\subsection{The Navier-Stokes equations and second-order statistics}
    \label{sec.NS-stats}

\begin{figure}
\begin{center}
         \includegraphics[width=6.15cm]{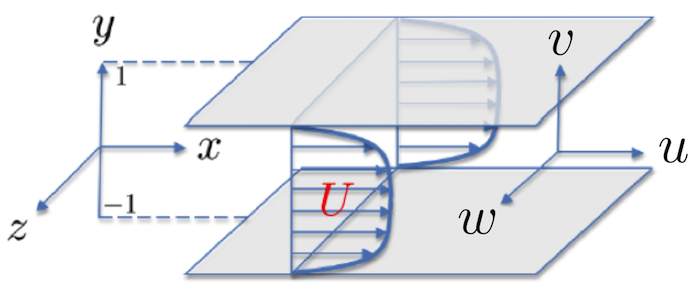}
\end{center}
\caption{Geometry of a pressure-driven turbulent channel flow.}
\label{fig.channel}
\end{figure}

The dynamics of incompressible Newtonian fluids are governed by the NS and continuity equations,
\begin{subequations}
	\label{eq.NScts}
	\begin{eqnarray}
    	\bu_t
    	&\; = \;&
    	- (\bu \cdot \nabla) \bu
    	\; - \;
    	\nabla P
    	\; + \; 
	\dfrac{1}{R_\tau} \, \Delta \bu,
    	\\
    	0
    	&\; = \;&
    	\nabla \cdot \bu,
	\end{eqnarray}
\end{subequations}
where $\bu$ is the velocity vector, $P$ is the pressure, $\nabla$ is the gradient, and $\Delta = \nabla\cdot\nabla$ is the Laplacian. In channel flow with geometry shown in figure~\ref{fig.channel}, the friction Reynolds number is $R_\tau = u_\tau h/\nu$, where $h$ is the channel half-height,  $u_{\tau} = \sqrt{\tau_w/\rho}$ is friction velocity, $\nu$ is kinematic viscosity, $\tau_w$ is wall-shear stress (averaged over wall-parallel directions and time), $\rho$ is the fluid density, and $t$ is time. In this formulation, spatial coordinates are non-dimensionalized by $h$, velocity by $u_\tau$, time by $h/u_\tau$, and pressure by $\rho u_\tau^2$. 

The velocity field in~(\ref{eq.NScts}) can be decomposed into the sum of mean, $\bar{\bu}$, and fluctuating parts, $\bv = [\,u\,~v\,~w\,]^T$,
	\be
	\bu 
	\; = \; 
	\bar{\bu} 
	\; + \; 
	\bv, 
	~~~~ 
	\bar{\bu} 
	\; = \; 
	\left< \bu \right>, 
	~~~~ 
	\left< \bv \right> 
	\; = \; 
	0.
	\ee
The components of the velocity fluctuation vector in the streamwise, $x$, wall-normal, $y$, and spanwise, $z$, directions are $u$, $v$, and $w$, and $\left< \cdot \right>$ is the {temporal} expectation operator.

For turbulent flows, the mean velocity field satisfies the Reynolds-averaged NS equations~\citep{mcc91,durpet00,pop00},
\begin{subequations}
	\label{eq.RANS}
	\begin{eqnarray}
		\bar{\bu}_t
		& \; = \; &
		- 
		\left(\bar{\bu} \cdot \nabla \right) \bar{\bu}
		\; - \;
		\nabla \langle P \rangle
       		\; + \;
       		\dfrac{1}{R_\tau} \, \Delta \bar{\bu}
        		\; - \;
		\nabla \cdot \left<\bv \bv^T \right>,
       		\\
       		0 
    		& \!\!\! = \!\!\! &
    		\nabla \cdot \bar{\bu},
	\end{eqnarray}
\end{subequations}
where $\left< \bv \bv^T \right>$ is the Reynolds stress tensor that arises from the second-order statistics of velocity fluctuations. Such statistics quantify the transport of momentum and they have profound influence on the mean velocity, and thereby on the resistance to flow motion~\citep{mcc91}. The difficulty in determining statistics of fluctuations comes from the nonlinearity in the NS equations which makes the $n$th velocity moment depend on the $(n+1)$th moment~\citep{mcc91}. Statistical theory of turbulence combines physical intuition and empirical observations with rigorous approximation of the flow equations to express the higher-order moments in terms of the lower-order moments~\citep{mcc91,durpet00,pop00}. For example, the turbulent viscosity hypothesis~\citep{pop00} relates turbulent fluxes to mean velocity gradients, thereby allowing approximate solutions of~\eqref{eq.RANS} to be computed. 


\subsection{Stochastically-forced linearized NS equations}
	\label{sec.LNSE}
	
Linearization around the turbulent mean velocity, $\bar{\bu}$, yields the equations that govern the dynamics of {velocity and pressure fluctuations,}
\begin{subequations}
	\label{eq.turblin}
	\begin{eqnarray}
    		\bv_t
    		& \; = \; &
    		-
    		\left( \nabla \cdot \bar{\bu} \right) \bv
     		\; - \; 
    		\left( \nabla \cdot \bv \right) \bar{\bu}
    		\; - \;
    		\nabla p
    		\; + \;
    		\dfrac{1}{R_\tau} \, \Delta \bv
    		\; + \;
    		\bd,
    		\\
    		0
    		&  \;= \; &
    		\nabla \cdot \bv,
	\end{eqnarray}
\end{subequations}
where $\bf d$ is an additive zero-mean stochastic body forcing. The presence of stochastic forcing can be justified in different ways and there is a rich literature on the subject \citep{farioa93,bamdah01,mj-phd04,jovbamJFM05}. For our purposes, turbulent flows have a well-recognized statistical signature which we want to reproduce, using perturbations around turbulent mean velocity, by postulating the stochastic model given by \eqref{eq.turblin}.

A standard conversion yields an evolution form of the linearized equations~\citep{schhen01}, with the state variable, $\bvarphi = [\,v\,~\eta\,]^T$, determined by the wall-normal velocity, $v$, and vorticity, $\eta = \partial_z u - \partial_x w$. In turbulent {channels the mean flow takes the form} $\bar{\bu} = [\,U(y)\,~0\,~0\,]^T$, thereby implying translational invariance of~(\ref{eq.turblin}) in the $x$ and $z$ directions. Application of the Fourier transform in the wall-parallel directions yields the evolution model,
\begin{subequations}
	\label{eq.lnse}
        	\begin{eqnarray}
            \bvarphi_t (y, \bk, t)
            & \!\!=\!\! &
            \left[ 
            \bA (\bk)
            \,
            \bvarphi ( \, \cdot \, , \bk, t)
            \right]
            (y)
            \; + \;
            \left[ 
            \bB (\bk)
            \,
            \bd ( \, \cdot \, , \bk, t)
            \right]
            (y)
            \\[0.05cm]
            \bv (y, \bk, t)
            & \;=\; &
            \left[
            \bC (\bk)
            \,
            \bvarphi ( \, \cdot \, , \bk, t)
            \right]
            (y),
        \end{eqnarray}
\end{subequations}
which is parameterized by the spatial wavenumber pair $\bk = (k_x,\,k_z)$. The operators $\bA$ and $\bC$ in~\eqref{eq.lnse} are given by
\be
	\label{eq.lnse_A-C}
	\ba{rcl}
	\bA (\bk)
	&\,=\,&
	\begin{bmatrix} \bA_{11} (\bk) &\!\! 0 \\[.15cm] \bA_{21} (\bk) & \bA_{22} (\bk)\end{bmatrix},~~\,\,
	\bC (\bk)
	\,=\,
	\begin{bmatrix}\bC_u (\bk) \\[.1cm] \bC_v (\bk) \\[.1cm] \bC_w (\bk) \end{bmatrix} 
	\,=\,
	 \dfrac{1}{k^2} \begin{bmatrix} \mri k_x \partial_y & - \mri k_z \\[.2cm] k^2 & 0 \\[.2cm] \mri k_z \partial_y & \mri k_x \end{bmatrix},
	\\[.7cm]
	\bA_{11} (\bk)
	& \, = \, &
	\Delta^{-1} \left( \left(1/R_\tau \right) \Delta^2 \,+\,  \mri k_x\, (U'' \,-\, U \Delta) \right),
	\\[0.15cm]
	\bA_{21} (\bk)
	& \, = \, &
	-\mri k_z U',
	\\[0.15cm]
	\bA_{22} (\bk)
	& \, = \, &
	\left( 1/R_\tau \right) \Delta \; - \; \mri k_x U,
	\ea
\ee
where prime denotes differentiation with respect to the wall-normal coordinate, $\mri$ is the imaginary unit, $\Delta = \partial_y^2 - k^2$, $\Delta^2 = \partial_y^4 - 2 k^2 \partial_y^2 + k^4$, and $k^2 = k_x^2 + k_z^2$. In addition, no-slip and no-penetration boundary conditions imply $v(\pm1, \bk, t) = v'(\pm1, \bk, t) = \eta(\pm1, \bk, t) = 0$. Here, $\bA_{11}$, $\bA_{22}$, and $\bA_{21}$ are the Orr-Sommerfeld, Squire, and coupling operators~\citep{schhen01}, and the operator $\bC (\bk)$, establishes a kinematic relationship between the components of $\bvarphi$ and the components of $\bv$. The operator $\bB$ specifies the way the external excitation $\bd$ affects the dynamics; see \cite{jovbamJFM05} for examples of $\bB$ in the case of channel-wide and near-wall excitations.

Finite-dimensional approximations of the operators in~\eqref{eq.lnse} are obtained using a pseudospectral scheme with $N$ Chebyshev collocation points in the {wall-normal} direction~\citep{weired00}. In addition, we use a change of coordinates to obtain a state-space representation in which the kinetic energy is determined by the Euclidean norm of the state vector{; see appendix~\ref{sec.coc}.} The resulting state-space model is given by
\begin{subequations}
	\label{eq.lnse1}
        \begin{eqnarray}
            \dot{\bpsi} (\bk, t)
            & \; = \; &
            A (\bk)
            \,
            \bpsi (\bk, t)
            \; + \;
            B (\bk)
            \,
            \bd (\bk, t),
            \\[0.05cm]
            \bv (\bk, t)
            & \; = \; &
            C (\bk)
            \,
            \bpsi (\bk,\, t),
        \end{eqnarray}
\end{subequations}
where $\bpsi (\bk, t)$ and $\bv (\bk, t)$ are vectors with complex-valued entries and $2N$ and $3N$ components, respectively, and state-space matrices $A(\bk)$, $B(\bk)$, and $C(\bk)$ are discretized versions of the corresponding operators that incorporate the aforementioned change of coordinates.

In statistical steady-state, the covariance matrix 
\be
	\Phi (\bk)
	\;=\;
	\lim_{t \, \to \, \infty}
	\left< 
	\bv(\bk,t)\, \bv^*(\bk,t) 
	\right>
\ee
of the velocity fluctuation vector, and
the covariance matrix
\be
	X (\bk)
	\;=\;
	\lim_{t \, \to \, \infty}\left< \bpsi(\bk,t)\, \bpsi^*(\bk,t) \right>
	\label{eq.statecovariance}
\ee
of the state in~(\ref{eq.lnse1}), are related as follows:
\be
	\Phi (\bk)
	\; = \;
	C (\bk)\, X (\bk)\, C^* (\bk),
	\label{eq.Xphi_relation}
\ee
where $*$ denotes complex-conjugate-transpose. The matrix $\Phi(\bk)$ contains information about all second-order statistics of the fluctuating velocity field, including the Reynolds stresses~\citep{moimos89}. The matrix $X (\bk)$ contains equivalent information and one can be computed from the other. Our interest in $X(\bk)$ stems from the fact that, as we explain next, the entries of $X(\bk)$ satisfy tractable algebraic relations that are dictated by the linearized dynamics \eqref{eq.lnse1} and the spectral content of the forcing $\bd(\bk,t)$.

\subsection{Second-order statistics of the linearized Navier-Stokes equations}
    \label{sec.velocity_correlations}

For the case where the stochastic forcing is zero-mean and white-in-time with covariance $W (\bk) = W^* (\bk) \succeq 0$, i.e.,
\be
	\left< {\bd} (\bk,t_1) \, {\bd}^* (\bk,t_2) \right> 
	\;=\; 
	W(\bk)
	\,
	\delta(t_1 \, - \, t_2),
	\label{eq.white_cov}
\ee
where $\delta$ is the Dirac delta function, the steady-state covariance of the state in \eqref{eq.lnse1} can be determined as the solution to the linear equation,
\be
	A(\bk)\, X (\bk)
	\;+\;
	X(\bk)\, A^*(\bk) 
	\;=\; 
	-B(\bk)\, W(\bk) \, B^*(\bk).
	\label{eq.standard_lyap}
\ee 
Equation~\eqref{eq.standard_lyap} is standard and it is known as the algebraic Lyapunov equation \citep[section~1.11.3]{kwasiv72}. It relates the statistics of the white-in-time forcing $W(\bk)$ to the covariance of the state $X(\bk)$ via the system matrices $A(\bk)$ and $B(\bk)$.

For the more general case, where the stochastic forcing is colored-in-time,
a corresponding algebraic relation was more recently developed by~\cite{geo02b,geo02a}. The new form is
\be
	A (\bk) \, X (\bk)
	\,+\, 
	X (\bk) \,A^* (\bk) 
	\;=\; 
	-\,B (\bk) \,H^* (\bk)
	\,-\, 
	H (\bk) \,B^* (\bk),
	\label{eq.lyap_BH}
\ee
where $H (\bk)$ {is a matrix that contains spectral information about the colored-in-time stochastic forcing and is related to the cross-correlation between the forcing and the state in~\eqref{eq.lnse1}; see section~\ref{sec.filter} and appendix~\ref{sec.H-interp} for details.}
For the special case where the forcing is white-in-time, 
{$H(\bk)=(1/2) B(\bk) W(\bk)$} and \eqref{eq.lyap_BH} reduces to the standard Lyapunov equation \eqref{eq.standard_lyap}. It should be noted that the right-hand-side of \eqref{eq.standard_lyap} is sign-definite, i.e., all eigenvalues of the matrix $B(\bk)\, W(\bk) \, B^*(\bk)$ are nonnegative. In contrast,  the right-hand-side of \eqref{eq.lyap_BH} is in general sign-indefinite. In fact, except for the case when the input is white noise, the matrix $Z (\bk)$ defined by
\begin{subequations}
	\label{eq.rhs_lyap}
	\begin{eqnarray}
	Z (\bk)
	& \; \DefinedAs \; &
	- \left(A(\bk)\,X(\bk)
	\,+\,
	X(\bk)\, A^*(\bk)
	\right)
	\\[0.1cm]
	& \;=\; &
	B(\bk)\, H^*(\bk)
	\,+\,
	H(\bk)\, B^*(\bk)
	\end{eqnarray}
\end{subequations}
may have both positive and negative eigenvalues.

Both equations, \eqref{eq.standard_lyap} and \eqref{eq.lyap_BH}, in the respective cases, are typically used to compute the state covariance $X(\bk)$ from the system matrices and forcing correlations. However, these same equations can be seen as linear algebraic constraints that restrict the values of the admissible covariances. It is in this sense that these algebraic constraints are used in the current paper. More precisely, while a state-covariance $X(\bk)$ is positive definite, not all positive-definite matrices can arise as state-covariances for the specific dynamical model \eqref{eq.lnse1}. As shown by~\cite{geo02b,geo02a}, the structure of state-covariances is an inherent property of the linearized dynamics.  Indeed, \eqref{eq.lyap_BH} provides necessary and sufficient conditions for a positive definite matrix $X(\bk)$ to be a state-covariance of \eqref{eq.lnse1}. Thus, given $X(\bk)$, \eqref{eq.lyap_BH} has to be solvable for $H(\bk)$. Equivalently, given $X(\bk)$, solvability of {\eqref{eq.lyap_BH}} amounts to the following rank condition:
	\be
	\label{eqn:rank Constraint on Sigma}
	\rank
	\left[
	\begin{matrix}
	A(\bk)X(\bk) \,+\, X(\bk) A^* (\bk) & B(\bk)
	\\
	B^* (\bk) & 0
	\end{matrix}
	\right]
	\, = \;
	\rank \left[\begin{matrix}
	0 & B(\bk)
	\\
	B^* (\bk) & 0
	\end{matrix}\right].
\ee
{This implies that any positive-definite matrix $X$ is admissible as a covariance of a linear time-invariant system if the input matrix $B$ is full row rank.}

In the next section, we utilize this framework to depart from white-in-time restriction on stochastic forcing and present a convex optimization framework for identifying colored-in-time excitations that account for partially available turbulent flow statistics. We also outline a procedure for designing a class of linear filters which generate the appropriate colored-in-time forcing.

\section{Completion of partially known turbulent flow statistics}
\label{sec.mcp}

In high-Reynolds-number flows, only a finite set of correlations is available due to experimental or numerical limitations. Ideally, one is interested in a more complete set of such correlations that provides insights into flow physics. This brings us to investigate the completion of the partially known correlations in a way that is consistent with perturbations of the flow field around turbulent mean velocity. The velocity fluctuations can be accounted for by stochastic forcing to the linearized equations. To this end, we seek stochastic forcing models of low complexity where complexity is quantified by the number of degrees of freedom that are directly influenced by stochastic forcing~\citep{chejovgeoCDC13, zarchejovgeoTAC16}. Such models arise as solutions to an inverse problem that we address using a  regularized maximum entropy formulation. Interestingly, the models we obtain can alternatively be interpreted as a low-rank perturbation to the original linearized dynamics.

	\vspace*{-2ex}
\subsection{Covariance completion problem}
\label{sec.ccp}

\begin{figure}
\begin{center}
         \includegraphics[width=6.15cm]{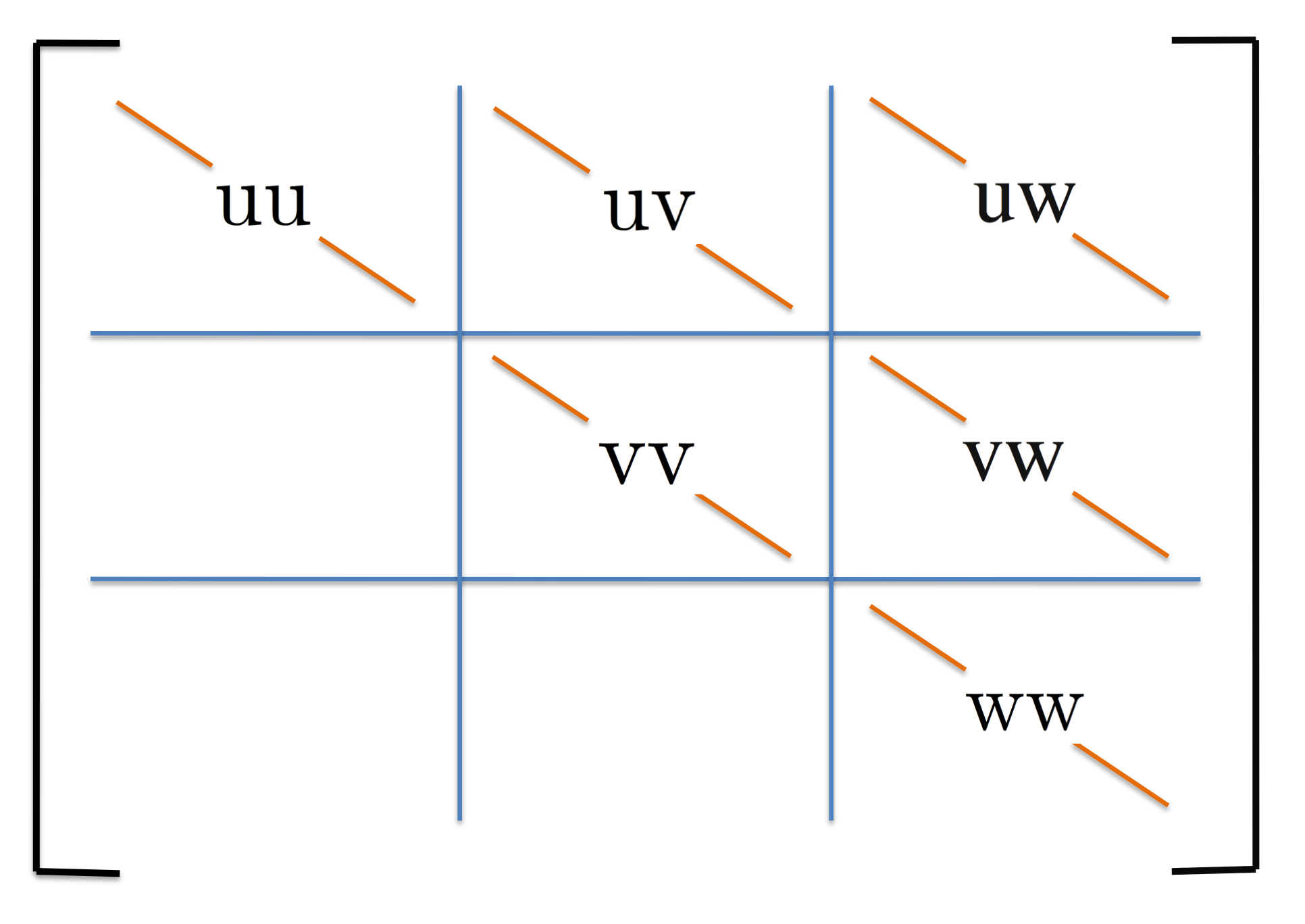}
\end{center}
\caption{Structure of the matrix $\Phi$ in optimization problem~\eqref{eq.CP}. At each pair of wavenumbers $\bk = (k_x,k_z)$, available second-order statistics are given by one-point correlations in the wall-normal direction, i.e., the diagonal entries of the blocks in the velocity covariance matrix $\Phi$. The data is obtained from {\sf http://torroja.dmt.upm.es/channels/data/}}
\label{fig.output_covariance}
\end{figure}

We begin with the Navier-Stokes equations linearized about the turbulent mean velocity profile \eqref{eq.lnse1}. As explained in \S~\ref{sec.velocity_correlations}, the covariance matrix $X$ of the state $\bpsi$ in \eqref{eq.lnse1}, in statistical steady-state, satisfies the Lyapunov-like linear equation
 \be
 	\label{eq.constraint_lyap}
 	A \, X 
	\;+\; 
	X A^* 
	\;+\; 
	Z  
	\;=\; 
	0,
 \ee
where $A$ is the generator of the linearized dynamics and $Z$ is the contribution of the stochastic excitation. For notational convenience, we omit the dependence on the wavenumber vector in this section. 
A subset of entries of the covariance matrix $\Phi$ of velocity fluctuations, namely $\Phi_{ij}$ for a selection of indices $(i,j)\in \mathcal I$, is assumed available. This yields an additional set of linear constraints for the matrix $X$,
\be
	\label{eq.constraint_observations}
	(CXC^*)_{ij}
	\;=\;
	\Phi_{ij}, ~~~~ (i,j)\in \mathcal I.
\ee
For instance, these known entries of $\Phi$ may represent one-point correlations in the wall-normal direction; see figure~\ref{fig.output_covariance} for an illustration.
Thus, our objective is to identify suitable choices of $X$ and $Z$ that satisfy the above constraints.

It is important to note that $X$ is a covariance matrix, and hence positive definite, while $Z$ is allowed to be sign indefinite. Herein, we follow a maximum entropy formalism {
and minimize $-\log\det (X)$ subject to the given constraints~\citep{goopay77}. Minimization of this logarithmic barrier function guarantees positive definiteness of the matrix $X$~\citep{boyvan04}.} 

The contribution of the stochastic excitation enters through the matrix $Z$, cf.~\eqref{eq.rhs_lyap}, which is of the form
\be
	Z
	\;=\; 
	B H^*
	\,+\,
	H B^*,
\ee
where color of the time-correlations and directionality of the forcing are reflected by the choices of $B$ and $H$. The matrix $B$ specifies {the preferred structure by which} stochastic excitation enters into the linearized evolution model while $H$ contains {spectral information about the colored-in-time stochastic forcing.} Trivially, when $B$ is taken to be the {full rank}, all degrees of freedom are excited and a forcing model that cancels the original linearized dynamics becomes a viable choice{; see remark~\ref{rem.obscure}}.
{Without additional restriction on the forcing model, minimization of $-\logdet(X)$ subject to the problem constraints yields a solution where the forcing excites all degrees of freedom in the linearized model. Such an approach may yield a solution that obscures important aspects of the underlying physics; see remark~\ref{rem.obscure}.} It is thus important to minimize the number of degrees of freedom that can be directly influenced by forcing. This can be accomplished by a suitable regularization, e.g., by minimizing the \mbox{rank of the matrix $Z$ \citep{chejovgeoCDC13,zarchejovgeoTAC16}.} 

Minimizing the rank, in general, leads to difficult non-convex optimization problems. Instead, the nuclear norm, i.e., the sum of singular values of a matrix,
\be
	\norm{Z}_{\star} 
	\; \DefinedAs \;
	\sum_{i} \sigma_i (Z),
	\label{eq.nuclearnorm}
\ee
can be used as a {\em convex\/} proxy for rank minimization~\citep{faz02,recfazpar10}. This leads to the following convex optimization problem
\begin{align} 
	\ba{cl}
	\minimize\limits_{X, \, Z}
	& 
	-\logdet\left(X\right) 
	\; + \; 
	\gamma \, \norm{Z}_\star
	\\[.1cm]
	\subject 
	&
	~A \, X \,+\, X A^* \,+\, Z  \;=\; 0
	\\[0.15cm]
	&
	\, (CXC^*)_{ij} \;=\;\Phi_{ij}, ~~~~ (i,j)\in \mathcal I,
	 \ea
	 \tag{CC}
	\label{eq.CP}
\end{align} 
where the matrices $A$ and $C$ as well as the available entries $\Phi_{ij}$ of the velocity covariance matrix represent problem data, the Hermitian matrices $X$, $Z \in \bbC^{n \times n}$ are the optimization variables, and the regularization parameter $\gamma>0$ reflects the relative weight specified for the nuclear norm objective. 
{
While minimizing $-\logdet(X)$ results in the maximum entropy solution, we also confine the complexity of the forcing model via nuclear norm minimization. The objective function in problem~\eqref{eq.CP} thus provides a trade-off between the solution to the maximum entropy problem and the complexity of the forcing model.}

Convexity of optimization problem~\eqref{eq.CP} follows from the convexity of the objective function (which contains entropy and nuclear norm terms) and the linearity of the constraint set. Convexity is important because it guarantees a unique globally optimal solution. In turn, this solution provides a choice for the completed covariance matrix $X$ and forcing contribution $Z$ that are consistent with the constraints.

Although optimization problem~\eqref{eq.CP} is convex, it is challenging to solve via conventional solvers for large-scale problems that arise in fluid dynamics. To this end, we have developed a scalable customized algorithm \citep{zarjovgeoACC15,zarchejovgeoTAC16}.

\subsection{Filter design: dynamics of stochastic forcing}
\label{sec.filter}

We now describe how the solution of optimization problem~\eqref{eq.CP} can be translated into a dynamical model for the colored-in-time stochastic forcing that is applied to the linearized NS equations.
We recall that, due to translational invariance in the channel flow geometry, optimization problem~\eqref{eq.CP} is fully-decoupled for different wavenumbers $\bk=(k_x,k_z)$. For each such pair, the solution matrices $X(\bk)$ and $Z(\bk)$ provide information about the temporal and wall-normal correlations of the stochastic forcing. We next provide the explicit construction of a linear dynamical model (filter) that generates the appropriate forcing. 

The class of linear filters that we consider is generically minimal in the sense that the state dimension of the filter coincides with the number of degrees of freedom in the finite-dimensional approximation of the linearized NS equations~\eqref{eq.lnse1}. The input to the filter represents a white-in-time excitation vector $\bw (\bk,t)$ with covariance $\Omega (\bk) \succ 0$. At each $\bk$ the filter dynamics are of the form
\begin{subequations}
\label{eq.filter}
\begin{eqnarray}
	   \dot{\bphi} (\bk, t)
	    & \; = \; &
	    A_f (\bk)
	    \,
	    \bphi (\bk,t)
	    \; + \;
	    B(\bk)
	    \,
	    \bw (\bk,t),
	    \\[0.05cm]
	    \bd (\bk, t)
	    & \; = \; &
	    C_f (\bk)
	    \,
	    \bphi (\bk,t)
	    \; + \;
	    \bw (\bk,t).
\end{eqnarray}
\end{subequations}
The generated output $ \bd (\bk, t)$ provides a suitable {\em colored-in-time\/} stochastic forcing to the linearized NS equations that reproduces the observed statistical signature of turbulent flow. As noted earlier, it is important to point out that white-in-time forcing to the linearized NS equations is often insufficient to explain the observed statistics.

The parameters of the filter are computed as follows
\begin{subequations}
\label{eq.filter-A-C}
\begin{eqnarray}
	A_f(\bk) 
	&\; = \;& 
	A(\bk) \,+\, B(\bk) \, C_f(\bk),
	\label{eq.filter-A}
	\\
	C_f(\bk)
	&\; = \; & 
	\left(
	H^*(\bk)  
	\; - \; 
	\dfrac{1}{2} \, \Omega (\bk) \, B^*(\bk) 
	\right)
	X^{-1}(\bk),
	\label{eq.filter-C}
	\end{eqnarray}
\end{subequations}
where the matrices $B(\bk)$ and $H(\bk)$ correspond to the factorization of $Z$ into $Z(\bk)=B(\bk) H^* (\bk) + H(\bk) B^*(\bk)$; see~\cite{chejovgeoCDC13, zarchejovgeoTAC16} for details. The spectral content of the excitation $\bd (\bk,t)$ is determined by the matrix-valued power spectral density
\be
	\Pi_f (\bk,\omega) 
	\; = \; 
	T_f(\bk,\omega)
	\,
	\Omega (\bk)
	\,
	T_f^* (\bk,\omega),
	\label{eq.power-spectrum}
\ee
where $T_f(\bk,\omega)$ is the frequency response of the filter, namely,
\be
	T_f(\bk,\omega)
	\; = \;
	C_f(\bk) \left( \mri \omega I 
	\, - \, 
	A_f(\bk) \right)^{-1} B(\bk) 
	\; + \;
	I,
	\label{eq.filter-tf}
\ee
and $I$ is the identity matrix. 

\begin{figure}
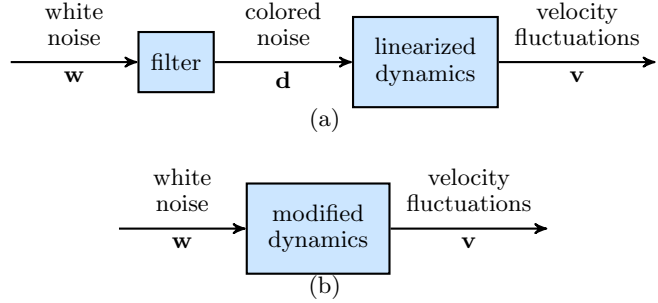

	\begin{center}
	\begin{tabular}{c}
		\subfigure[]{
%
%
%
%
%
%
\input{figures/Tikz_common_styles}
%
%
\noindent
\begin{tikzpicture}[scale=1, auto, >=stealth']
  
    \small

    
     \node[block, minimum height = .8cm, top color=RoyalBlue!20, bottom color=RoyalBlue!20] (sys1) {filter};
     
     \node[block, minimum height = 1.2cm, top color=RoyalBlue!20, bottom color=RoyalBlue!20] (sys2) at ($(sys1.east) + (2.8cm,0)$) {$\ba{c} \mbox{linearized} \\ \mbox{dynamics}\ea$};
     
     \node[] (output-node) at ($(sys2.east) + (2.2cm,0)$) {};
     
     \node[] (input-node) at ($(sys1.west) - (1.8cm,0)$) {}; 
      
%
%
%
%
%
%
%


    \draw [connector] (input-node) -- node [midway, above] {$\ba{c} \mbox{white} \\ \mbox{noise} \ea$} node [midway, below] {$\bw$} (sys1.west);
    \draw [connector] (sys2.east) -- node [midway, above] {$\ba{c} \mbox{velocity} \\ \mbox{fluctuations} \ea$} node [midway, below] {$\bv$} (output-node);
    \draw [connector] (sys1.east) -- node [midway, above] {$\ba{c} \mbox{colored} \\ \mbox{noise} \ea$} node [midway, below] {$\bd$} (sys2.west);
\end{tikzpicture}
		         \label{fig.filter}
		         }
		        \\
		\subfigure[]{
%
%
%
%
%
%
\input{figures/Tikz_common_styles}
%
%
\noindent
\begin{tikzpicture}[scale=1, auto, >=stealth']
  
    \small

    
     \node[block, minimum height = 1.2cm, top color=RoyalBlue!20, bottom color=RoyalBlue!20] (sys) {$\ba{c} \mbox{modified} \\ \mbox{dynamics}\ea$};
     
     \node[] (output-node) at ($(sys.east) + (2.2cm,0)$) {};
     
     \node[] (input-node) at ($(sys.west) - (1.8cm,0)$) {}; 
      
%
%
%
%
%
%
%


    \draw [connector] (input-node) -- node [midway, above] {$\ba{c} \mbox{white} \\ \mbox{noise} \ea$} node [midway, below] {$\bw$} (sys.west);
    \draw [connector] (sys.east) -- node [midway, above] {$\ba{c} \mbox{velocity} \\ \mbox{fluctuations} \ea$} node [midway, below] {$\bv$} (output-node);
\end{tikzpicture}
		         \label{fig.sys-modified}
		         }
	\end{tabular}
	\end{center}
	\caption{(a) Spatio-temporal filter~\eqref{eq.filter} is designed to provide colored stochastic input to the linearized NS equations~\eqref{eq.lnse1} in order to reproduce partially available second-order statistics of turbulent channel flow. The dynamics of this cascade connection are governed by the evolution model~\eqref{eq.cascade}; (b) An equivalent reduced-order representation of~\eqref{eq.cascade} is given by~\eqref{eq.feedback_dyn}.}
	\label{fig.filter_sys-feedback}
\end{figure}

As illustrated in figure~\ref{fig.filter}, the output $\bd (\bk, t)$ of the linear filter~\eqref{eq.filter} is the input to the linearized NS equations~\eqref{eq.lnse1}. This cascade connection can be represented via the evolution model
 \begin{subequations}
 	\label{eq.cascade}
 	\begin{eqnarray}
	\!\!\!\!\!\!
             	\tbo{\dot{\bpsi}(\bk,t)}{\dot{\bphi}(\bk,t)}
            	& = &
            	\tbt{A(\bk)}{B(\bk)\,C_f(\bk)}{0}{A(\bk) + B(\bk) \, C_f(\bk)}
            	\tbo{\bpsi(\bk,t)}{\bphi(\bk,t)}
            	\, + \,
            	\tbo{B(\bk)}{B(\bk)}
		\bw(\bk,t)
		\\[0.05cm]
		\!\!\!\!\!\!
		\bv(\bk,t)
		& = &
		\obt{C(\bk)}{0}
		\tbo{\bpsi(\bk,t)}{\bphi(\bk,t)},
	\end{eqnarray}
 \end{subequations}
which has twice as many degrees of freedom as the spatial discretization of the original linearized NS model. 
{
As shown by~\cite{zarchejovgeoTAC16}, due to the presence of uncontrollable modes in~\eqref{eq.cascade}, the coordinate transformation
	\beq
	\label{eq.psi-phi-chi}
	\tbo{\bpsi(\bk,t)}{\bchi(\bk,t)}
	\; = \;
	\tbt{\phantom{-}I}{0}{-I}{I}
	\tbo{\bpsi(\bk,t)}{\bphi(\bk,t)},
	\eeq
can be used to bring system~\eqref{eq.cascade} into the following form
	\beq             	
	\ba{rcl}
	\tbo{\dot{\bpsi}(\bk,t)}{\dot{\bchi}(\bk,t)}
            	& \!\!\! = \!\!\!  &
            	\tbt{A(\bk) + B(\bk) \, C_f(\bk)}{B(\bk)\,C_f(\bk)}{0}{A(\bk)}
            	\tbo{\bpsi(\bk,t)}{\bchi(\bk,t)}
            	\, + \,
            	\tbo{B(\bk)}{0}
		\bw(\bk,t)
		\\[0.35cm]
		\bv(\bk,t)
		& \!\!\! = \!\!\!  &
		\obt{C(\bk)}{0}
		\tbo{\bpsi(\bk,t)}{\bchi(\bk,t)}.
	\ea
 	\eeq
Clearly, the input $\bw (\bk,t)$ does not enter into the equation that governs the evolution of $\bchi (\bk,t)$. Thus, the reduced-order representation
\be
	\label{eq.feedback_dyn}
	\ba{rcl}
        \dot{\bpsi} 
        (\bk, t)
        & \!\! = \!\! &
        \left( A(\bk) \, + \, B(\bk) \, C_f(\bk) \right) 
        \bpsi (\bk, t)
        \;+\; 
        B(\bk) \, \bw (\bk, t),
        \\[0.15cm]
        \bv(\bk,t)
	& \!\! = \!\! &
	C(\bk) \,
	\bpsi(\bk,t),
	\ea
\ee
which has the same number of degrees of freedom as~\eqref{eq.lnse1}, completely captures the influence of $\bw (\bk,t)$ on $\bpsi (\bk,t)$; see figure~\ref{fig.sys-modified} for an illustration. Furthermore, stability of $A(\bk) + B(\bk) C_f(\bk)$ (see remark~\ref{rem.stability}) implies that the initial conditions $\bpsi(\bk,0)$ and $\bphi(\bk,0)$ only influence the transient response and do not have any impact on the steady-state statistics. The corresponding algebraic Lyapunov equation in conjunction with~\eqref{eq.filter-C} yields
\be
	\label{eq.modified-dyn-lyap}
	\ba{l}
	\left( A \, + \, B\,  C_f \right) X 
	\;+\;
	X \left( A \, + \, B \, C_f \right) ^*
	\;+\;
	B\,\Omega\, B^*
	\\[.15cm]
	\hspace{1cm}
	=\;
	A\,X \;+\, X\,A^* \,+\, B\, \Omega\, B^* \,+\, B\,C_f\,X \,+\, X\, C_f^*\,B^*
	\\[.15cm]
	\hspace{1cm}
	=\;
	A\,X \;+\; X\,A^* \;+\; B\,H^* \;+\; H\, B^*
	\\[.15cm]
	\hspace{1cm}
	=\; 0,
	\ea
\ee
which shows that~\eqref{eq.filter} generates a stochastic process $\bd(\bk,t)$ that is consistent with $X(\bk)$.} In what follows, without loss of generality, we choose the covariance matrix of the white noise $\bw (\bk,t)$ to be the identity matrix, $\Omega = I$.

\begin{figure}
	\begin{center}
%
%
%
%
%
%
\input{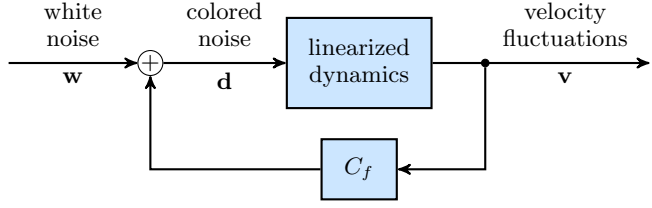}
%
%
\noindent
\begin{tikzpicture}[scale=1, auto, >=stealth']
  
    \small

    
     \node[block, minimum height = 1.2cm, top color=RoyalBlue!20, bottom color=RoyalBlue!20] (sys1) {$\ba{c} \mbox{linearized} \\ \mbox{dynamics}\ea$};
     
     \node[block, minimum height = .8cm, top color=RoyalBlue!20, bottom color=RoyalBlue!20] (sys4) at ($(sys1.south) - (0cm,.8cm)$) {$C_f$};
     
     \node[] (input-node) at ($(sys1.west) - (3.8cm,0)$) {}; 
     
     \node[] (output-node) at ($(sys1.east) + (3.0cm,0)$) {};
     
     \node[] (mid-node1) at ($(sys1.east) + (1cm,1cm)$) {}; 
          
     \node[sum] (esum1) at ($(sys1.west) - (1.8cm,0)$) {$+$};
           
     \node[branch] (R) at ($(sys1.east) + (.7cm,0.0cm)$){};
     
	

	
    \draw [connector] (input-node) -- node [midway, above] {$\ba{c} \mbox{white} \\ \mbox{noise} \ea$} node [midway, below] {$\bw$} (esum1.west);
                    	
    \draw [line] (sys1.east) -- (R);
    \draw [connector] (R.west) -- node [midway, above] {$\ba{c} \mbox{velocity} \\ \mbox{fluctuations} \ea$} node [midway, below] {$\bv$} (output-node);
    
    \draw [connector] (R.south) |- (sys4.east);
    
    \draw [connector] (sys4.west) -| (esum1.south);
    
    \draw [connector] (esum1.east) -- node [midway, above] {$\ba{c} \mbox{colored} \\ \mbox{noise} \ea$} node [midway, below] {$\bd$} (sys1.west);
    
    
\end{tikzpicture}
	\end{center}
	\caption{An equivalent feedback representation of the cascade connection in figure~\ref{fig.filter}.}
	\label{fig.sys-feedback-interp}
\end{figure}

\begin{remark}{\em The compact representation~\eqref{eq.feedback_dyn} allows for alternative interpretations of colored-in-time forcing and, at the same time, offers advantages from a computational standpoint. First, the structure of \eqref{eq.feedback_dyn} suggests that the colored-in-time forcing realized by \eqref{eq.filter} can be equivalently interpreted as a modification of the dynamical generator of the linearized NS equations due to state-feedback interactions; see figure~\ref{fig.sys-feedback-interp}.
This interpretation allows seeking suitable ``feedback gains'' $C_f(\bk)$ that may now be optimal with respect to alternative design criteria~\citep*{zarjovgeoCDC16}.
Moreover, the term $B(\bk)C_f(\bk)$ can be seen as a low-rank modification of the dynamical generator $A(\bk)$ of the linearized NS equations.
Finally, time-domain simulations require numerical integration of system~\eqref{eq.feedback_dyn} which has half the number of states as compared to system~\eqref{eq.cascade}, thereby offering computational speedup.}
\end{remark}

	\begin{remark}
{\em Important aspects of the underlying physics may be obscured when the forcing is allowed to excite all degrees of freedom in the linearized model. As discussed above, if the nuclear norm of $Z = BH^* + H B^*$ is not accounted for in~\eqref{eq.CP}, the resulting input matrix $B$ will be of full rank. In this case, without loss of generality, we can choose $B = I$ which simplifies equation~\eqref{eq.lyap_BH},
	\be
	A (\bk) \, X (\bk)
	\; + \; 
	X (\bk) \,A^* (\bk) 
	\;=\; 
	-H^* (\bk)
	\; - \;
	H (\bk).
	\label{eq.lyap_B_identity}
	\ee
Clearly, this equation is satisfied with $H^* (\bk) = -A (\bk) \, X (\bk)$. With this choice of $H (\bk)$, the reduced-order representation~\eqref{eq.feedback_dyn} is given by
	\be
	\dot{\bpsi} (\bk, t)
        ~ = \;
        - \dfrac{1}{2}
        \,
        X^{-1} (\bk)
        \, \bpsi (\bk, t)
        \;+\; 
        \bw (\bk, t).
        \ee
This demonstrates that colored-in-time forcing of the linearized NS equations which excites all degrees of freedom can lead to the complete cancelation of the linearized dynamical generator $A (\bk)$. It is thus crucial to restrict the number of input channels via the nuclear norm penalty in the objective function of optimization problem~\eqref{eq.CP}.}
	\label{rem.obscure}
	\end{remark}

\begin{remark}
{{\em It is known that the linearized NS equations around the turbulent mean velocity profile are stable~\citep{mal56,reytie67}. Interestingly and independently of this fact, the modified dynamical generator, $A(\bk) + B(\bk) C_f(\bk)$ in~\eqref{eq.feedback_dyn}, can be shown to be stable by standard Lyapunov theory. More specifically, substituting the expression for the matrix $H(\bk)$ from~\eqref{eq.filter-C} into equation~\eqref{eq.lyap_BH} yields
    \[
    	\left(A(\bk)+B(\bk)\,C_f(\bk)\right) X(\bk) \;+\; X(\bk) \left(A(\bk)+B(\bk)\,C_f(\bk)\right)^*
	\;=\;
	-B(\bk)\,\Omega(\bk)\, B^*(\bk).
    \]
This is a standard Lyapunov equation. 
Since $(A, B)$ is a controllable pair, so is $(A + B\,C_f, B)$, and therefore $(A + B\,C_f, B \, \Omega^{1/2})$ is controllable as well.
Standard Lyapunov theory implies that the positive semi-definiteness of $B(\bk)\Omega(\bk)B^*(\bk)$ is sufficient to guarantee that all eigenvalues of $A(\bk) + B(\bk)C_f(\bk)$ are in the left-half of the complex plane.}}
	\label{rem.stability}
	\end{remark}

	
\section{Application to turbulent channel flow}
    \label{sec.application_turbchannel}

In this section, we utilize the modeling and optimization framework developed in \S~\ref{sec.mcp} to account for partially observed second-order statistics of turbulent channel flow. In our setup, the mean velocity profile and one-point velocity correlations in the wall-normal direction at various wavenumber pairs $\bk$ are obtained from DNS with $R_\tau=186$~\citep{kimmoimos87,moskimman99,deljim03,deljimzanmos04}; see figure~\ref{fig.output_covariance} for an illustration. We show that stochastically-forced linearized NS equations can be used to exactly reproduce the available statistics and to complete unavailable two-point correlations of the turbulent velocity field. The {\em colored-in-time\/} forcing with the {\em identified\/} power spectral density is generated by linear filters that introduce low-rank perturbations to the linearization around turbulent mean velocity; cf.~\eqref{eq.feedback_dyn}. As a result of this modification to the linearized NS equations, all one-point correlations are {\em perfectly matched\/} and the one-dimensional energy spectra is {\em completely reconstructed\/}. In addition, we show that two-point velocity correlations compare favorably with the result of DNS. As aforementioned, the modified dynamics that result from our modeling framework have the same number of degrees of freedom as the finite-dimensional approximation of the linearized NS dynamics and are thus convenient for the purpose of conducting linear stochastic simulations. We utilize these simulations to verify the ability of our model to account for the statistical signatures of turbulent channel flow. Finally, we close this section by showing that our framework can be also used to capture the velocity correlations at higher Reynolds numbers.

\subsection{Necessity for the colored-in-time forcing}
    \label{sec.color_necessity}

	For homogeneous isotropic turbulence,~\cite{jovgeoAPS10} showed that the steady-state velocity correlation matrices can be exactly reproduced by the linearized NS equations. This can be achieved with white-in-time solenoidal forcing whose second-order statistics are proportional to the turbulent energy spectrum~\cite[for additional details see][Appendix~C]{rashad-phd12}. For turbulent channel flow, however, we next show that the matrix
    $
    A (\bk) X_{\mathrm{dns}} (\bk)
    +
    X_{\mathrm{dns}} (\bk) A^* (\bk)
    $
can fail to be negative semi-definite for numerically-generated covariances $X_{\mathrm{dns}} (\bk)$ of the state $\bpsi$. Here, $A (\bk)$ is the generator of the linearized dynamics around the turbulent mean velocity profile and $X_{\mathrm{dns}} (\bk)$ is the steady-state covariance matrix resulting from DNS of turbulent channel flow. 

\begin{figure}
	\begin{center}
	\begin{tabular}{rc}
	\begin{tabular}{c}
	\vspace{4.8cm}
	\normalsize{\rotatebox{90}{
	$
	\lambda_i \left(
	A \, X_{\mathrm{dns}} \, + \, X_{\mathrm{dns}} \, A^* 
	\right)$}}
	\end{tabular}
	&
	\hspace{-.4cm}
        \includegraphics[width=6.5cm]{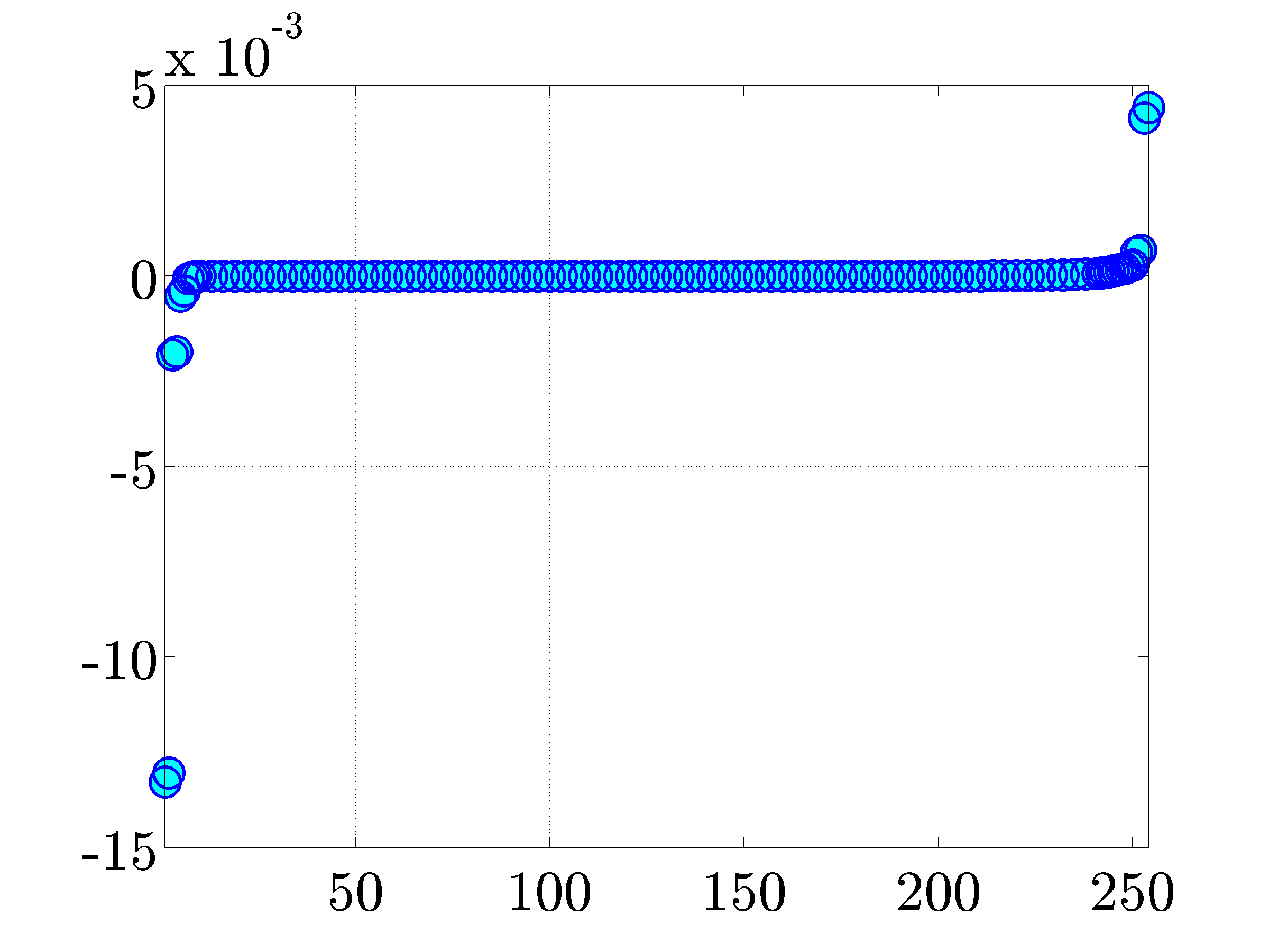}
        \\[-4cm] 
        &
        \normalsize{$i$}
        \end{tabular} 
	\end{center}
\caption{Positive eigenvalues of the matrix 
	$
    A (\bk) X_{\mathrm{dns}} (\bk)
    +
    X_{\mathrm{dns}} (\bk) A^* (\bk),
    $
for channel flow with $R_\tau=186$ and $\bk = (2.5,7)$, indicate that turbulent velocity covariances cannot be reproduced by the linearized NS equations with white-in-time stochastic forcing; cf.\ equation~\eqref{eq.standard_lyap}.}
\label{fig.Forcing_evalues}
\end{figure}

Figure~\ref{fig.Forcing_evalues} shows the eigenvalues of the matrix 
	$
    A (\bk) X_{\mathrm{dns}} (\bk)
    +
    X_{\mathrm{dns}} (\bk) A^* (\bk)
    $
for channel flow with $R_\tau=186$ and $\bk = (2.5,7)$. The presence of both positive and negative eigenvalues indicates that the second-order statistics of turbulent channel flow cannot be exactly reproduced by the linearized NS equations with white-in-time stochastic excitation. As we show in the next subsection, this limitation can be overcome by departing from the white-in-time restriction.
	
\subsection{Reproducing available and completing unavailable velocity correlations}
\label{sec.feasibility}

We next employ the optimization framework of \S~\ref{sec.mcp} to account for second-order statistics of turbulent channel flow with $R_\tau=186$ via a low-complexity model. We use $N=127$ collocation points in the wall-normal direction and show that all one-point velocity correlations can be exactly reproduced using the linearized NS equations with colored-in-time forcing. {Grid convergence is ensured by doubling the number of collocation points.} In addition, we demonstrate that an appropriate choice of the regularization parameter $\gamma$ provides good completion of two-point correlations that are {\em not\/} used as problem data in optimization problem~\eqref{eq.CP}. Appendix~\ref{sec.gamma} offers additional insight into the influence of this parameter on the quality of completion.

\begin{figure}
	\begin{tabular}{cccc}
		\hspace{.15cm}\subfigure[]{\label{fig.Averaged_uu}}&&\hspace{-.4cm}\subfigure[]{\label{fig.Averaged_uv}}
		\\[-.8cm]
		&
		\hspace{-.04cm}
		\begin{tabular}{c}
			\includegraphics[width=6.15cm]{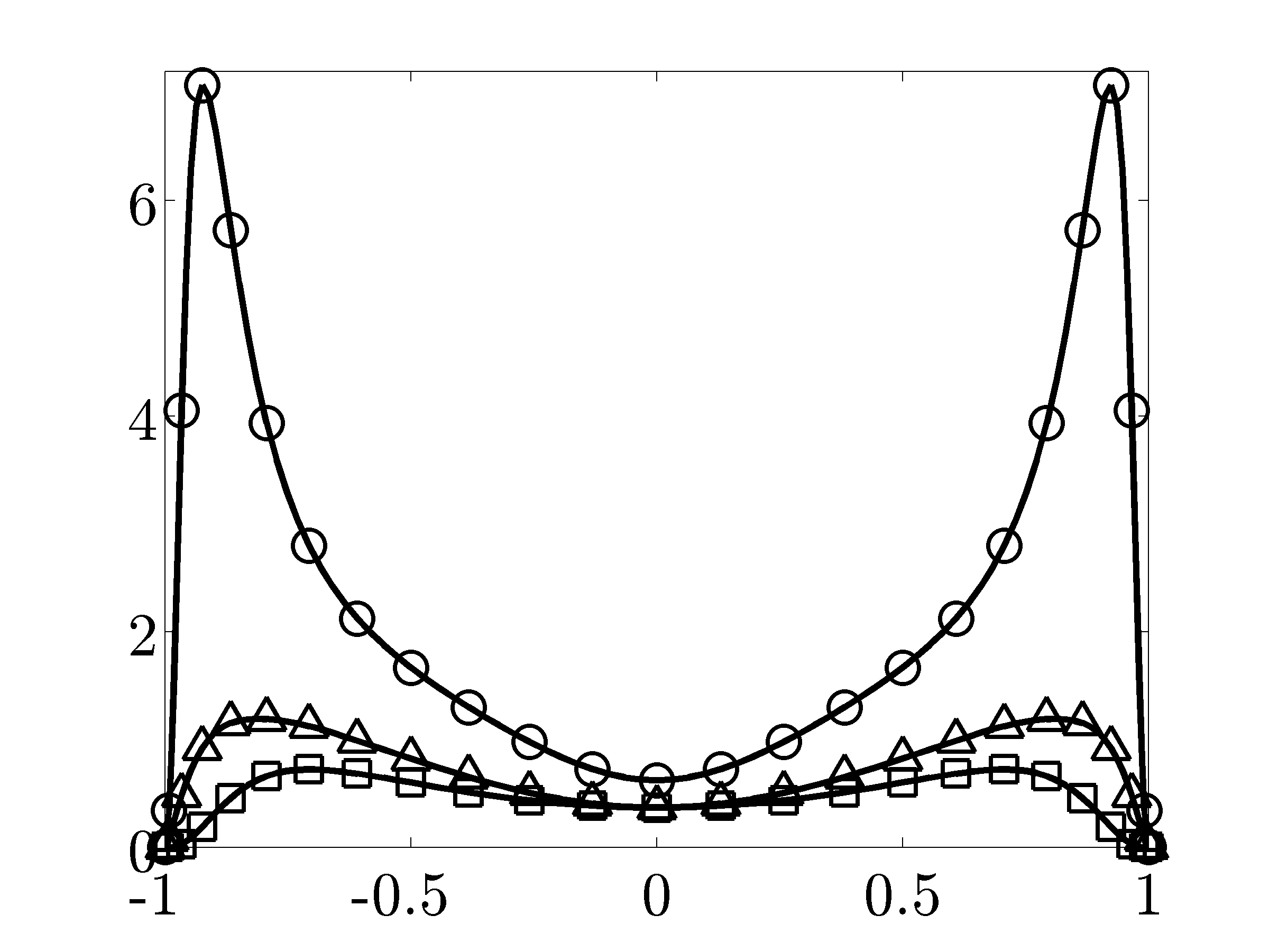}
			\\[-.1cm]
			\hspace{.1cm}
			{\normalsize $y$}
		\end{tabular}
		&&
		\hspace{-.2cm}
		\begin{tabular}{c}
			\includegraphics[width=6.15cm]{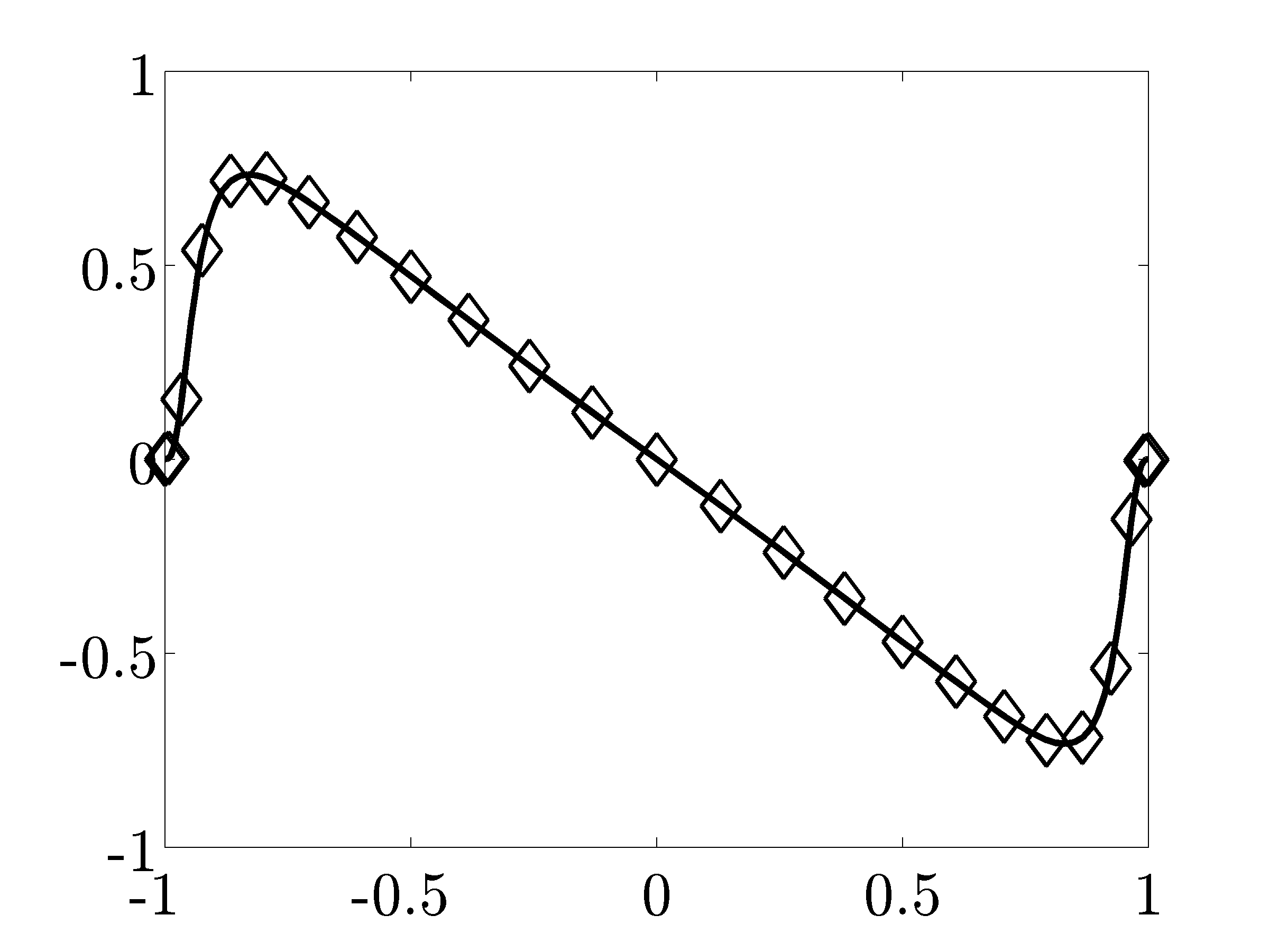}
			\\[-.1cm]
			\hspace{.1cm}
			{\normalsize $y$}
		\end{tabular}
	\end{tabular}
	\caption{(a) Correlation profiles of normal and (b) shear stresses resulting from DNS of turbulent channel flow with $R_\tau=186$ (--) and from the solution to~\eqref{eq.CP}; $uu$ ($\Circle$), $vv$ ($\Box$), $ww$ ($\triangle$), $-uv$ ($\Diamond$). {We observe perfect matching of all one-point velocity correlations that result from integration over wall-parallel wavenumbers. Note: plot markers are sparse for data presentation purposes and do not indicate grid resolution.}
	}
	\label{fig.averaged_profiles}
\end{figure}

\begin{figure*}
	\begin{tabular}{cccc}
	\hspace{-.5cm} \subfigure[]{\label{fig.E_uu_lxyp_Re186}}
	& \normalsize{$k_x E_{uu} (y^+, \lambda_x^+)$} &
	\hspace{-0.8cm} \subfigure[]{\label{fig.E_uu_lzyp_Re186}}
	&
	\normalsize{$k_z E_{uu} (y^+, \lambda_z^+)$}
	\\[-.38cm]
	\begin{tabular}{c}
		\vspace{.4cm}
		\hspace{-.1cm}
		\normalsize{$\lambda_x^+$}
	\end{tabular}
	&
	\hspace{-.5cm}
	\begin{tabular}{c}
		\includegraphics[width=6.15cm, height=4.1cm]{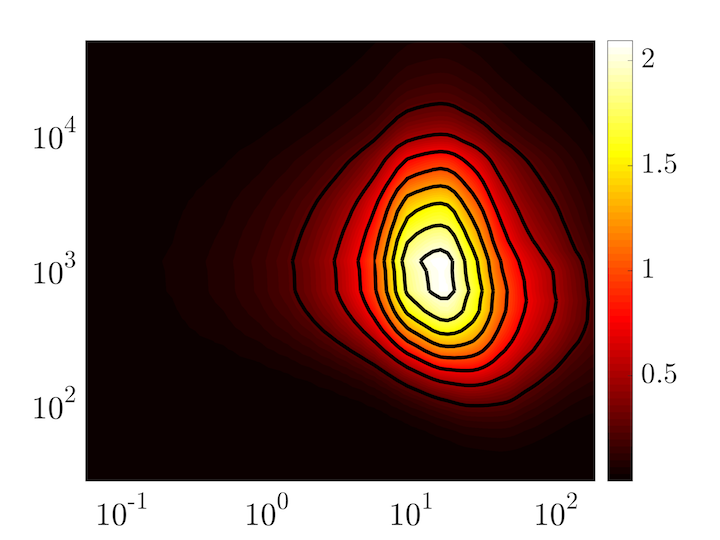}
	\end{tabular}
	&
	\begin{tabular}{c}
		\vspace{.4cm}
		\hspace{-.4cm}
		\normalsize{$\lambda_z^+$}
	\end{tabular}
	&
	\hspace{-.5cm}
	\begin{tabular}{c}
		\includegraphics[width=6.15cm, height=4.1cm]{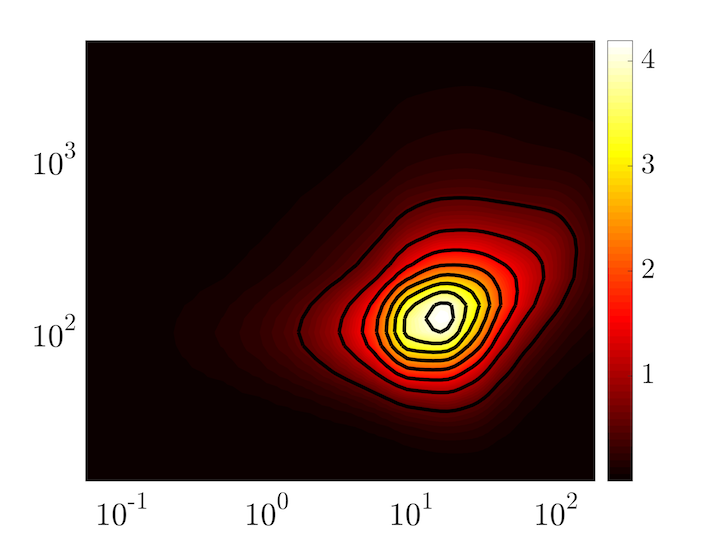}
	\end{tabular}
	\\[2cm]
	\hspace{-.5cm} \subfigure[]{\label{fig.E_vv_lxyp_Re186}}
	& \normalsize{$k_x E_{vv} (y^+, \lambda_x^+)$} &
	\hspace{-.8cm} \subfigure[]{\label{fig.E_vv_lzyp_Re186}}
	&
	\normalsize{$k_z E_{vv} (y^+, \lambda_z^+)$}
	\\[-.39cm]
	\begin{tabular}{c}
		\vspace{.4cm}
		\hspace{-.1cm}
		\normalsize{$\lambda_x^+$}
	\end{tabular}
	&
	\hspace{-.5cm}
	\begin{tabular}{c}
		\includegraphics[width=6.15cm, height=4.1cm]{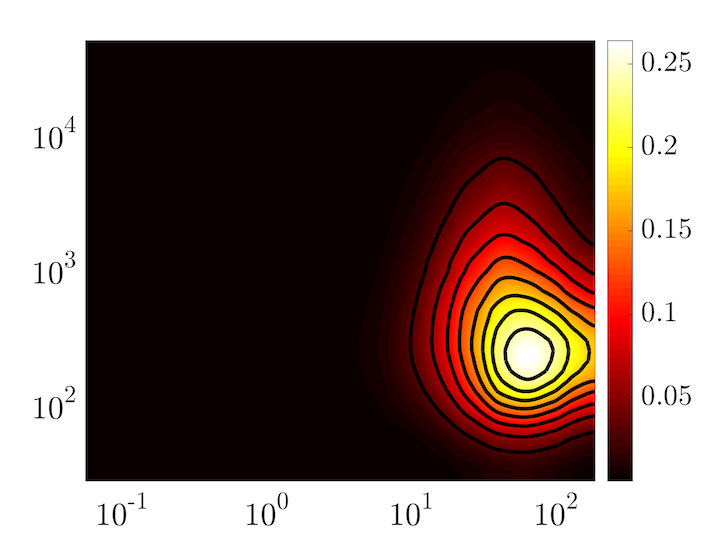}
	\end{tabular}
	&
	\begin{tabular}{c}
		\vspace{.4cm}
		\hspace{-.4cm}
		\normalsize{$\lambda_z^+$}
	\end{tabular}
	&
	\hspace{-.5cm}
	\begin{tabular}{c}
		\includegraphics[width=6.15cm, height=4.1cm]{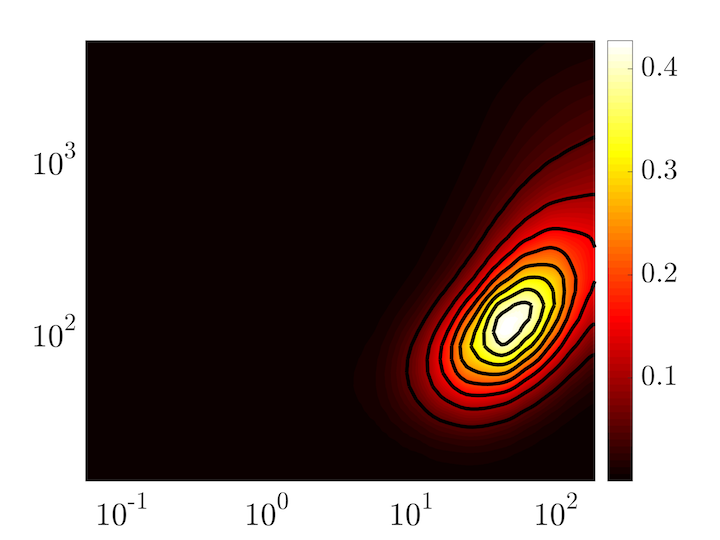}
	\end{tabular}
	\\[2cm]
	\hspace{-.5cm} \subfigure[]{\label{fig.E_ww_lxyp_Re186}}
	& \normalsize{$k_x\, E_{ww} (y^+, \lambda_x^+)$} &
	\hspace{-.8cm} \subfigure[]{\label{fig.E_ww_lzyp_Re186}}
	&
	\normalsize{$k_z E_{ww} (y^+, \lambda_z^+)$}
	\\[-.39cm]
	\begin{tabular}{c}
		\vspace{.4cm}
		\hspace{-.1cm}
		\normalsize{$\lambda_x^+$}
	\end{tabular}
	&
	\hspace{-.5cm}
	\begin{tabular}{c}
		\includegraphics[width=6.15cm, height=4.1cm]{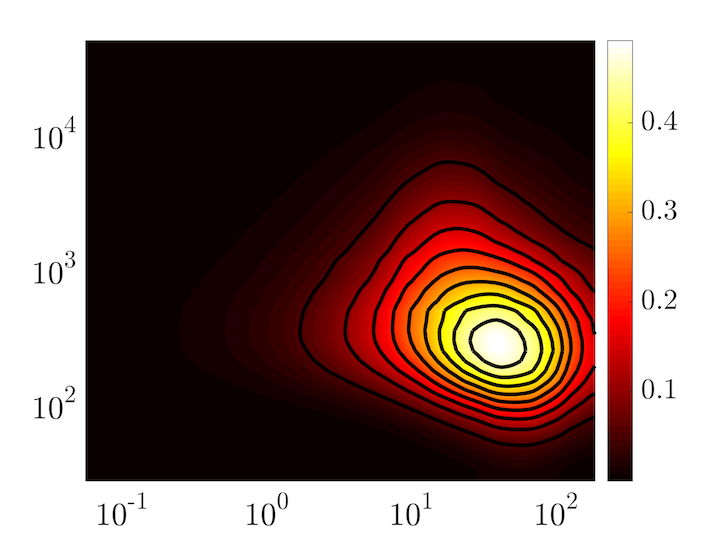}
	\end{tabular}
	&
	\begin{tabular}{c}
		\vspace{.4cm}
		\hspace{-.4cm}
		\normalsize{$\lambda_z^+$}
	\end{tabular}
	&
	\hspace{-.5cm}
	\begin{tabular}{c}
		\includegraphics[width=6.15cm, height=4.1cm]{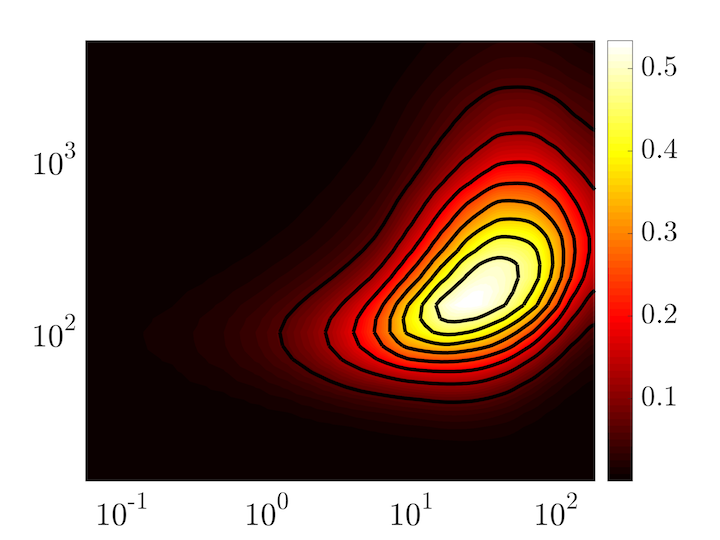}
	\end{tabular}
	\\[2cm]
	\hspace{-.5cm} \subfigure[]{\label{fig.E_uv_lxyp_Re186}}
	& \normalsize{$k_x\, E_{uv} (y^+, \lambda_x^+)$} &
	\hspace{-.8cm} \subfigure[]{\label{fig.E_uv_lzyp_Re186}}
	&
	\normalsize{$k_z\, E_{uv} (y^+, \lambda_z^+)$}
	\\[-.39cm]
	\begin{tabular}{c}
		\vspace{.4cm}
		\hspace{-.1cm}
		\normalsize{$\lambda_x^+$}
	\end{tabular}
	&
	\hspace{-.5cm}
	\begin{tabular}{c}
		\includegraphics[width=6.15cm, height=4.1cm]{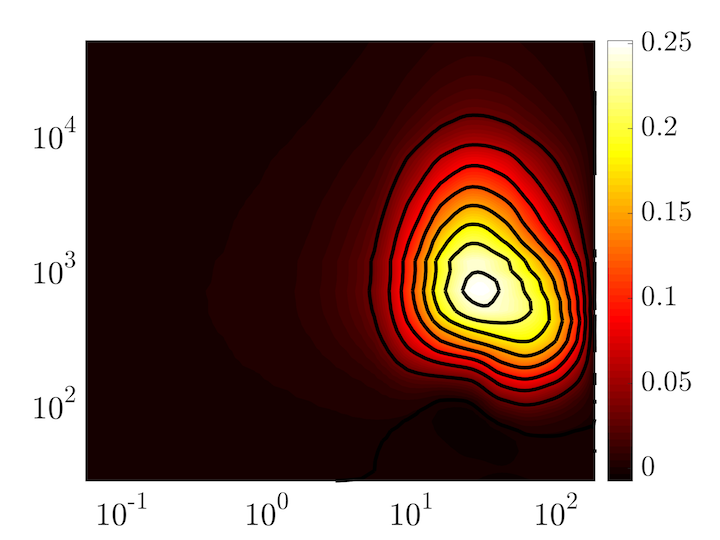}
		\\
		\normalsize{$y^+$}
	\end{tabular}
	&
	\begin{tabular}{c}
		\vspace{.4cm}
		\hspace{-.4cm}
		\normalsize{$\lambda_z^+$}
	\end{tabular}
	&
	\hspace{-.5cm}
	\begin{tabular}{c}
		\includegraphics[width=6.15cm, height=4.1cm]{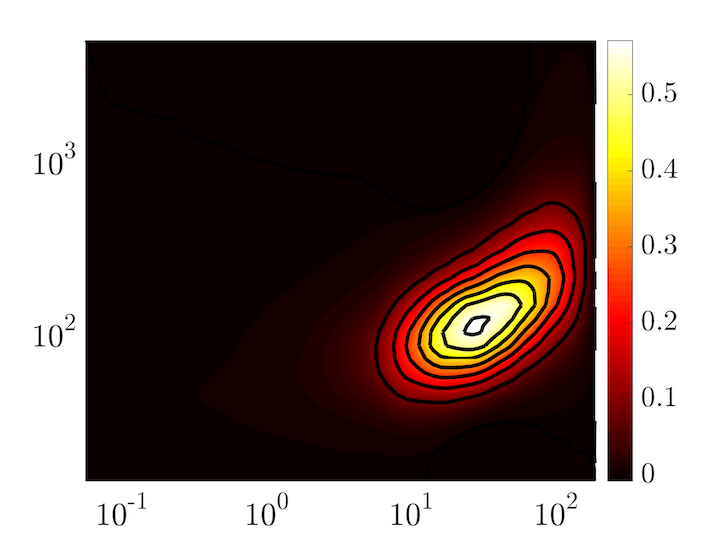}
		\\
		\normalsize{$y^+$}
	\end{tabular}
	\end{tabular}
	\caption{Pre-multiplied one-dimensional energy spectrum of streamwise (a,b), wall-normal (c,d), spanwise (e,f) velocity fluctuations, and the Reynolds stress co-spectrum (g,h) in terms of streamwise (left) and spanwise (right) wavelengths and the wall-normal coordinate (all in inner units). Color plots: DNS-generated spectra of turbulent channel flow with $R_\tau=186$. Contour lines: spectra resulting from the solution to~\eqref{eq.CP}.}
	\label{fig.velocity_corr_spectrum}
\end{figure*}

Figures~\ref{fig.averaged_profiles} and~\ref{fig.velocity_corr_spectrum} show that the solution to optimization problem~\eqref{eq.CP} {\em exactly reproduces\/} available one-point velocity correlations resulting from DNS at various wavenumbers. At each $\bk$, the constraint~\eqref{eq.constraint_observations} restricts all feasible solutions of problem~\eqref{eq.CP} to match available one-point correlations. Our computational experiments demonstrate feasibility of optimization problem~\eqref{eq.CP} at each $\bk$. Thus, regardless of the value of the regularization parameter $\gamma$, {\em all available one-point correlations\/} of turbulent flow can be {\em recovered\/} by a stochastically-forced linearized model. 

Figures~\ref{fig.Averaged_uu} and~\ref{fig.Averaged_uv} display perfect matching of all one-point velocity correlations that result from integration over wall-parallel wavenumbers. Since problem~\eqref{eq.CP} is not feasible for $Z \succeq 0$, this cannot be achieved with white-in-time stochastic forcing; see \S~\ref{sec.color_necessity}. In contrast, colored-in-time forcing enables recovery of the one dimensional energy spectra of velocity fluctuations resulting from DNS; in figure~\ref{fig.velocity_corr_spectrum}, pre-multiplied spectra are displayed as a function of the wall-normal coordinate, streamwise (left plots), and spanwise (right plots) wavelengths. All of these are given in inner (viscous) units with $y^+ = R_\tau(1+y)$, $\lambda_x^+ = 2 \pi R_\tau/k_x$, and $\lambda_z^+ = 2 \pi R_\tau/k_z$. 

Our results should be compared and contrasted to~\cite{moajovtroshamckPOF14}, {where} a gain-based low-order decomposition {was} used to  approximate the velocity spectra of turbulent channel flow. Twelve optimally-weighted resolvent modes approximated the Reynolds shear stress, streamwise, wall-normal, and spanwise intensities with $25\%$, $20\%$, $17\%$, and $6\%$ error, respectively. While the results presented here are at a lower Reynolds number ($R_\tau=186$ vs.\ $R_\tau=2003$), our computational experiments demonstrate feasibility of optimization problem~\eqref{eq.CP} at all wavenumber pairs. Thus, all one-point correlations can be perfectly matched with colored-in-time stochastic forcing. As we show in \S~\ref{sec.highRe}, this holds even at higher Reynolds numbers.

\begin{figure*}
		\begin{tabular}{cccc}
	\hspace{-.4cm} \subfigure[]{\label{fig.Ruu_DNS}}
	& \normalsize{$\Phi_{uu, \mathrm{dns}}$} &
	\hspace{-0.2cm} \subfigure[]{\label{fig.Ruu_NM}}
	&
	\normalsize{$\Phi_{uu}$}
	\\[-.4cm]
	\begin{tabular}{c}
		\vspace{.4cm}
		\hspace{-.1cm}
		\normalsize{$y$}
	\end{tabular}
	&
	\hspace{-.4cm}
	\begin{tabular}{c}
		\includegraphics[width=6.15cm, height=4.15cm]{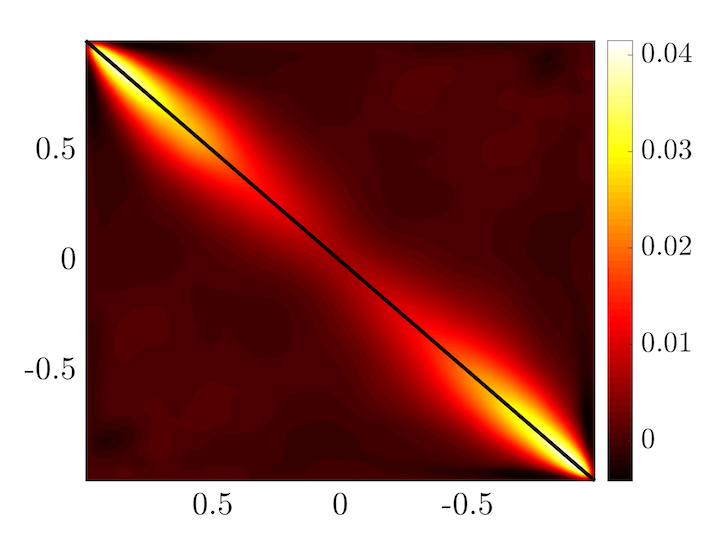}
	\end{tabular}
	&
	&
	\hspace{-.05cm}
	\begin{tabular}{c}
		\includegraphics[width=6.15cm, height=4.15cm]{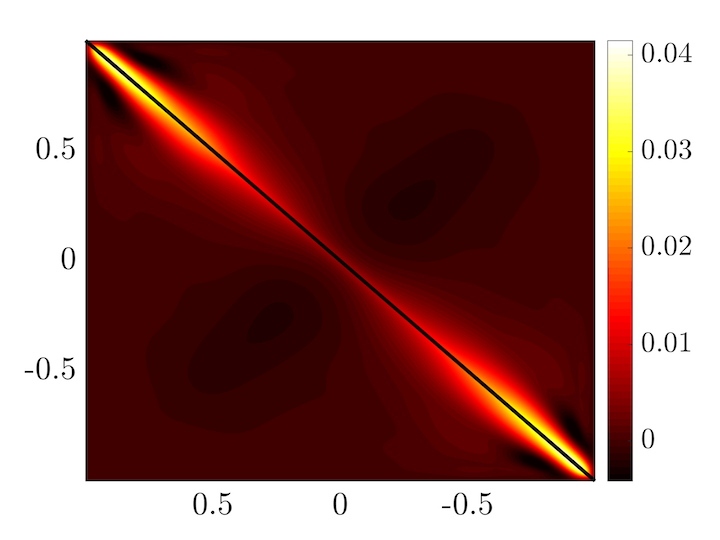}
	\end{tabular}
	\\[2cm]
	\hspace{-.4cm} \subfigure[]{\label{fig.Rvv_DNS}}
	& \normalsize{$\Phi_{vv, \mathrm{dns}}$} &
	\hspace{-.2cm} \subfigure[]{\label{fig.Rvv_NM}}
	&
	\normalsize{$\Phi_{vv}$}
	\\[-.4cm]
	\begin{tabular}{c}
		\vspace{.4cm}
		\hspace{-.1cm}
		\normalsize{$y$}
	\end{tabular}
	&
	\hspace{-.4cm}
	\begin{tabular}{c}
		\includegraphics[width=6.15cm, height=4.15cm]{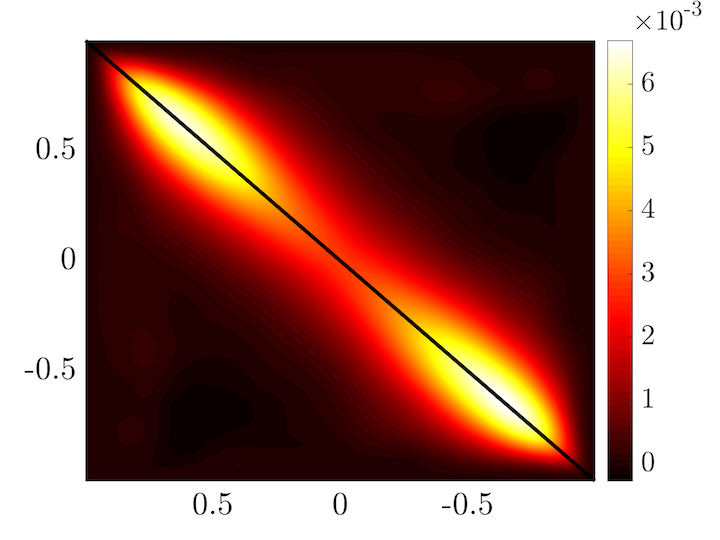}
	\end{tabular}
	&
	&
	\hspace{-.05cm}
	\begin{tabular}{c}
		\includegraphics[width=6.15cm, height=4.15cm]{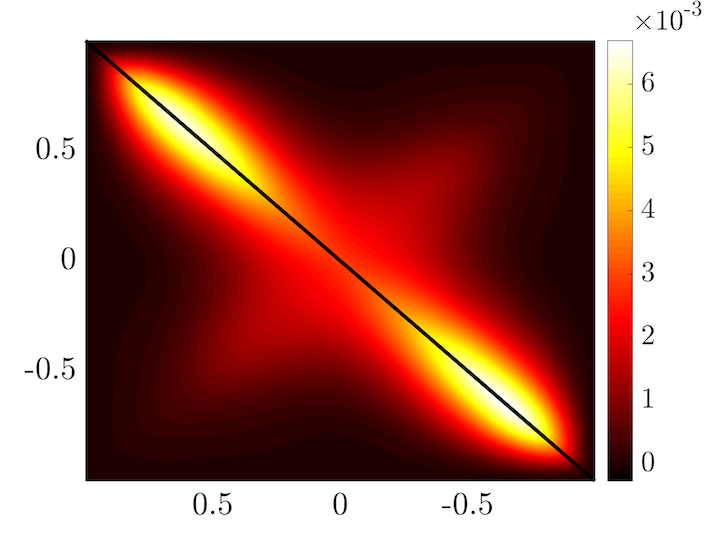}
	\end{tabular}
	\\[2cm]
	\hspace{-.4cm} \subfigure[]{\label{fig.Rww_DNS}}
	& \normalsize{$\Phi_{ww, \mathrm{dns}}$} &
	\hspace{-.2cm} \subfigure[]{\label{fig.Rww_NM}}
	&
	\normalsize{$\Phi_{ww}$}
	\\[-.4cm]
	\begin{tabular}{c}
		\vspace{.4cm}
		\hspace{-.1cm}
		\normalsize{$y$}
	\end{tabular}
	&
	\hspace{-.4cm}
	\begin{tabular}{c}
		\includegraphics[width=6.15cm, height=4.15cm]{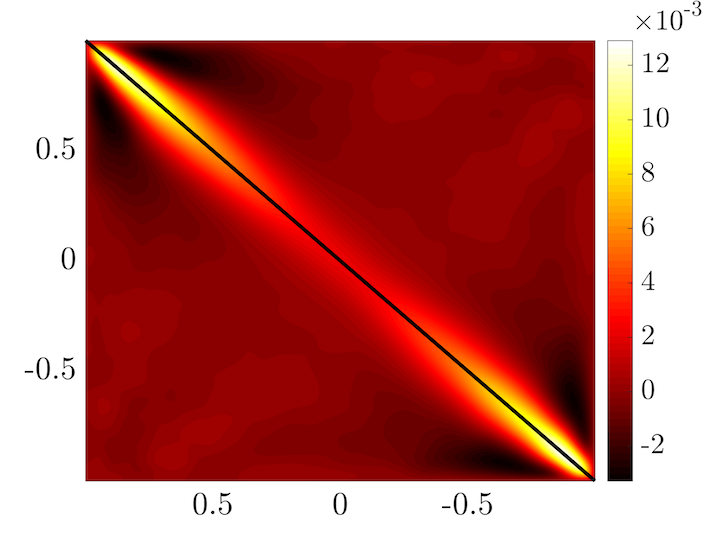}
	\end{tabular}
	&
	&
	\hspace{-.05cm}
	\begin{tabular}{c}
		\includegraphics[width=6.15cm, height=4.5cm]{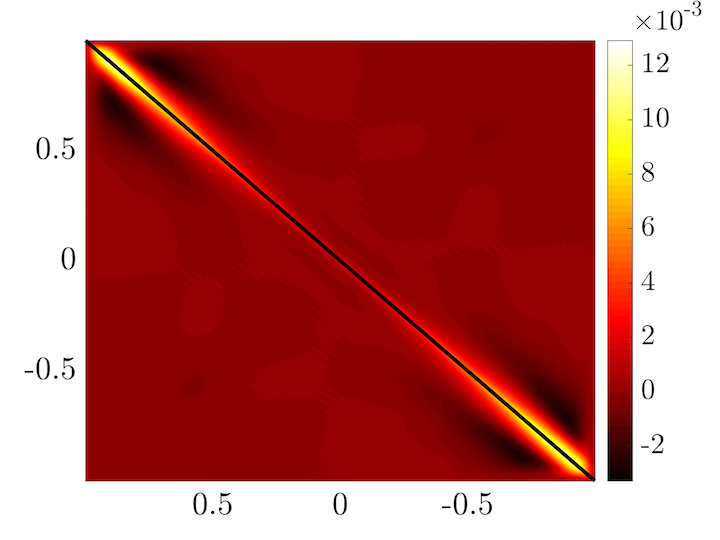}
	\end{tabular}
	\\[2cm]
	\hspace{-.4cm} \subfigure[]{\label{fig.Ruv_DNS}}
	& \normalsize{$\Phi_{uv, \mathrm{dns}}$} &
	\hspace{-.2cm} \subfigure[]{\label{fig.Ruv_NM}}
	&
	\normalsize{$\Phi_{uv}$}
	\\[-.4cm]
	\begin{tabular}{c}
		\vspace{.4cm}
		\hspace{-.1cm}
		\normalsize{$y$}
	\end{tabular}
	&
	\hspace{-.4cm}
	\begin{tabular}{c}
		\includegraphics[width=6.15cm, height=4.15cm]{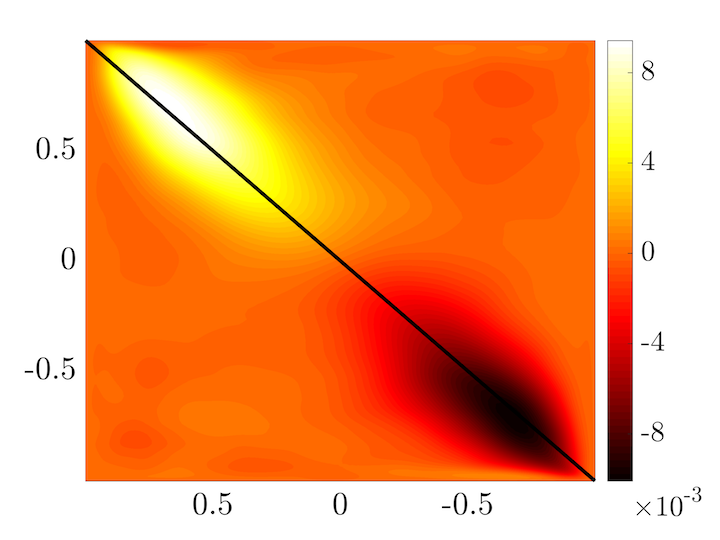}
		\\
		\normalsize{$y$}
	\end{tabular}
	&
	&
	\hspace{-.05cm}
	\begin{tabular}{c}
		\includegraphics[width=6.15cm, height=4.15cm]{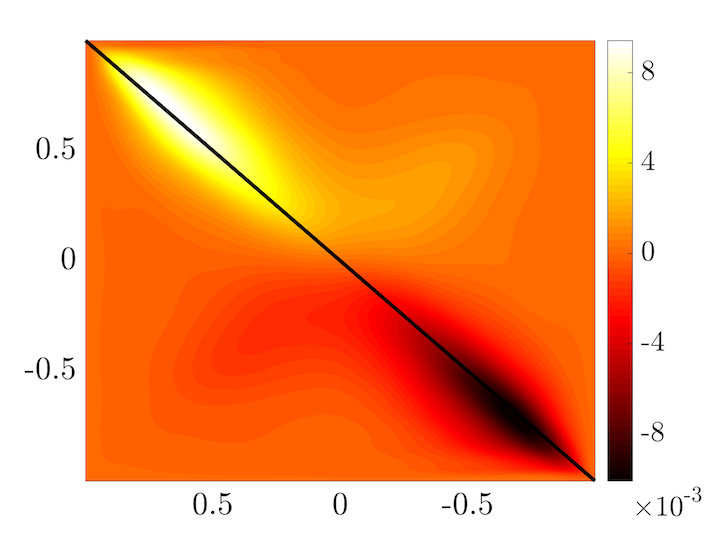}
		\\
		\normalsize{$y$}
	\end{tabular}
	\end{tabular}
		\caption{Covariance matrices resulting from DNS of turbulent channel flow with $R_\tau=186$ (left plots); and the solution to optimization problem~\eqref{eq.CP} with $\gamma = 300$ (right plots). (a, b) Streamwise $\Phi_{uu}$, (c, d) wall-normal $\Phi_{vv}$, (e, f) spanwise $\Phi_{ww}$, and the streamwise/wall-normal $\Phi_{uv}$ two-point correlation matrices at $\bk = (2.5, 7)$. The one-point correlation profiles that are used as problem data in~\eqref{eq.CP} are marked by black lines along the main diagonals.}
	\label{fig.covariance_DNS_NM}
\end{figure*}

\begin{figure*}
		\begin{tabular}{cccccc}
	\hspace{-.4cm} \subfigure[]{\label{fig.Rww_DNS_zoomed}}
	& 
	\hspace{-.4cm} \small{$\Phi_{ww, \mathrm{dns}}$} 
	&
	\hspace{-.8cm} \subfigure[]{\label{fig.Rww_NM_zoomed}}
	&
	\hspace{-.4cm} \small{$\Phi_{ww}$}
	&
	\hspace{-.77cm} \subfigure[]{\label{fig.Rww_yp15_NM_zoomed}}
	&
	\hspace{-.39cm} \small{$\Phi_{ww}(y^+=15,:)$}
	\\[-.4cm]
	\hspace{-.1cm}
	\begin{tabular}{c}
		\vspace{.4cm}
		\hspace{-.1cm}
		\small{$y$}
	\end{tabular}
	&
	\hspace{-.4cm}
	\begin{tabular}{c}
		\includegraphics[width=4.1cm]{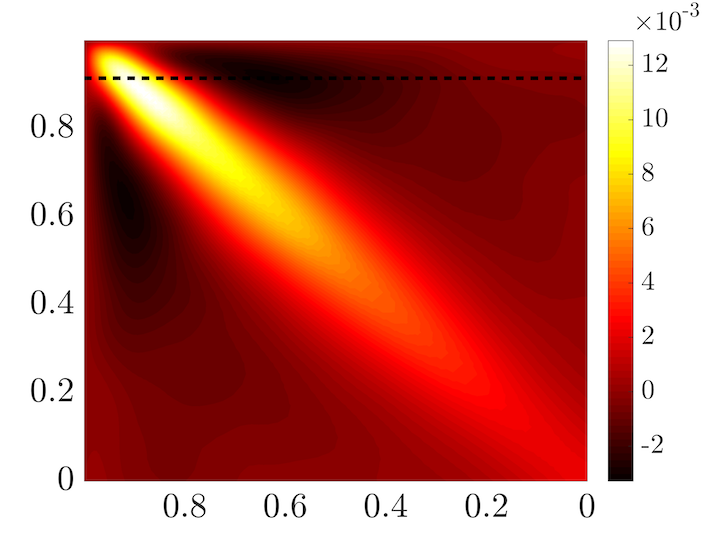}
		\\[-.1cm]
		\small{$y$}
	\end{tabular}
	&
	\hspace{-.65cm}
	\begin{tabular}{c}
		\vspace{.4cm}
		\hspace{.1cm}
		\small{$y$}
	\end{tabular}
	&
	\hspace{-.4cm}
	\begin{tabular}{c}
		\includegraphics[width=4.1cm]{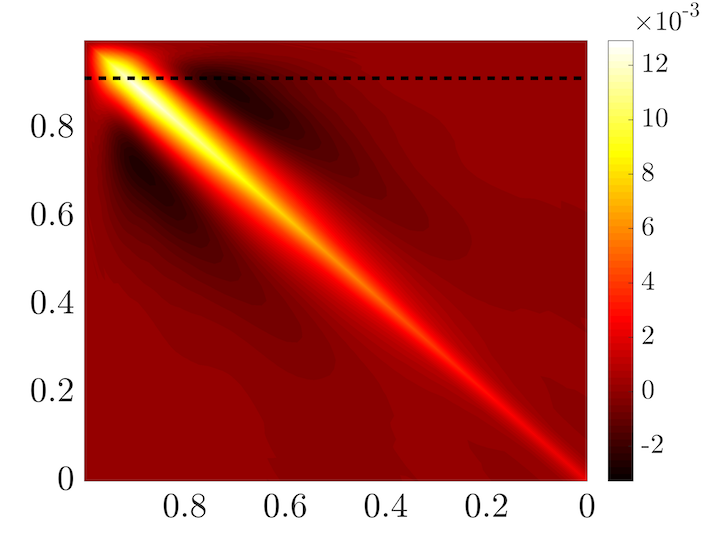}
		\\[-.1cm]
		\small{$y$}
	\end{tabular}
	&&
	\hspace{-.39cm}
	\begin{tabular}{c}
		\includegraphics[width=4.1cm]{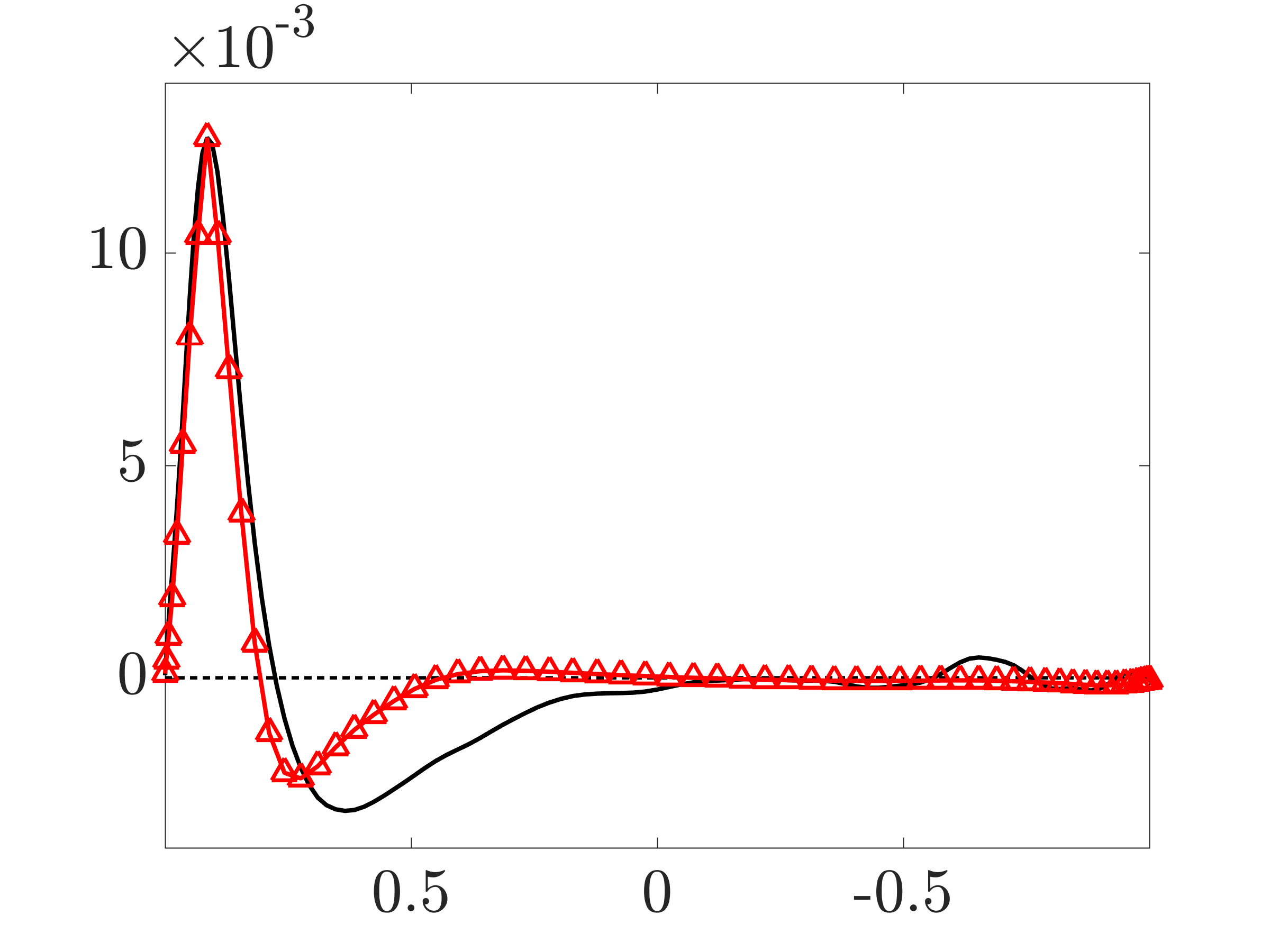}
		\\[-.1cm]
		\small{$y$}
	\end{tabular}
	\end{tabular}
		\caption{Quadrant II of the spanwise covariance matrices resulting from (a) DNS of turbulent channel flow with $R_\tau=186$, and (b) the solution to optimization problem~\eqref{eq.CP} with $\gamma = 300$ at $\bk = (2.5, 7)$. The horizontal black lines mark $y^+=15$. (c) Comparison of the two-point correlation $\Phi_{ww}$ at $y^+=15$ with other wall-normal locations: DNS (--); solution of~\eqref{eq.CP} ($\triangle$).}
	\label{fig.2point_correlation_ww_DNS_NM}
\end{figure*}

We next demonstrate that the solution to optimization problem~\eqref{eq.CP} also provides good recovery of two-point velocity correlations. We examine the wavenumber pair $\bk = (2.5,7)$ at which the premultiplied energy spectrum at $R_\tau = 186$ peaks. The left column in figure~\ref{fig.covariance_DNS_NM} displays the streamwise $\Phi_{uu}$, wall-normal $\Phi_{vv}$, spanwise $\Phi_{ww}$, and the streamwise/wall-normal $\Phi_{uv}$ covariance matrices resulting from DNS. The right column in figure~\ref{fig.covariance_DNS_NM} shows the same covariance matrices that are obtained from the solution to optimization problem~\eqref{eq.CP}. Although only diagonal elements of these matrices (marked by black lines in figure~\ref{fig.covariance_DNS_NM}) were used as data in~\eqref{eq.CP}, we have good recovery of the off-diagonal entries as well. In particular, for $\gamma=300$, we observe approximately $60\%$ recovery of the DNS-generated two-point correlation matrix $\Phi_{\mathrm{dns}}(\bk)$. The quality of approximation is assesed using~(see appendix~\ref{sec.gamma}),
\be
	\dfrac{\norm{\Phi(\bk) \,-\, \Phi_{\mathrm{dns}}(\bk)}_F}{\norm{\Phi_{\mathrm{dns}}(\bk)}_F},
\ee
where $\| \cdot \|_F$ denotes the Frobenius norm of a given matrix and $\Phi(\bk) = C(\bk)X(\bk)C^*(\bk)$ represents the two-point correlation matrix of the velocity fluctuations resulting from our optimization framework. 

We note that the solution of optimization problem~\eqref{eq.CP} also captures the presence of negative correlations in the covariance matrix of spanwise velocity; cf.\ figures~\ref{fig.Rww_DNS} and~\ref{fig.Rww_NM}. Figures~\ref{fig.Rww_DNS_zoomed} and~\ref{fig.Rww_NM_zoomed} show the second quadrants of the covariance matrices $\Phi_{ww, \mathrm{dns}}$ and $\Phi_{ww}$. In addition to matching the diagonal entries, i.e., one-point correlations of the spanwise velocity, the essential trends of two-point correlations resulting from DNS are also recovered. Figure~\ref{fig.Rww_yp15_NM_zoomed} illustrates this by showing the dependence of the auto-correlation of the spanwise velocity at $y^+ =15 $ on the wall-normal coordinate. This profile is obtained by extracting the corresponding row of $\Phi_{ww}$ and is marked by the black dashed line in figures~\ref{fig.Rww_DNS_zoomed} and~\ref{fig.Rww_NM_zoomed}. Clearly, the solution to optimization problem~\eqref{eq.CP} recovers the basic features (positive-negative-positive) of the DNS results. These features are indicators of coherent structures that reside at various wall-normal locations in the channel flow~\citep{monstewilcho07,smimckmar11}. 

It is worth noting that such {\em high-quality recovery of two-point correlations\/} would not have been possible {\em without incorporating the physics of the linearized NS equations as the structural constraint~\eqref{eq.constraint_lyap} into optimization problem~\eqref{eq.CP}.} 

\begin{figure}
\begin{center}
	\begin{tabular}{cccc}
	\hspace{-.3cm} \subfigure[]{\label{fig.rankZ_gamma300}}
       & 
       &
       \hspace{-.7cm} \subfigure[]{\label{fig.rankZ_gamma1e4}}
	&
	\\[-.35cm]
	\hspace{.05cm}
	\begin{tabular}{c}
		\vspace{.4cm}
		{\normalsize \rotatebox{90}{$\sigma_i(Z)$}}
	\end{tabular}
	&
	\hspace{-.4cm}
	\begin{tabular}{c}
		\includegraphics[width=6.15cm]{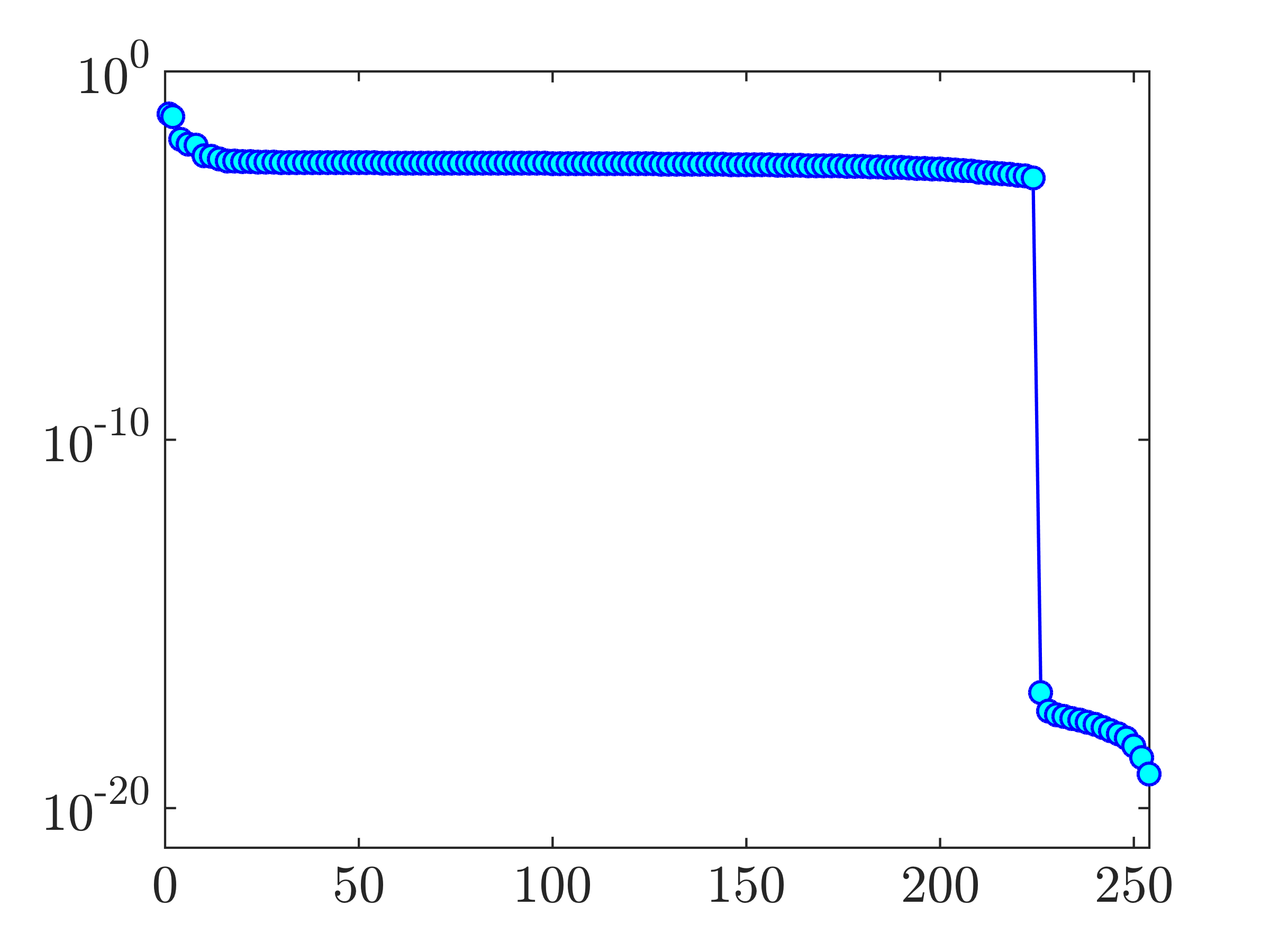}
		\\[-.1cm]
		\hspace{-.3cm}{\normalsize $i$}
	\end{tabular}
	&
	&
	\hspace{-.3cm}
	\begin{tabular}{c}
		\includegraphics[width=6.15cm]{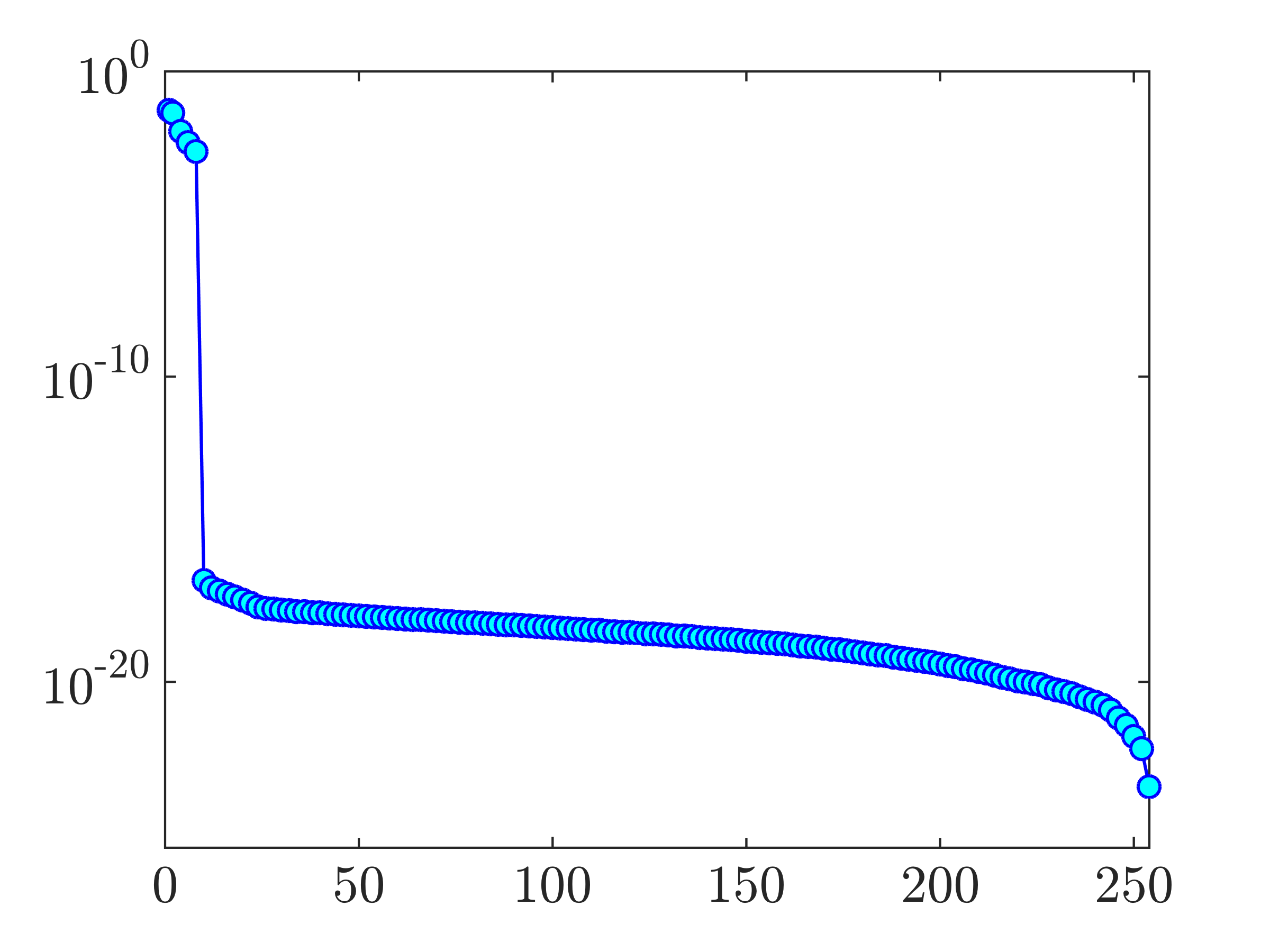}
		\\[-.1cm]
		~~{\normalsize $i$}
	\end{tabular}
	\end{tabular}
\end{center}
\caption{Singular values of the solution $Z$ to~\eqref{eq.CP} in turbulent channel flow with $R_\tau = 186$, $\bk = (2.5, 7)$, and $N = 127$ for (a) $\gamma=300$; and (b) $\gamma=10^4$.}
\label{fig.svdZ_kx2p5_kz7}
\end{figure}

\begin{remark}
	{ \em In optimization problem~\eqref{eq.CP}, the regularization parameter $\gamma$ determines the importance of the nuclear norm relative to the logarithmic barrier function. Larger values of $\gamma$ yield $Z(\bk)$ of lower rank, but may compromise quality of completion of two-point correlations; see appendix~\ref{sec.gamma}. For turbulent channel flow with $R_\tau=186$ and $\bk = (2.5, 7)$, figure~\ref{fig.svdZ_kx2p5_kz7} shows the singular values of $Z$ for two values of $\gamma$, $\gamma = 300$ and $10^4$. Clearly, the higher value of $\gamma$ results in a much lower rank of the matrix $Z$, with $6$ positive and $2$ negative eigenvalues. \cite{chejovgeoCDC13} showed that the maximum number of positive or negative eigenvalues of the matrix $Z$ bounds the number of inputs into the linearized NS model~\eqref{eq.lnse1}. This implies that partially available statistics can be reproduced with $6$ colored-in-time inputs. However, as discussed in appendix~\ref{sec.gamma}, the quality of completion is best for $\gamma=300$. In this case, the matrix $Z$ has $225$ non-zero eigenvalues, $221$ positive and $4$ negative. Thus, for $\gamma = 300$ and a spatial discretization with $N = 127$ collocation points in $y$, $221$ colored-in-time inputs are required to account for partially available statistics.}
	\end{remark}

\subsection{Verification in stochastic linear simulations}
	\label{sec.stochastic_sim}

We next conduct stochastic simulations of the linearized flow equations and compare the resulting statistics with DNS at $R_\tau=186$. Filter~\eqref{eq.filter} that generates colored-in-time forcing $\bf{d} (\bk,t)$ is obtained from the solution to~\eqref{eq.CP} with $\bk = (2.5, 7)$ and $\gamma = 10^4$. This filter in conjunction with the linearized dynamics~\eqref{eq.lnse1} yields representation~\eqref{eq.feedback_dyn} which is driven by white-in-time Gaussian process $\bw (\bk,t)$ with zero mean and unit variance. We recall that this reduced-order representation is equivalent to the original NS equations subject to the colored-in-time stochastic forcing $\bd (\bk,t)$ with properly identified power spectrum. As shown in \S~\ref{sec.filter}, system~\eqref{eq.cascade} can be equivalently represented by a low-rank modification to the linearized NS dynamics~\eqref{eq.feedback_dyn}, which has the same number of degrees of freedom as the finite-dimensional approximation of the original linearized NS dynamics. Here, we consider a spatial discretization with $N=127$ collocation points in the wall-normal direction. Thus, at each wavenumber pair, linear system~\eqref{eq.feedback_dyn} that results from our modeling framework has $254$ degrees of freedom.

\begin{figure}
\begin{center}
	\begin{tabular}{cc}
	\begin{tabular}{c}
		\vspace{.2cm}
		{\normalsize \rotatebox{90}{kinetic energy}}
	\end{tabular}
	&
	\hspace{-.5cm}
	\begin{tabular}{c}
		\includegraphics[width=6.5cm]{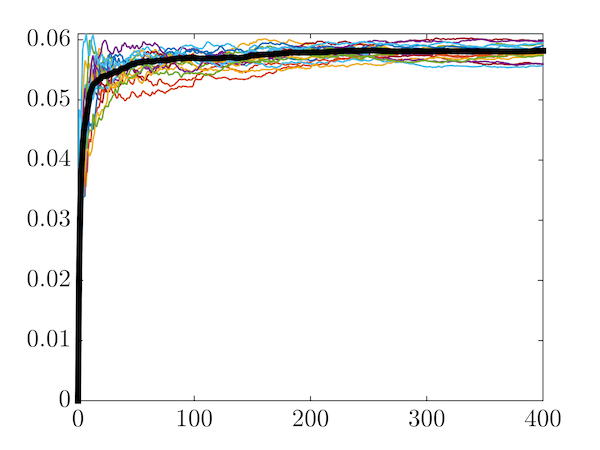}
		\\[-.1cm]
		{\normalsize$t$}
	\end{tabular}
	\end{tabular}
\end{center}
\caption{Time evolution of fluctuation's kinetic energy for twenty realizations of the forcing to the modified linearized dynamics~\eqref{eq.feedback_dyn} with $R_\tau=186$ and $\bk = (2.5, 7)$; the energy averaged over all simulations is marked by the thick black line.}
\label{fig.stochastic_sim}
\end{figure}

Stochastic linear simulations that we present next confirm that one-point correlations can indeed be recovered by stochastically-forced linearized dynamics. Since the proper comparison with DNS or experiments requires ensemble-averaging, rather than comparison at the level of individual stochastic simulations, we have conducted twenty simulations of system~\eqref{eq.feedback_dyn}. The total simulation time was set to 400 viscous time units.

\begin{figure}
\begin{center}
	\begin{tabular}{cccc}
		\hspace{-.4cm} \subfigure[]{\label{fig.uu_sim}} && \hspace{-.5cm} \subfigure[]{\label{fig.vv_sim}} &
		\\[-.5cm]
		\begin{tabular}{c}
			\vspace{.45cm}
			{\normalsize \rotatebox{90}{$uu$}}
		\end{tabular}
		&
		\hspace{-.4cm}
		\begin{tabular}{c}
		        \includegraphics[width=6.15cm]{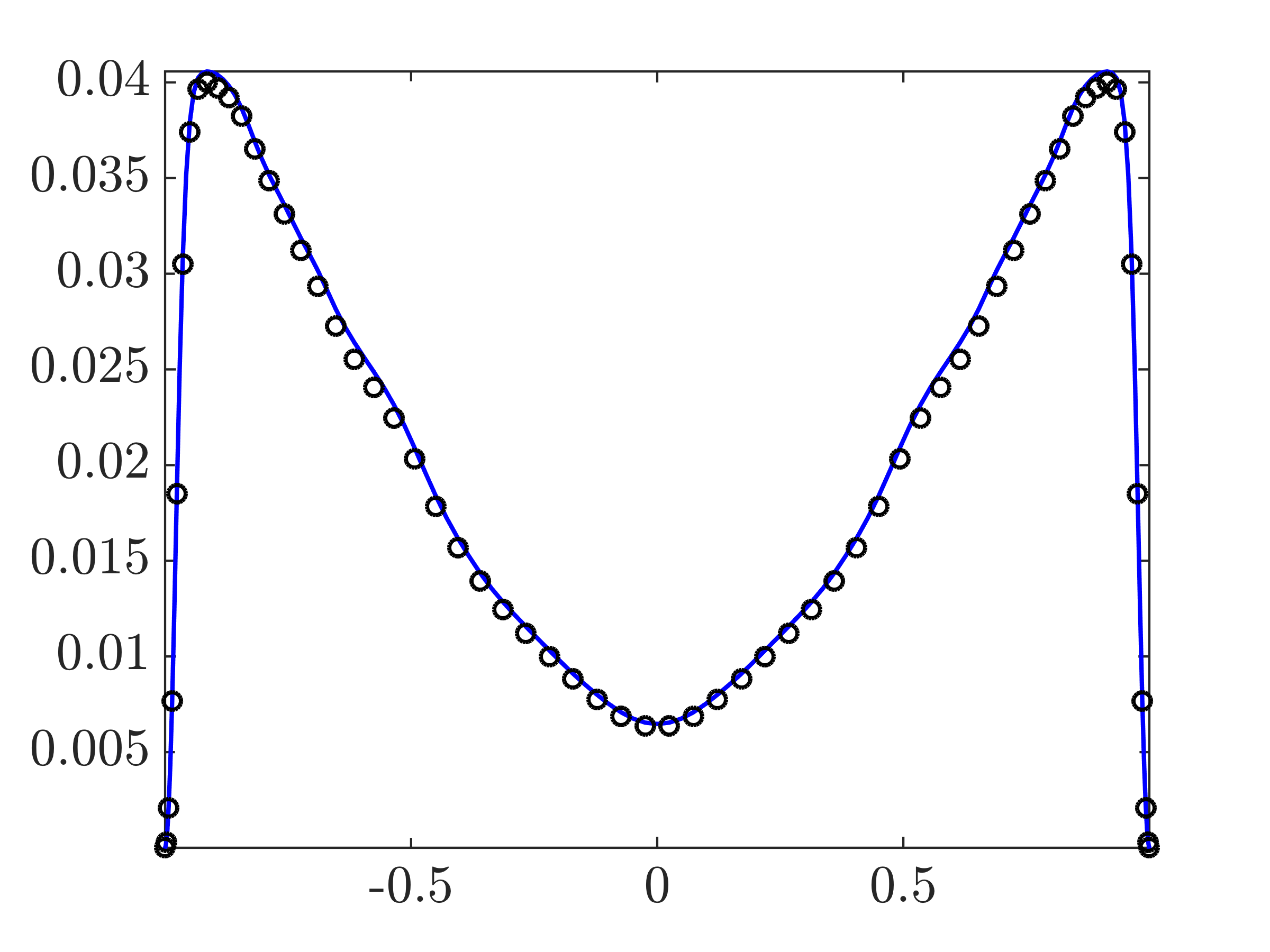}
	        	\\[-.1cm]
		        {\normalsize$y$}
	        \end{tabular}
	        &
	        \begin{tabular}{c}
	        	\vspace{.4cm}
	        	{\normalsize \rotatebox{90}{$vv$}}
		\end{tabular}
		&
		\hspace{-.45cm}
	        \begin{tabular}{c}
	        	\includegraphics[width=6.15cm]{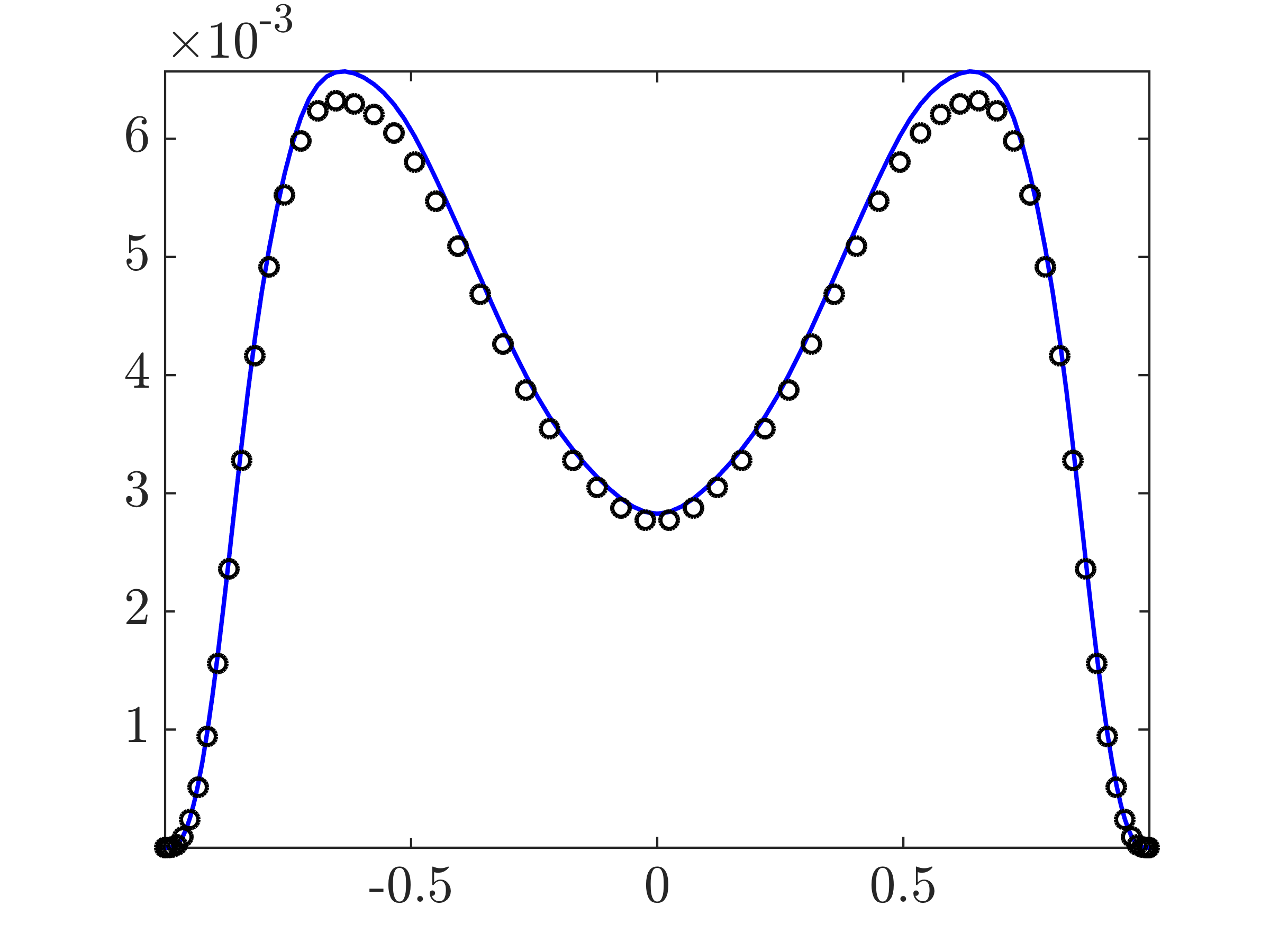}
			\\[-.1cm]
	        	{\normalsize $y$}
	        \end{tabular}
	        \\[.2cm]
	        \hspace{-.4cm} \subfigure[]{\label{fig.ww_sim}} && \hspace{-.5cm} \subfigure[]{\label{fig.uv_sim}} &
		\\[-.5cm]
		\begin{tabular}{c}
			\vspace{.45cm}
			{\normalsize \rotatebox{90}{$ww$}}
		\end{tabular}
		&
		\hspace{-.4cm}
		\begin{tabular}{c}
		        \includegraphics[width=6.15cm]{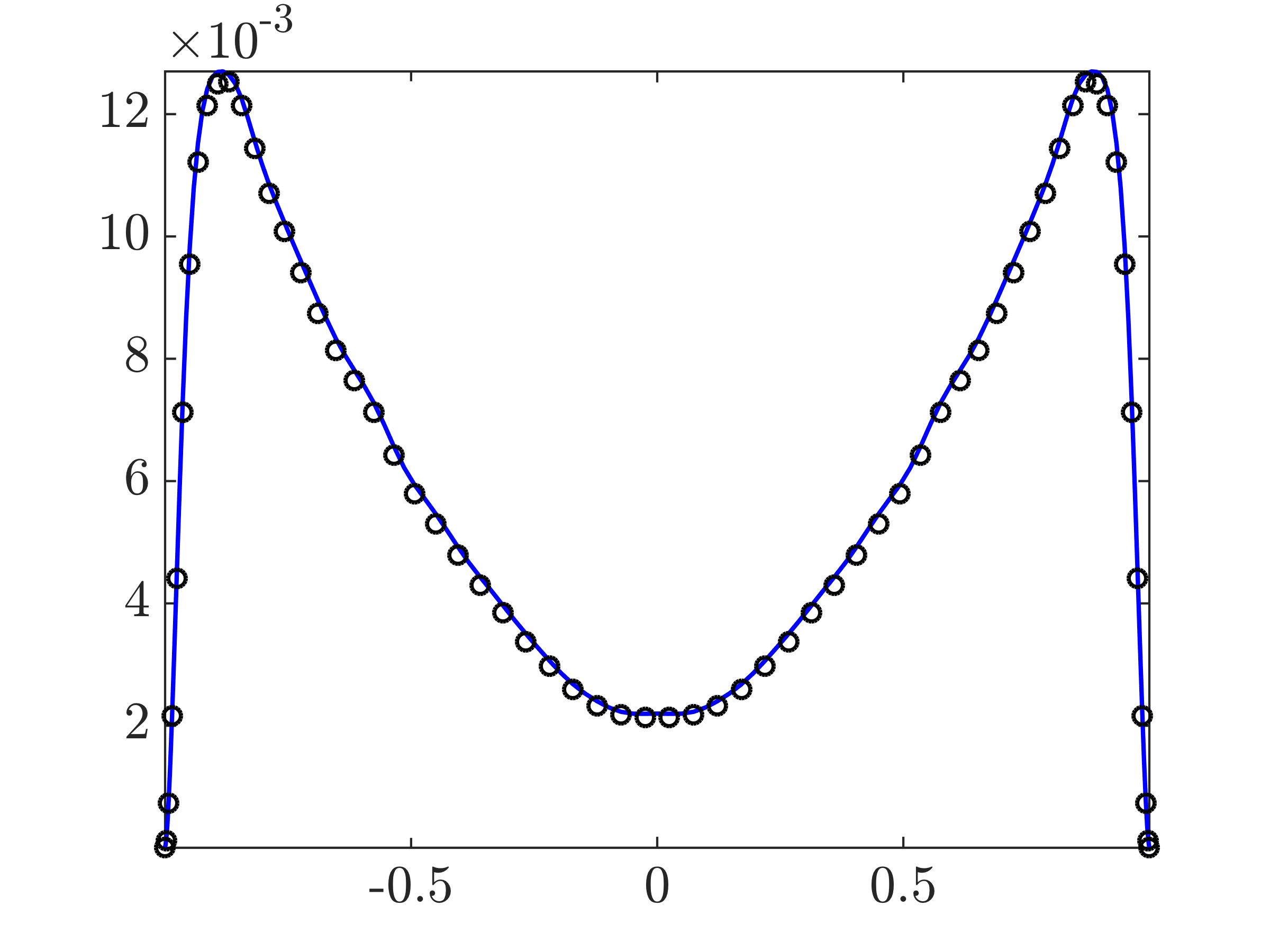}
	        	\\[-.1cm]
		        {\normalsize$y$}
	        \end{tabular}
	        &
	        \begin{tabular}{c}
	        	\vspace{.4cm}
	        	{\normalsize \rotatebox{90}{$uv$}}
		\end{tabular}
		&
		\hspace{-.45cm}
	        \begin{tabular}{c}
	        	\includegraphics[width=6.15cm]{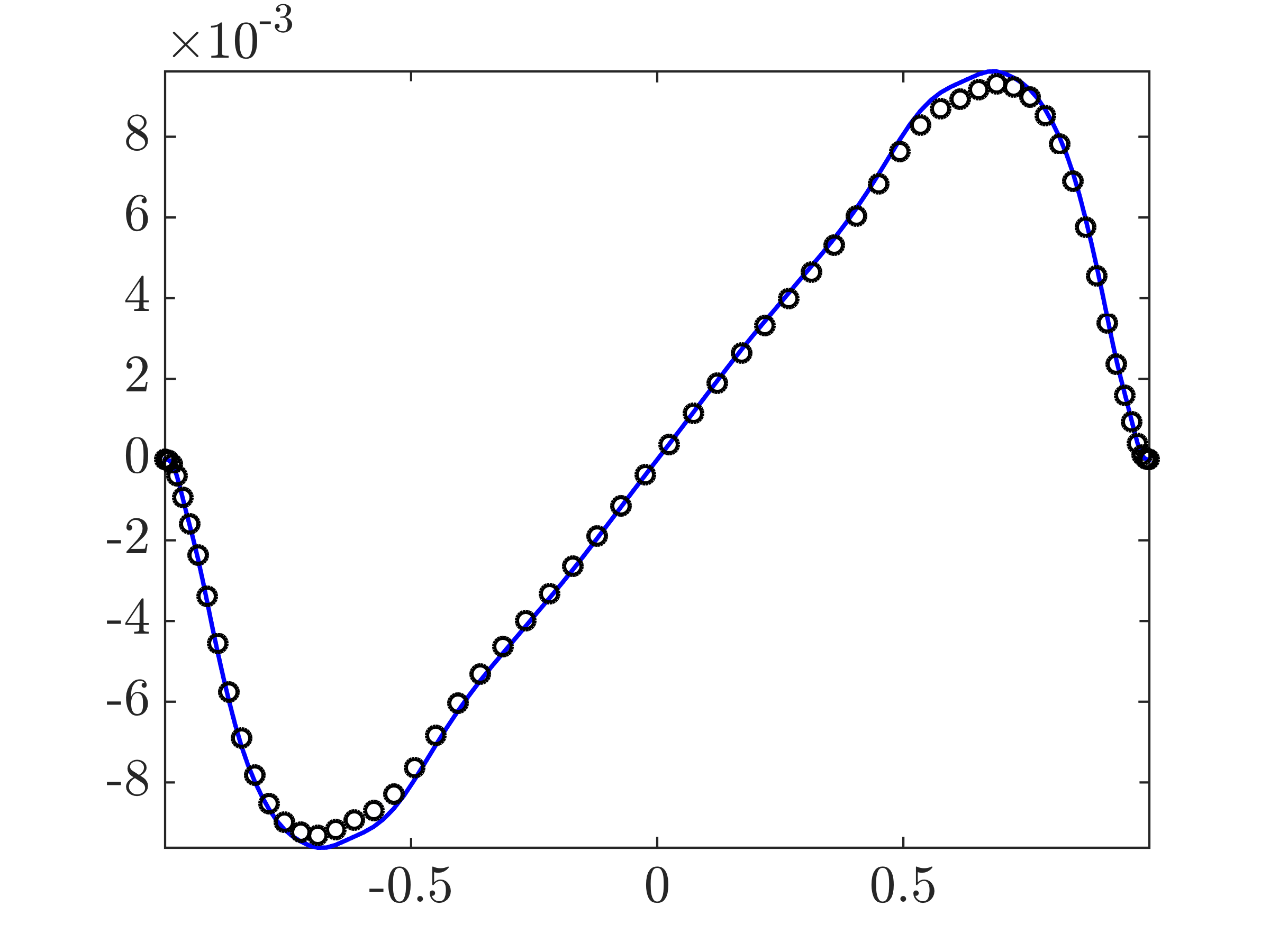}
			\\[-.1cm]
	        	{\normalsize $y$}
	        \end{tabular}
        \end{tabular}
\end{center}
\caption{Normal stress profiles in the (a) streamwise, (b) wall-normal, (c) spanwise direction, and (d) shear stress profile resulting from DNS of turbulent channel flow with $R_\tau=186$ at $\bk=(2.5,7)$ (--) and stochastic linear simulations ($\Circle$).}
\label{fig.sim_results}
\end{figure}

Figure~\ref{fig.stochastic_sim} shows the time evolution of the energy (variance) of velocity fluctuations, for twenty realizations of white-in-time forcing $\bw (\bk,t)$ to system~\eqref{eq.feedback_dyn}. The variance averaged over all simulations is marked by the thick black line. Even though the responses of individual simulations differ from each other, the average of twenty sample sets asymptotically approaches the correct value of turbulent kinetic energy in the statistical steady-state, $\trace \, (\Phi (\bk))$. Figure~\ref{fig.sim_results} displays the normal and shear stress profiles resulting from DNS and from stochastic linear simulations. We see that the averaged output of twenty simulations of the linearized dynamics agrees well with DNS results. This close agreement can be further improved by running additional linear simulations and by increasing the total simulation times.

\subsection{Reproducing statistics at higher Reynolds numbers}
    \label{sec.highRe}

\begin{figure}
	\begin{tabular}{cccc}
		\hspace{.15cm}\subfigure[]{\label{fig.uu_547}}&&\hspace{-.2cm}\subfigure[]{\label{fig.uv_547}}
		\\[-.8cm]
		&
		\hspace{-.03cm}
		\begin{tabular}{c}
			\includegraphics[width=6.15cm]{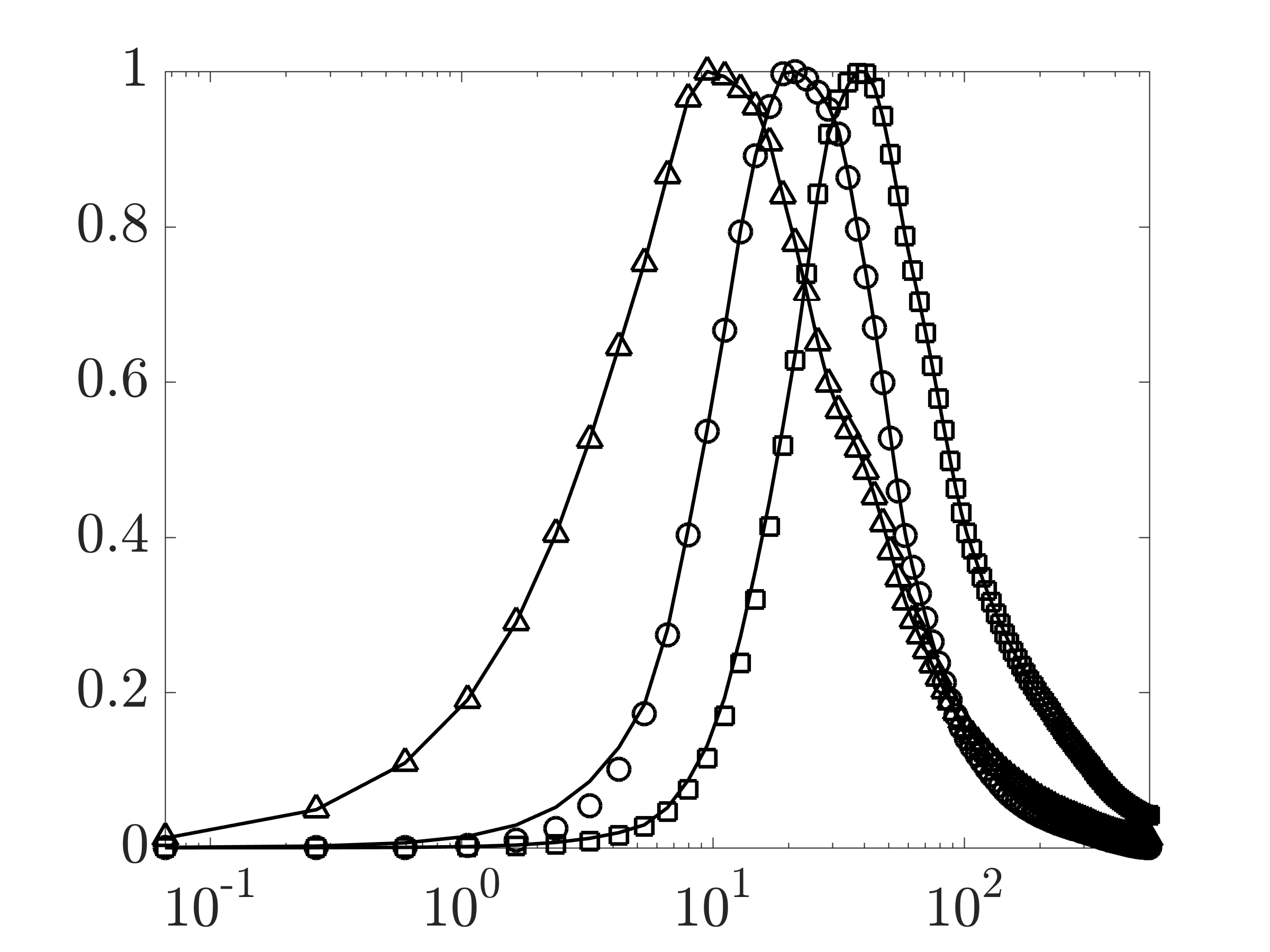}
		\end{tabular}
		&&
		\hspace{-.1cm}
		\begin{tabular}{c}
			\includegraphics[width=6.15cm]{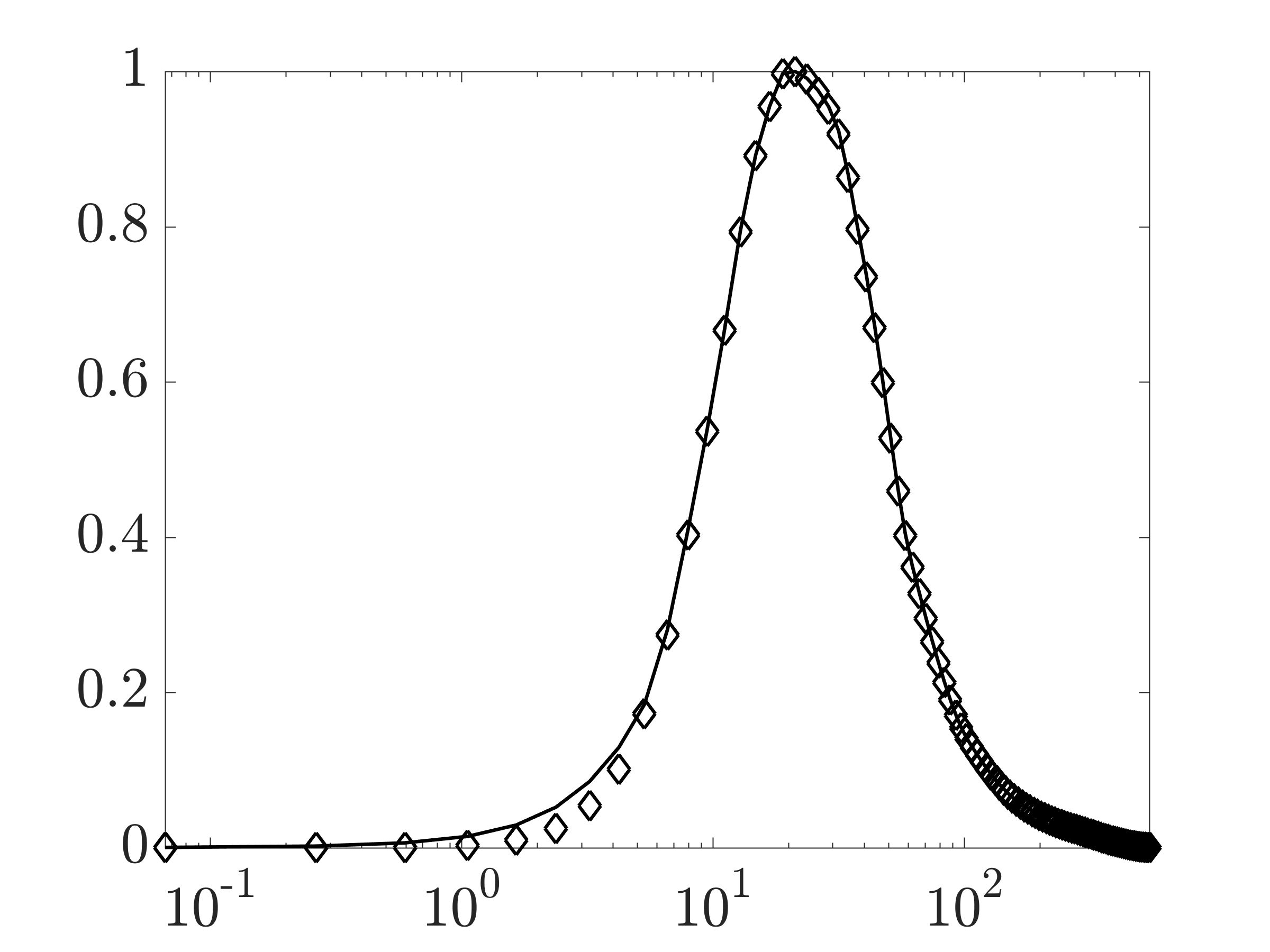}
		\end{tabular}
	\\[.2cm]
		\hspace{.15cm}\subfigure[]{\label{fig.uu_934}}&&\hspace{-.2cm}\subfigure[]{\label{fig.uv_934}}
		\\[-.8cm]
		&
		\hspace{-.03cm}
		\begin{tabular}{c}
			\includegraphics[width=6.15cm]{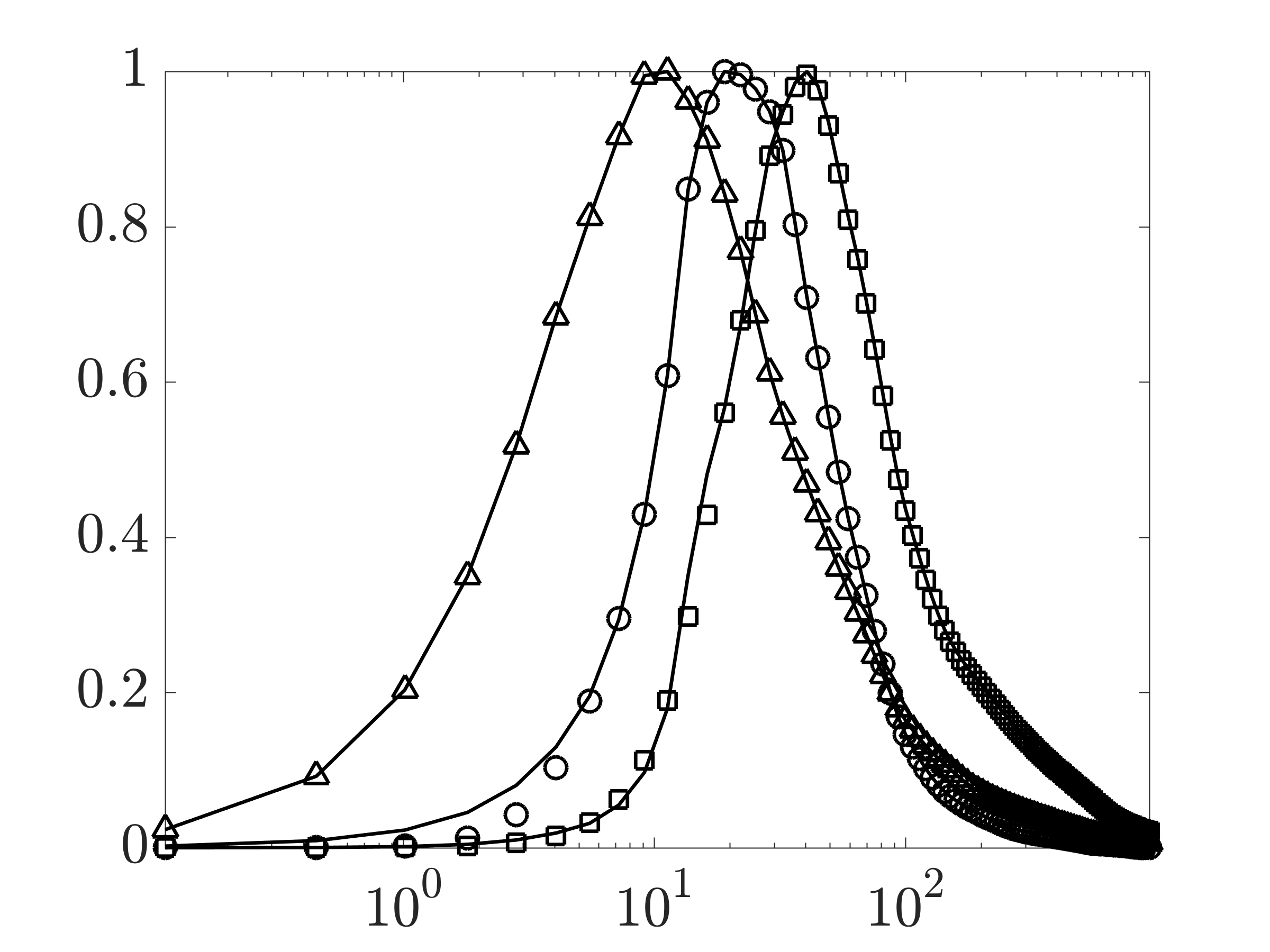}
		\end{tabular}
		&&
		\hspace{-.1cm}
		\begin{tabular}{c}
			\includegraphics[width=6.15cm]{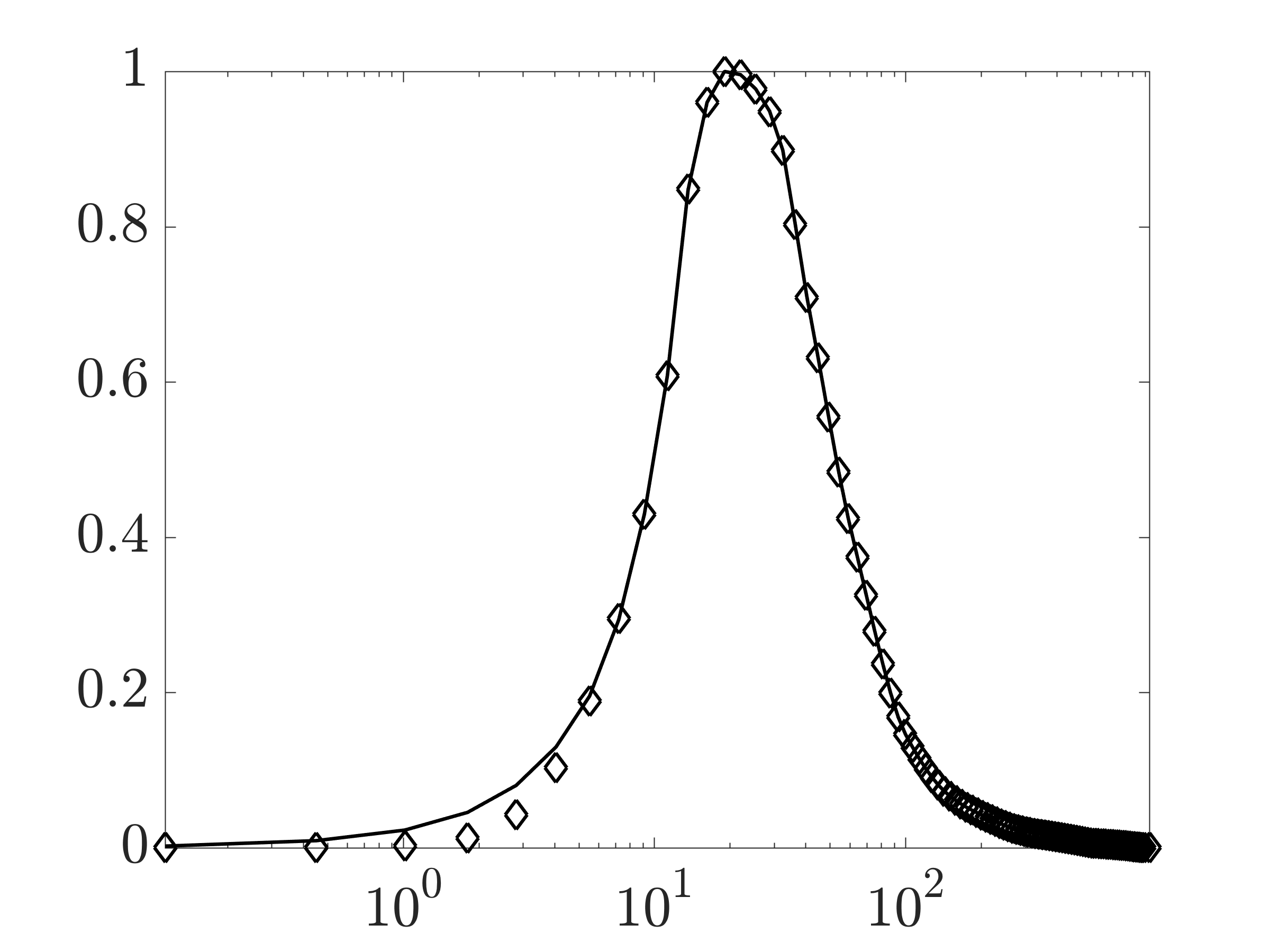}
		\end{tabular}
	\\[.2cm]
		\hspace{.15cm}\subfigure[]{\label{fig.uu_2003}}&&\hspace{-.2cm}\subfigure[]{\label{fig.uv_2003}}
		\\[-.8cm]
		&
		\hspace{-.03cm}
		\begin{tabular}{c}
			\includegraphics[width=6.15cm]{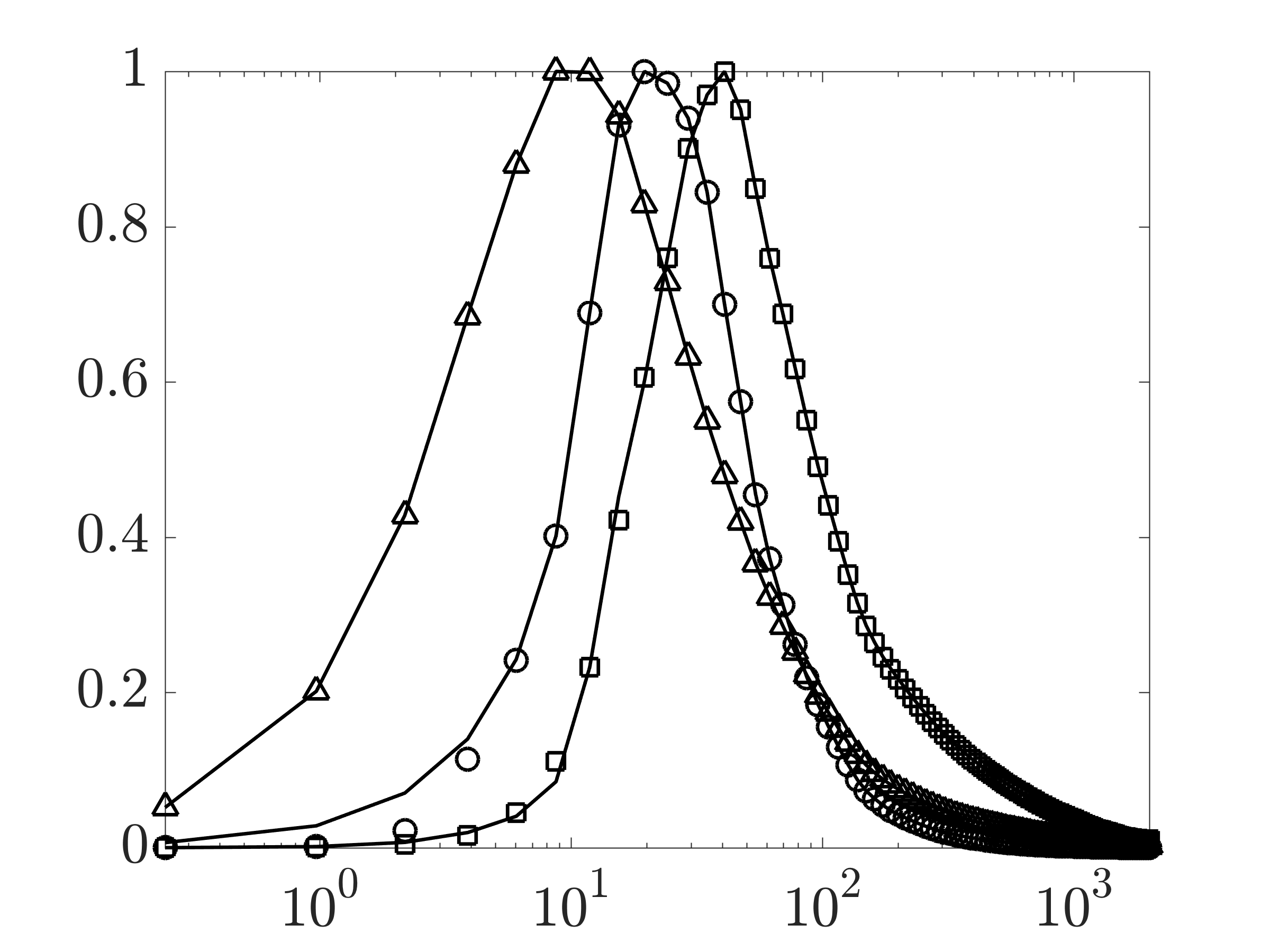}
			\\[-.1cm]
			\hspace{.1cm}
			{\normalsize $y^+$}
		\end{tabular}
		&&
		\hspace{-.1cm}
		\begin{tabular}{c}
			\includegraphics[width=6.15cm]{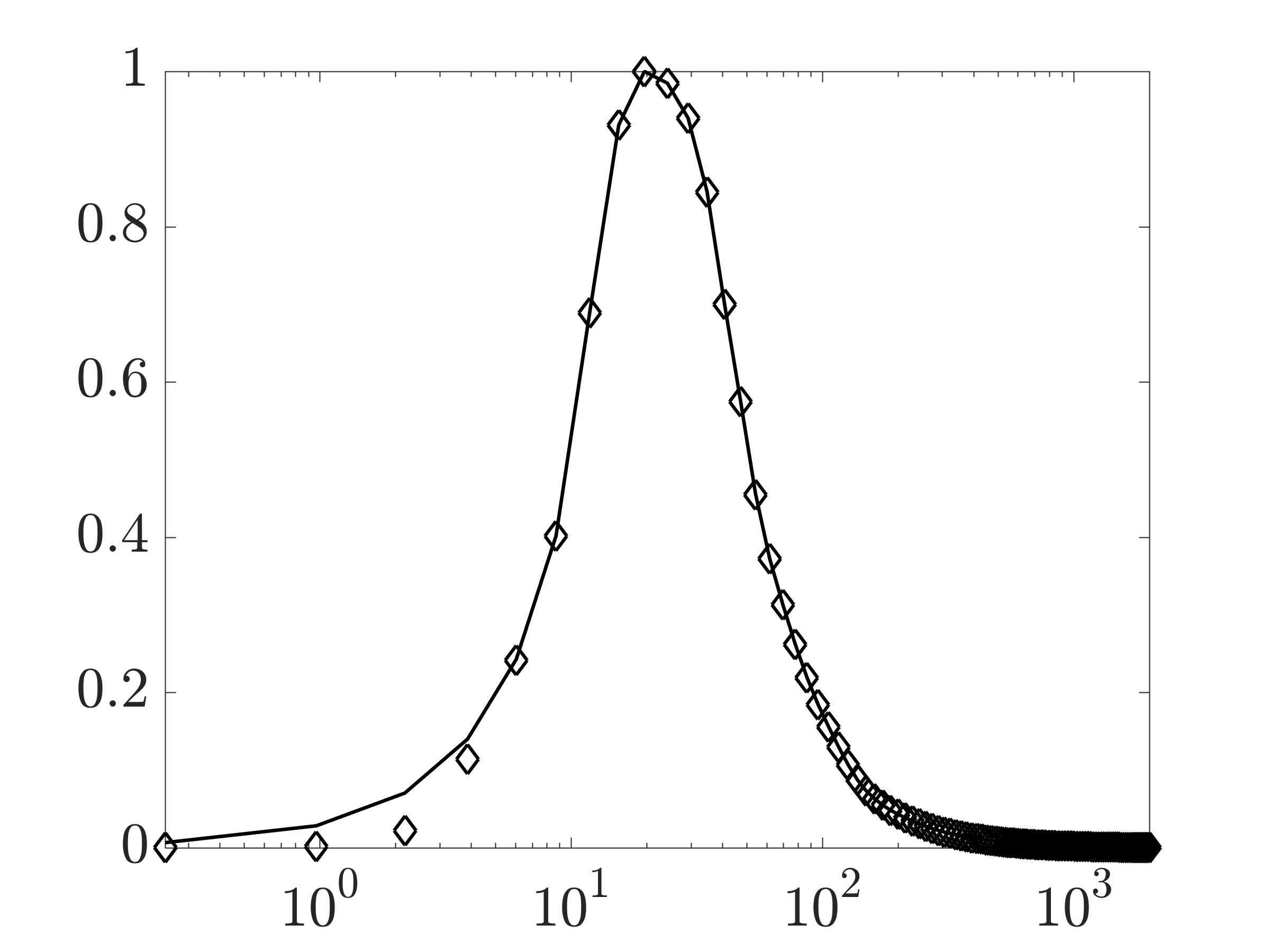}
			\\[-.1cm]
			\hspace{.1cm}
			{\normalsize $y^+$}
		\end{tabular}
	\end{tabular}
	\caption{Normalized normal (left) and shear (right) stress profiles resulting from DNS (--) and from the solution to~\eqref{eq.CP} with $\gamma=300$ at $\lambda_x^+ = 1000$ and $\lambda_z^+ = 100$ {(in inner units)}; $uu$ ($\Circle$), $vv$ ($\Box$), $ww$ ($\triangle$), $-uv$ ($\Diamond$). (a, b) $R_\tau=547$; (c, d) $R_\tau=934$; (e, f) $R_\tau=2003$.}
	\label{fig.norm_shear_profiles_HighRe}
\end{figure}

We next apply our optimization framework to channel flows with higher Reynolds numbers~\citep{deljim03,deljimzanmos04,hoyjim06}. We use $N=201$ collocation points to discretize differential operators for turbulent flows with $R_\tau = 549$, $934$, and $2003$. We focus on a pair of wavelengths that are relevant to the study of near-wall structures, i.e., $\lambda_x^+=1000$ and $\lambda_z^+=100$. This wavelength pair is associated with the near-wall system of quasi-streamwise streaks and counter-rotating vortices which is responsible for large production of turbulent kinetic energy~\citep{klireyschrun67,smimet83}. For all Reynolds numbers, optimization problem~\eqref{eq.CP} is solved with $\gamma=300$ and up to the same accuracy.

Figure~\ref{fig.norm_shear_profiles_HighRe} shows the normal and shear stress profiles for the aforementioned Reynolds numbers and selected wavelength pair. For illustration, these profiles have been normalized by their largest values and are presented in inner units. We see that the solution to optimization problem~\eqref{eq.CP} achieves perfect recovery of all \mbox{one-point velocity correlations.}

\begin{figure}
\begin{center}
	\begin{tabular}{cc}
	\begin{tabular}{c}
		\vspace{.2cm}
		{\normalsize \rotatebox{90}{$\sigma_i(Z)$}}
	\end{tabular}
	&
	\hspace{-.5cm}
	\begin{tabular}{c}
		\includegraphics[width=6.5cm]{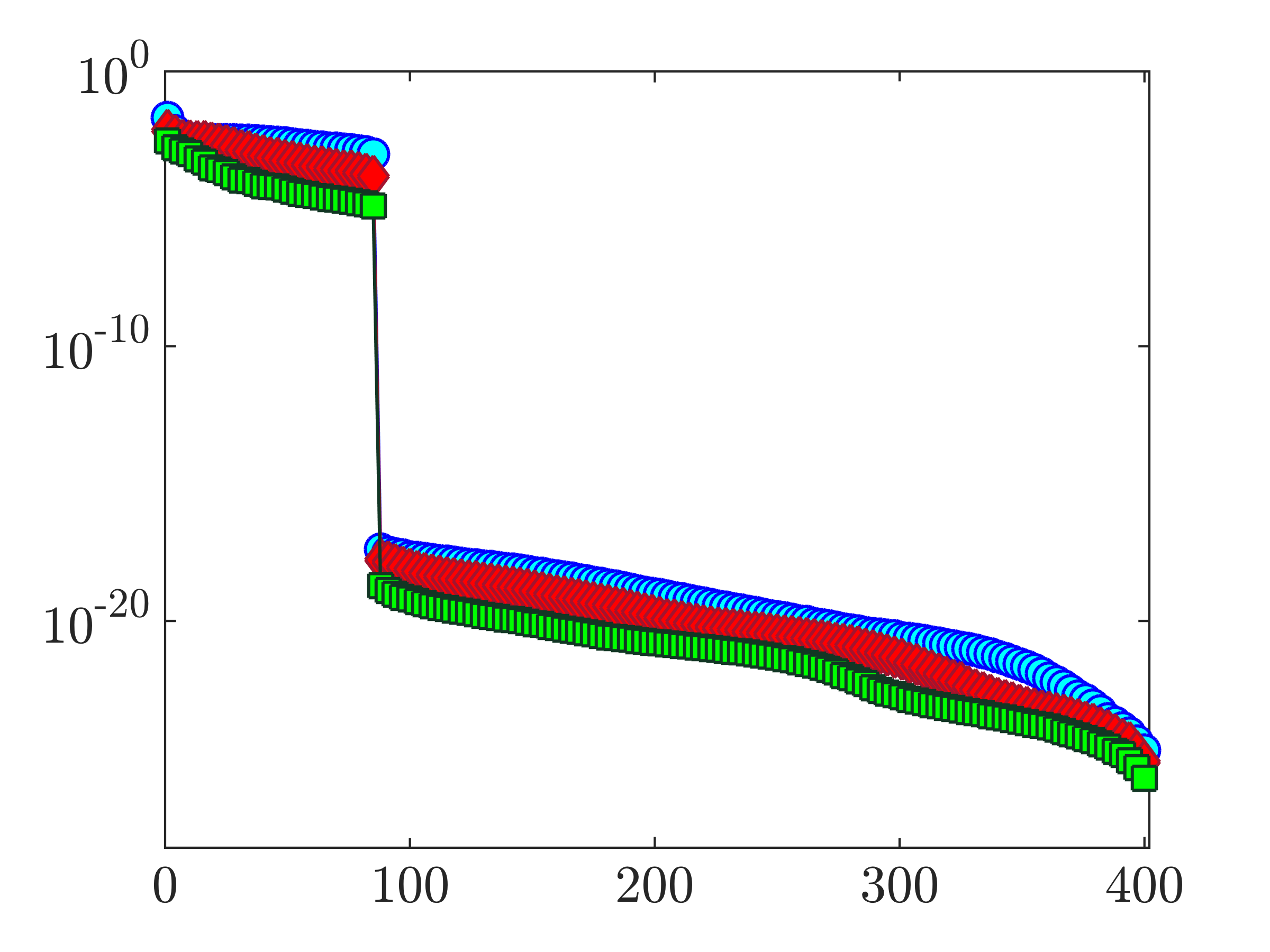}
		\\[-.1cm]
		{\normalsize $i$}
	\end{tabular}
	\end{tabular}
\end{center}
\caption{Singular values of the matrix $Z$ resulting from the solution to~\eqref{eq.CP} with $\gamma=300$ at $R_\tau = 549$ ($\Circle$), $934$ ($\Diamond$), and $2003$ ($\square$).}
\label{fig.HighRe_svd}
\end{figure}

Figure~\ref{fig.HighRe_svd} shows the singular values of the matrix $Z$ resulting from the solution to~\eqref{eq.CP}. At this pair of wall-parallel wavelengths, we observe that higher Reynolds numbers result in matrices $Z$ of similar rank. For $R_\tau=547$, $934$, and $2003$, matrix $Z$ has $84$, $80$, and $76$ significant positive and $2$, $5$, and $8$ significant negative eigenvalues, respectively. We thus conclude that, at higher Reynolds numbers, a similar number of inputs can be utilized to recover turbulent statistics by the linearized NS equations with colored-in-time stochastic forcing. Equivalently, the modification to the dynamical generator of the linearized NS equations which is required to capture partially available second-order statistics at higher Reynolds numbers is of similar rank.

	\vspace*{-2ex}
\section{Spatio-temporal analysis of linear model}
    \label{sec.frequency-anal}
	
	System~\eqref{eq.feedback_dyn} provides a linear model that captures second-order statistics of turbulent channel flow in statistical steady-state. As illustrated in \S~\ref{sec.stochastic_sim}, this model can be advanced in time by conducting linear stochastic simulations. More importantly, it can be analyzed using tools from linear systems theory. For example, dominant spatio-temporal flow structures can be easily identified and two-point correlations in time can be readily computed. These tools have provided useful insight into the dynamics of both laminar~\citep{butfar92,farioa93,redhen93,redschhen93,tretrereddri93,bamdah01,mj-phd04,jovbamJFM05} and turbulent~\citep{deljimjfm06,pujgarcosdep09,hwacosJFM10a,hwacosJFM10b,mcksha10,shamck13,moashatromck13} wall-bounded shear flows. 
	
	Application of the temporal Fourier transform on system~\eqref{eq.feedback_dyn} yields
\be
	\label{eq.input-output}
	\bv(\bk,\omega)
	\;=\;
	T_{\bv \bw}(\bk,\omega)\,
	\bw(\bk,\omega),
\ee
where $\omega$ is the temporal frequency and $T_{\bv \bw}(\bk,\omega)$ is the spatio-temporal frequency response,
\be
	\label{eq.freq_res}
	T_{\bv \bw} (\bk, \omega)
	\;=\;
	-
	C(\bk)
	\left(
	\mri \omega \bI \,+\, A_f(\bk)
	\right)^{-1}
	B(\bk).
\ee
Here, $A_f(\bk)$ is the generator of linear dynamics~\eqref{eq.feedback_dyn} which result from the modeling and optimization framework of \S~\ref{sec.mcp}. Equation~\eqref{eq.input-output} facilitates decomposition of the fluctuating velocity field $\bv(\bk,\omega)$ into the sum of spatio-temporal Fourier modes which correspond to physical structures with streamwise and spanwise wavelengths $\lambda_x = 2\pi/k_x$ and $\lambda_z = 2\pi/k_z$. These structures convect at speed $c=\omega/k_x$ in the streamwise direction. Since the dominant waves in turbulent channel flow travel downstream~\citep{mcksha10}, the sign of the temporal frequency in~\eqref{eq.freq_res} is changed relative to the convention used in~\eqref{eq.filter-tf}. With proper definition of the matrices $A$, $B$, and $C$ the spatio-temporal frequency response analysis can be conducted for different linear approximations of the NS equations, e.g., the original linearized NS model~\eqref{eq.lnse1} or an eddy-viscosity-enhanced linearized NS model~\citep{reyhus72-3,deljimjfm06,cospujdep09,pujgarcosdep09,hwacosJFM10b}.

Singular value decomposition of the frequency response~\eqref{eq.freq_res} brings input-output representation~\eqref{eq.input-output} into the following form, 
\be
	\label{eq.inout_analysis}
	\ba{rcl}
	\bv (\bk,\omega)
	&  \!\!=\!\!  &
	T_{\bv \bw}(\bk,\omega)\, \bw(\bk,\omega)
	\; = \;
	\ds{\sum^{r}_{j \, = \,1}} \, \sigma_j(\bk,\omega)\, a_j(\bk,\omega)\, \tilde{\xi}_j(\bk, \omega),
	\ea
\ee
where $\sigma_1 \geq \sigma_2 \geq \cdots \geq \sigma_r > 0$ are the singular values of $T_{\bv \bw}(\bk,\omega)$, $\tilde{\xi}_j(\bk,\omega)$ is the $j$th left singular vector of $T_{\bv \bw}(\bk,\omega)$, and $a_j(\bk,\omega)$ is the projection of the forcing $\bw(\bk,\omega)$ onto the $j$th right singular vector. The left and right singular vectors provide insight into coherent structures of velocity and forcing fluctuations~\citep{sch07}. In particular, symmetries in the wall-parallel directions can be used to express velocity components as
\begin{subequations}
	\label{eq.physical_uvw}
\begin{eqnarray}
	u_j(x,z,t)
	&\;=\;&
	4 \cos(k_z z)\, \mathrm{Re} \left(\tilde{u}_j(\bk,\omega) \, \mre^{\mri(k_x x \,-\, \omega t)} \right),
	\label{eq.physical_u}
	\\
	v_j(x,z,t)
	&\;=\;&
	4 \cos(k_z z)\, \mathrm{Re} \left(\tilde{v}_j(\bk,\omega) \, \mre^{\mri(k_x x \,-\, \omega t)} \right),
	\label{eq.physical_v}
	\\
	w_j(x,z,t)
	&\;=\;&
	- 4 \sin(k_z z)\, \mathrm{Im} \left(\tilde{w}_j(\bk,\omega) \, \mre^{\mri(k_x x \,-\, \omega t)} \right).
	\label{eq.physical_w}
\end{eqnarray}
\end{subequations}
Here, $\mathrm{Re}$ and $\mathrm{Im}$ denote real and imaginary parts, and $\tilde{u}_j(\bk,\omega)$, $\tilde{v}_j(\bk,\omega)$, and $\tilde{w}_j(\bk,\omega)$ are the streamwise, wall-normal, and spanwise components of the $j$th left singular vector $\tilde{\xi}_j(\bk, \omega)$ in~\eqref{eq.inout_analysis}. 

The power spectral density (PSD) of $\bv(\bk,\omega)$ quantifies amplification of white-in-time stochastic forcing $\bw (\bk,t)$, across temporal frequencies $\omega$ and spatial wavenumbers $\bk$, 
\be
	\label{eq.PSD}
	\Pi_\bv (\bk,\omega)
	\;=\;
	\trace \left( T_{\bv \bw} (\bk, \omega)\, T_{\bv \bw}^* (\bk, \omega) \right)
	\; = \;
	\sum_i
	\, 
	\sigma_i^2 (T_{\bv \bw} (\bk, \omega)).
\ee
The integration of $\Pi_\bv (\bk,\omega)$ over temporal frequency yields the $H_2$ norm or, equivalently, the energy spectrum as a function of wavenumbers $\bk$~\citep{jovbamJFM05}.
While the PSD is given by the sum of squares of the singular values, the maximum singular value of $T_{\bv \bw} (\bk, \omega)$ quantifies the worst-case amplification of finite energy disturbances,
\be
	G_\bv(\bk,\omega)
	\; \DefinedAs \;
	\sup \limits_{\norm{\bw}^2\leq1}\, \dfrac{\norm{\bv(\bk,\omega)}^2}{\norm{\bw(\bk,\omega)}^2}
	\; = \;
	\sigma^2_{\max} (T_{\bv\bw}(\bk,\omega)).
\ee
Here, $\norm{\cdot}^2$ is the standard energy norm and the largest amplification over temporal frequencies determines the $H_{\infty}$ norm~\citep{zhodoyglo96}, 
	$
	\sup_\omega \sigma_{\max}(T_{\bv\bw}(\bk,\omega)).
	$
For any $\bk$, the $H_{\infty}$ norm quantifies the worst-case amplification of purely harmonic (in $x$, $z$, and $t$) deterministic (in $y$) disturbances~\citep{mj-phd04}.

Temporal two-point correlations of linear model~\eqref{eq.feedback_dyn} can be also computed without running stochastic simulations. For example, the autocovariance of streamwise velocity \mbox{fluctuations is given by}
\be
	\Phi_{uu}(\bk,\tau)
	\;=\;
	\lim\limits_{t \, \to \, \infty}
	\left<u(\bk,t+\tau)\,u^*(\bk,t)\right>,
\ee
where $u(\bk,t+\tau)$ is computed from~\eqref{eq.feedback_dyn},
\be
	u(\bk,t+\tau)
	\;=\;
	C_u(\bk)\, \mre^{A_f(\bk) \tau}
	\bpsi(\bk,t)
	\,+\,
	\ds{\int_t^{t+\tau}} C_u(\bk)\,\mre^{A_f(\bk)(t+\tau-\zeta)}\, B(\bk)\, \bw(\bk,\zeta)\,\mrd \zeta.
\ee
Since the state $\bpsi(\bk,t)$ and the white-in-time input $\bw(\bk,t)$ are not correlated, we have
\be
	\ba{rcl}
	\Phi_{uu}(\bk,\tau)
	&\!\! = \!\!&
	\lim\limits_{t \, \to \, \infty}
	\left<
	C_u(\bk)\,\mre^{A_f(\bk) \tau}
	\bpsi(\bk,t)\,
	\bpsi^*(\bk,t)\, C_u^*(\bk)
	\right>
	\\[.2cm]
	&\!\! = \!\!&
	C_u(\bk)\,\mre^{A_f(\bk) \tau} X(\bk) \,C_u^*(\bk),
	\ea
\ee
where $X(\bk)$ is the steady-state covariance matrix of $\bpsi$. Correlations between other velocity components can be obtained in a similar way. Note that, at any $\bk$ and $\tau$, the diagonal entries of the matrix $\Phi_{uu}(\bk,\tau)$ provide information about two-point temporal correlations at various wall-normal locations. 

\subsection{Spatio-temporal frequency responses}
	\label{sec.stoch_anal}

\begin{figure}
\begin{center}
	\begin{tabular}{cccc}
	\hspace{-.3cm} \subfigure[]{\label{fig.FreqResp_PSD_ArtAedAf_R186_kx2p5_kz7}}
       &&
       \hspace{-.8cm} \subfigure[]{\label{fig.FreqResp_maxsig_ArtAedAf_R186_kx2p5_kz7}}
	&
	\\[-.5cm]
	\hspace{.05cm}
	\begin{tabular}{c}
		\vspace{.2cm}
		{\normalsize \rotatebox{90}{$\Pi_\bv (\bk,\omega)$}}
	\end{tabular}
	&
	\hspace{-.5cm}
	\begin{tabular}{c}
		\includegraphics[width=6.15cm]{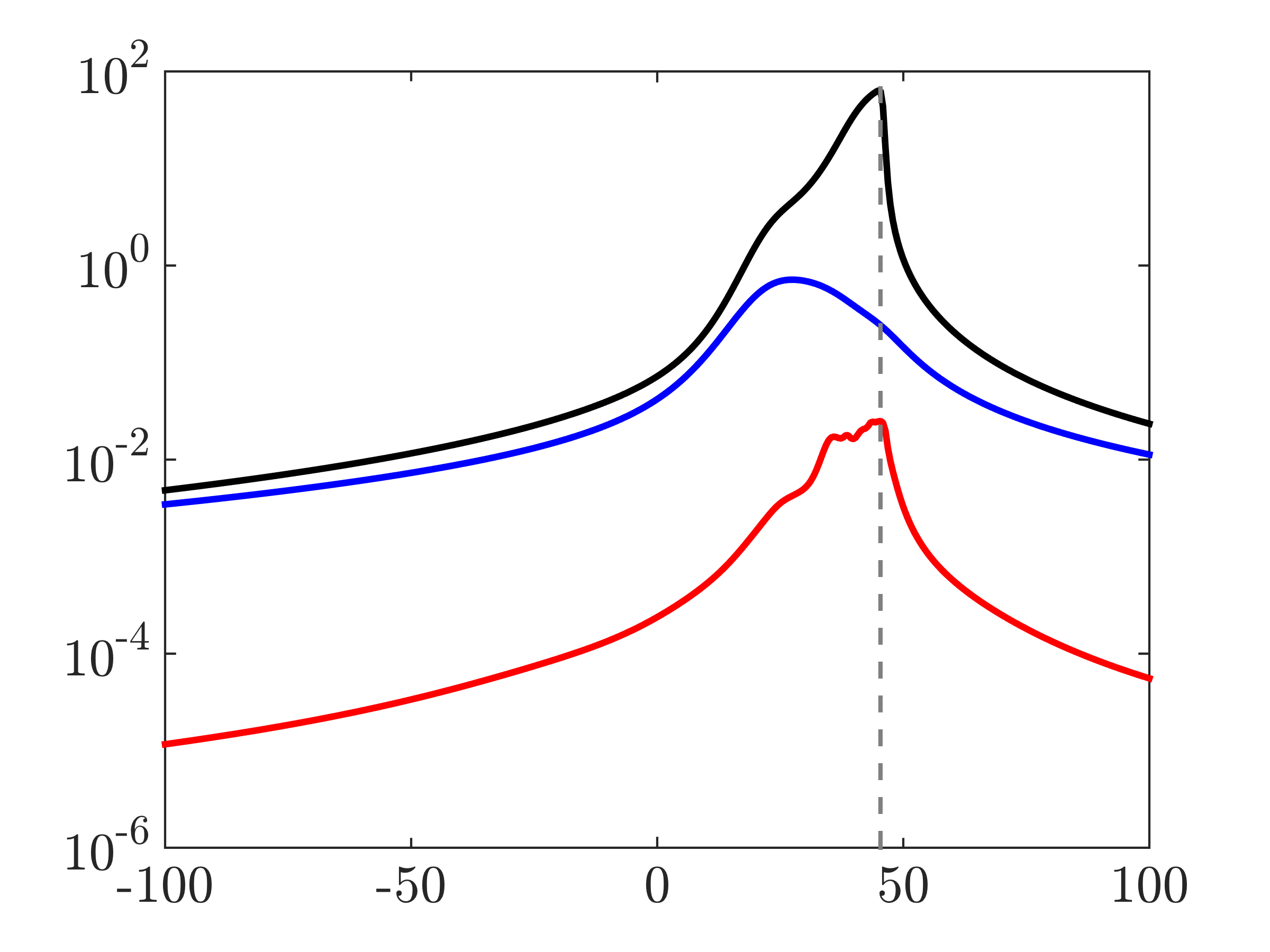}
		\\[-.1cm]
		~\,{\normalsize $\omega$}
	\end{tabular}
	&
	\hspace{-.35cm}
	\begin{tabular}{c}
		\vspace{.2cm}
		{\normalsize \rotatebox{90}{$G_\bv (\bk,\omega)$}}
	\end{tabular}
	&
	\hspace{-.5cm}
	\begin{tabular}{c}
		\includegraphics[width=6.15cm]{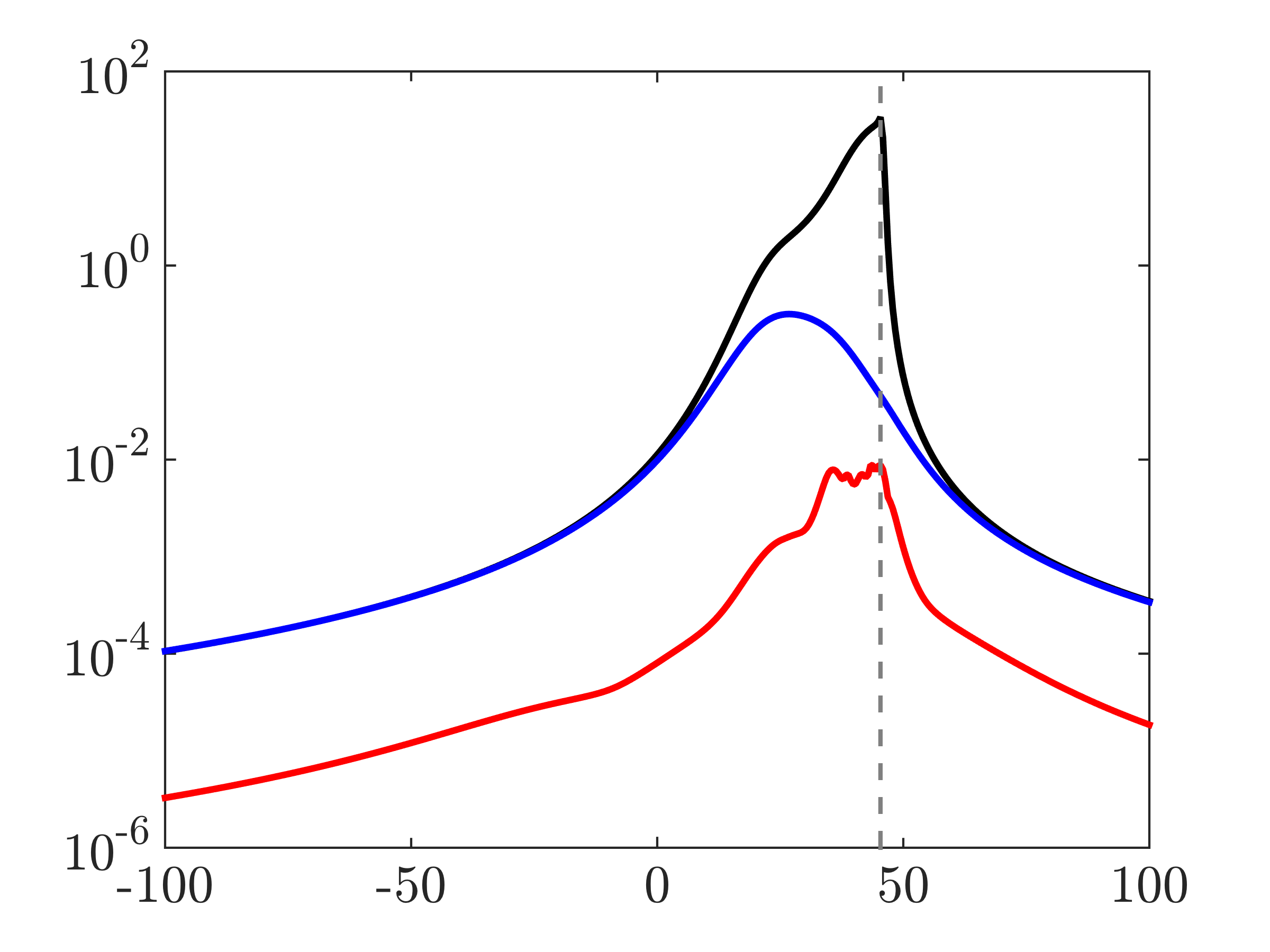}
		\\[-.1cm]
		~\,{\normalsize $\omega$}
	\end{tabular}
	\end{tabular}
\end{center}
\caption{(a) Power spectral density $\Pi_\bv (\bk,\omega)$ and (b) worst-case amplification $G_\bv (\bk,\omega)$ as a function of temporal frequency $\omega$ in turbulent channel flow with $R_\tau=186$ and $\bk=(2.5,7)$, resulting from the linearized NS model~\eqref{eq.lnse1} (black curve), an eddy-viscosity-enhanced linearized NS model (blue curve), and the modified linearized dynamics~\eqref{eq.feedback_dyn} (red curve).}
\label{fig.FreqResp_ArtAedAf}
\end{figure}

Figure~\ref{fig.FreqResp_PSD_ArtAedAf_R186_kx2p5_kz7} compares the power spectral densities of three linear approximations of the NS equations in turbulent flow with $R_\tau = 186$ and $\bk=(2.5,7)$. These are respectively given by the linearized NS equations~\eqref{eq.lnse1}, an eddy-viscosity-enhanced linearized NS equations, and the low-rank modification of the linearized dynamics~\eqref{eq.feedback_dyn}. For the standard and eddy-viscosity-enhanced linearizations, the input matrix $B(\bk)$ excites all degrees of freedom in the evolution model~\citep{jovbamJFM05}; for the modified dynamics~\eqref{eq.feedback_dyn}, the input matrix $B(\bk)$ comes from the framework of \S~\ref{sec.mcp}. While the temporal frequency at which the PSD peaks is similar for the linearized NS equations and the modified dynamics ({$\omega=45$}), it is smaller for the eddy-viscosity-enhanced model ({$\omega=27$}). Compared to the linearization around turbulent mean profile, both the eddy-viscosity-enhanced model and our model attenuate amplification of disturbances at all temporal frequencies. Thus, the low-rank modification of the linearized NS equation introduces eddy-viscosity-like features and provides additional damping across temporal frequencies. 

Figure~\ref{fig.FreqResp_maxsig_ArtAedAf_R186_kx2p5_kz7} illustrates similar trends for the worst-case amplification to harmonic forcing, $G_\bv(\bk,\omega)$. We recall that the PSD quantifies the total energy amplification, $\sum_{i} \sigma_i^2$, and that the worst case amplification is determined by the largest singular value of the frequency response, $\sigma_{\max}^2$. Clearly, in both cases, the low-rank modification reduces amplification of disturbances relative to the linearization around turbulent mean velocity but does not modify the temporal frequency at which the energy amplification peaks.

\begin{figure}
	\begin{tabular}{cccccc}
            	\hspace{-.5cm} \subfigure[]{\label{fig.Evv_Art}}
            	&
		\hspace{-.3cm}
		{\small linearized NS}
		&
            	\hspace{-.75cm} \subfigure[]{\label{fig.Evv_Aed}}
            	&
		\hspace{-.3cm}
		{\small eddy-viscosity model}
		&
            	\hspace{-.75cm} \subfigure[]{\label{fig.Evv_Af}}
            	&
		\hspace{-.3cm}
		{\small modified dynamics}
            	\\[-.35cm]
            	\hspace{-.2cm}
            	\begin{tabular}{c}
            		\vspace{.4cm}
            		\hspace{-.1cm}
            		\small{$c$}
            	\end{tabular}
            	&
            	\hspace{-.4cm}
            	\begin{tabular}{c}
            		\includegraphics[width=4.1cm]{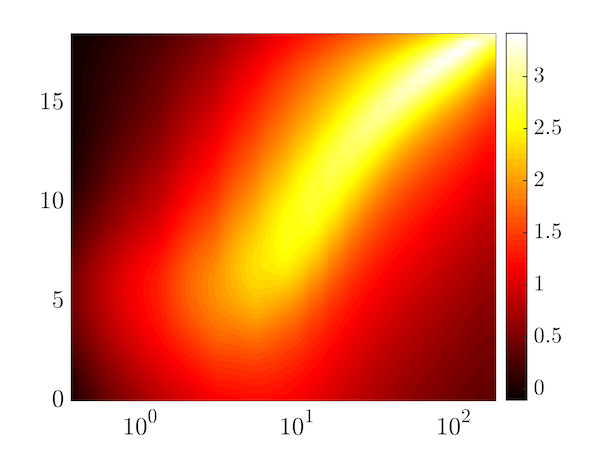}
            	\end{tabular}
            	&&
            	\hspace{-.34cm}
            	\begin{tabular}{c}
            		\includegraphics[width=4.1cm]{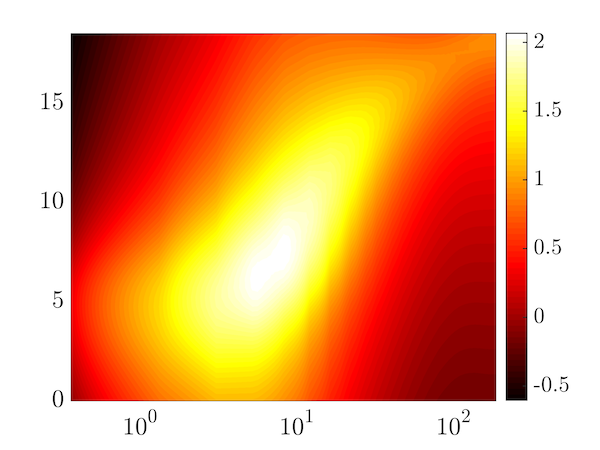}
            	\end{tabular}
            	&&
            	\hspace{-.36cm}
            	\begin{tabular}{c}
            		\includegraphics[width=4.1cm]{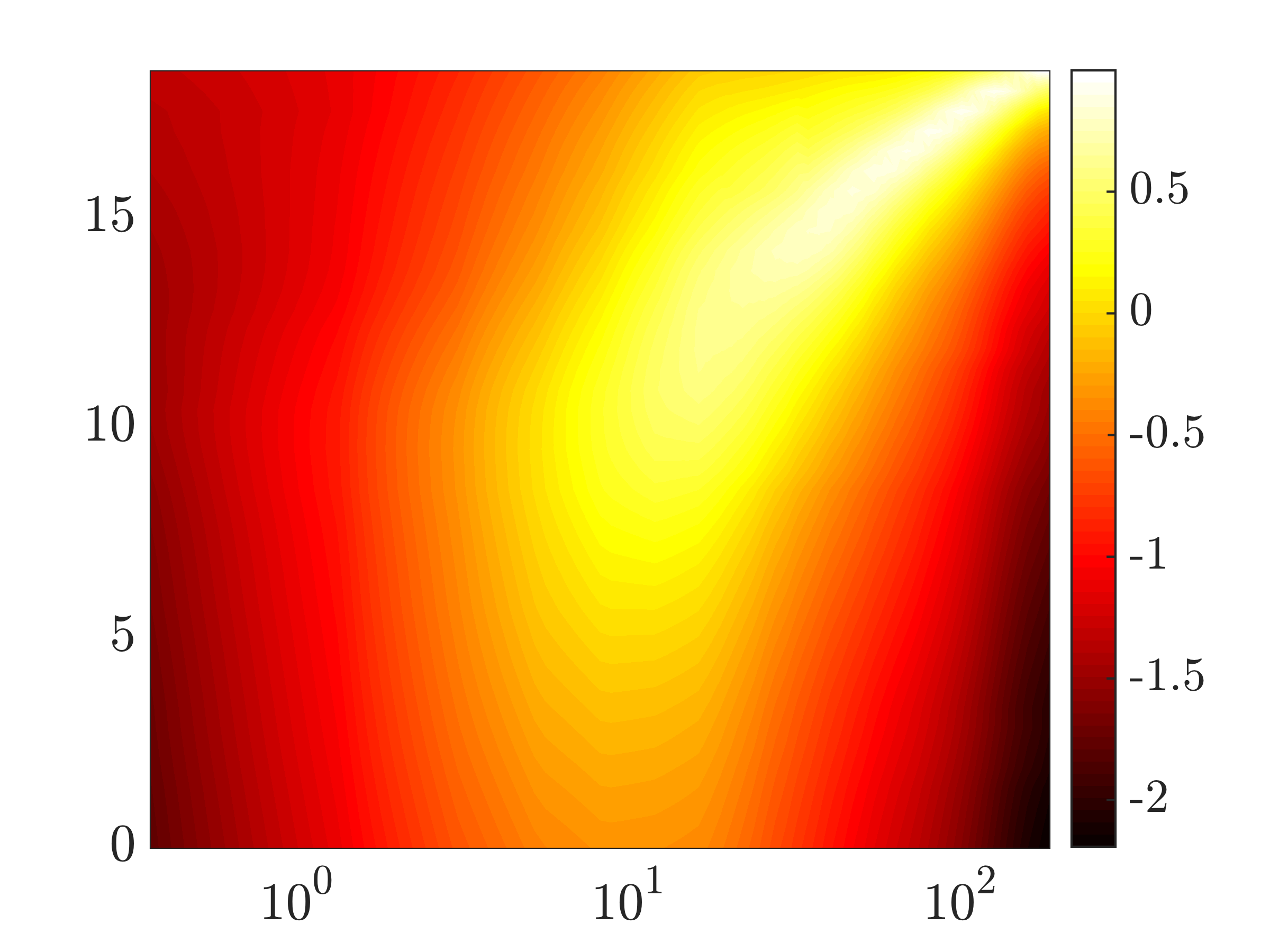}
            	\end{tabular}
		\\[-.2cm]
		\hspace{-.5cm} \subfigure[]{\label{fig.Evv_Art_normalized}}
            	&&
            	\hspace{-.75cm} \subfigure[]{\label{fig.Evv_Aed_normalized}}
            	&&
            	\hspace{-.75cm} \subfigure[]{\label{fig.Evv_Af_normalized}}
            	&
            	\\[-.5cm]
            	\hspace{-.2cm}
            	\begin{tabular}{c}
            		\vspace{.4cm}
            		\hspace{-.1cm}
            		\small{$c$}
            	\end{tabular}
            	&
            	\hspace{-.4cm}
            	\begin{tabular}{c}
            		\includegraphics[width=4.1cm]{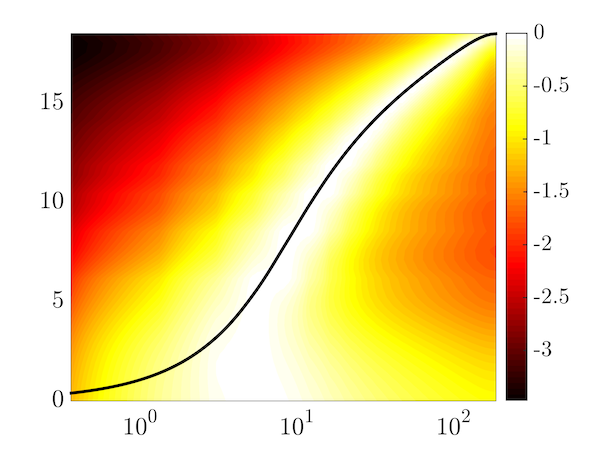}
            		\\[-.1cm]
            		\small{$y^+$}
            	\end{tabular}
            	&&
            	\hspace{-.34cm}
            	\begin{tabular}{c}
            		\includegraphics[width=4.1cm]{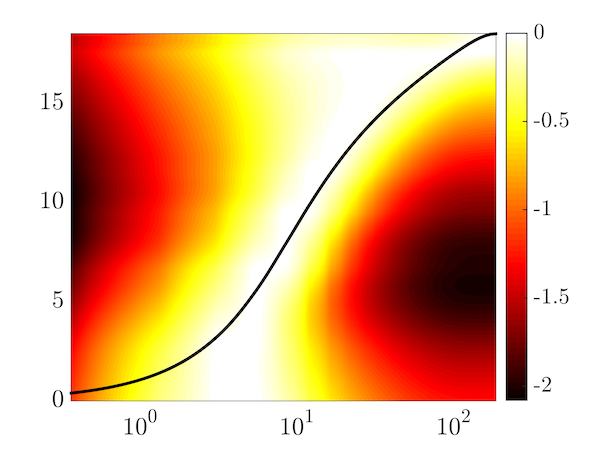}
            		\\[-.1cm]
            		\small{$y^+$}
            	\end{tabular}
            	&&
            	\hspace{-.36cm}
            	\begin{tabular}{c}
            		\includegraphics[width=4.1cm]{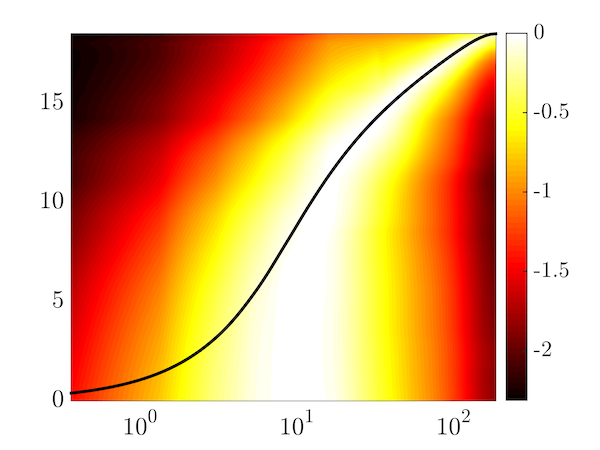}
            		\\[-.1cm]
            		\small{$y^+$}
            	\end{tabular}
	\end{tabular}
		\caption{The one-dimensional energy density $E_{\bv\bv}$ as a function of $y^+$ and $c$ computed using the linearized NS equations~\eqref{eq.lnse1} (a, d), an eddy-viscosity-enhanced linearized NS model (b, e), and the modified dynamics~\eqref{eq.feedback_dyn} (c, f) for turbulent channel flow with $R_\tau=186$. Plots (d-f) show the energy density normalized by its maximum value over $y$ for fixed values of $c$. The colors are in logarithmic scale. The turbulent mean velocity is marked by the black curve in (d-f).}
	\label{fig.1D_energydensity}
\end{figure}

Figure~\ref{fig.1D_energydensity} shows the one-dimensional energy density $E_{\bv\bv}$ as a function of propagation speed $c=\omega/k_x$ and $y^+$ for turbulent channel flow with $R_\tau=186$. This quantity is obtained by integrating $\diag \left( T_{\bv \bw} (\bk, \omega)\, T_{\bv \bw}^* (\bk, \omega) \right)$ over {
$50\times51$ logarithmically spaced wavenumbers with $0< k_x < k_{x,\max}$ and $0< k_z < k_{z,\max}$, and} for a range of wave speeds $0 < c < U_{\mathrm{c}}$, where $U_{\mathrm{c}}$ is the mean centerline velocity. 
{Here, $k_{x,\max}=42.5$ and $k_{z,\max}=84.5$ are the largest wavenumbers used in the DNS of~\cite{deljim03} and~\cite{deljimzanmos04}; they capture the energetically significant portion of the premultiplied turbulent energy spectrum of channel flow with $R_\tau=186$.} In contrast to {the} PSD, $E_{\bv\bv}$ provides insight into the wall-normal variation of the energy amplification in stochastically-forced flows. For a fixed value of $c$, the energy density is localized in a narrow wall-normal region; see figures~\ref{fig.1D_energydensity}(a-c). To highlight this localization, we normalize the energy density by its maximum value over $y$ for fixed values of $c$. As shown in figures~\ref{fig.1D_energydensity}(d-f), the normalized energy density peaks in the vicinity of the wall-normal location where the turbulent mean velocity, marked by the thick black lines, equals the wave speed; \cite{mcksha10} argued that the emergence of critical layers is because the resolvent norm peaks for $c \approx U(y)$. Our observations are in agreement with~\cite{moashatromck13} where the contribution of the principal resolvent mode to the streamwise energy density was studied for the NS equations linearized around turbulent mean velocity in channel flow with $R_\tau=2003$. 

Figure~\ref{fig.1D_energydensity} shows that at each wall-normal location $y$ the modes that convect at the critical speed $c=U(y)$ are most amplified. This observation holds in almost the entire channel and, with slight disparity, is valid for all three models. Based on Taylor's frozen turbulence hypothesis~\citep{tay38}, flow structures in turbulent flows propagate downstream at a speed that is close to the local mean velocity. For a large extent of the channel height our observation is in agreement with this hypothesis. However, for all three models, the scatter in the energy density increases as the wall is approached. In addition, {near-wall modes peak at $y^+ \approx 10$ and they travel at speeds that are smaller than the local mean velocity.} Similar observations were made in analytical~\citep{moashatromck13}, experimental~\citep{moncho09,lehguamck11} and numerical studies~\citep{kimhus93,deljim09}, thereby suggesting that application of Taylor's hypothesis can yield inaccurate energy spectra close to the wall. 

From figure~\ref{fig.Evv_Aed} it is evident that while eddy-viscosity enhancement reduces the scatter close to the wall, the energy density resulting from this model is less concentrated away from the wall. Even though the low-rank modification does not significantly alter the general trend in the energy density, figure~\ref{fig.Evv_Af_normalized} illustrates that, for $y^+<15$, the scatter in the normalized energy density increases. Namely, the highest energy amplification no longer occurs at velocities that are close to the local mean velocity near the wall.

 \subsubsection*{Principal output and forcing directions}
    \label{sec.turb_structures}
	
	We now utilize the singular value decomposition to analyze the principal output and forcing directions for flow with $R_\tau=2003$ and $\bk=(1,6)$. This wavenumber pair is associated with {the energetic length scale of VLSMs in canonical flows~\citep{monstewilcho07,hutmar07a,monhutngmarcho09}} and it has been previously considered in the study of coherent structures in turbulent pipes~\citep{shamck13}. 
	
	For $N=201$ wall-normal collocation points, the matrix $Z(\bk)$ that results from optimization problem~\eqref{eq.CP} with $\gamma=10^3$ has $7$ positive and $5$ negative eigenvalues. Therefore, the input matrix $B(\bk)$, which introduces colored-in-time forcing into the linearized NS dynamics, has $7$ columns. This choice of $\gamma$ provides a balance between the number of colored-in-time inputs and quality of completion of the two-point correlations. We examine temporal frequencies $\omega=21.4$ and $26.5$ which correspond to the streamwise propagation speed $c=\omega/k_x$ of structures that reside in the middle of the logarithmic region~\citep{marmonhulsmi13} and the peak of PSD (not shown here), respectively.
	
\begin{figure}
\begin{center}
	\begin{tabular}{cccc}
	\hspace{-.7cm} \subfigure[]{\label{fig.singularvalues_Twv_omega21p4}}
       &&
       \hspace{-.7cm} \subfigure[]{\label{fig.singularvalues_Twv_omega26p5}}
	&
	\\[-.5cm]
	\hspace{-.35cm}
	\begin{tabular}{c}
		\vspace{.2cm}
		{\normalsize \rotatebox{90}{$\sigma_j^2/(\ds{\sum_{i=1}^r \sigma_i^2)}$}}
	\end{tabular}
	&
	\hspace{-.5cm}
	\begin{tabular}{c}
		\includegraphics[width=6.15cm]{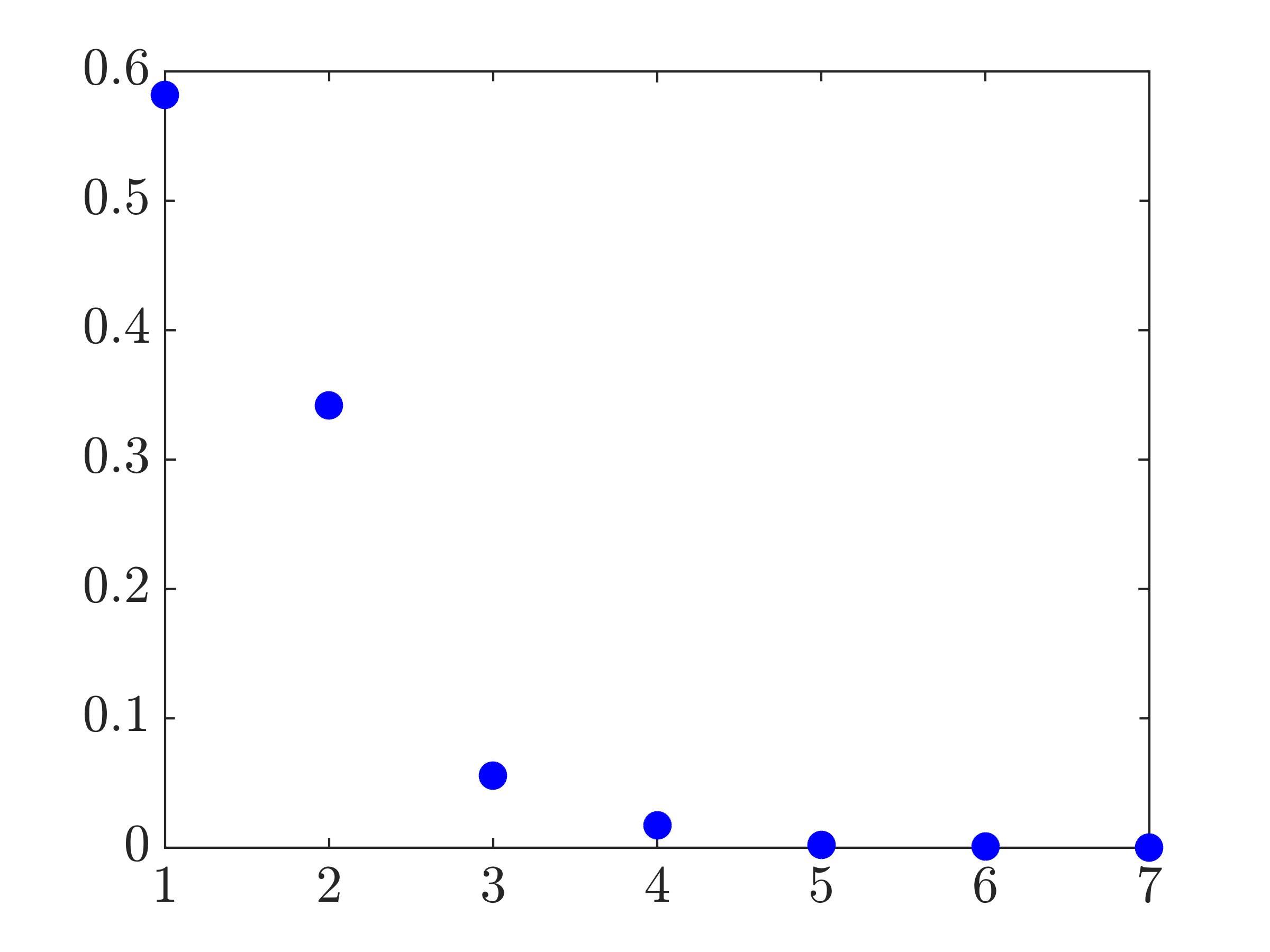}
		\\[-.1cm]
		~\,{\normalsize $j$}
	\end{tabular}
	&
	&
	\hspace{-.3cm}
	\begin{tabular}{c}
		\includegraphics[width=6.15cm]{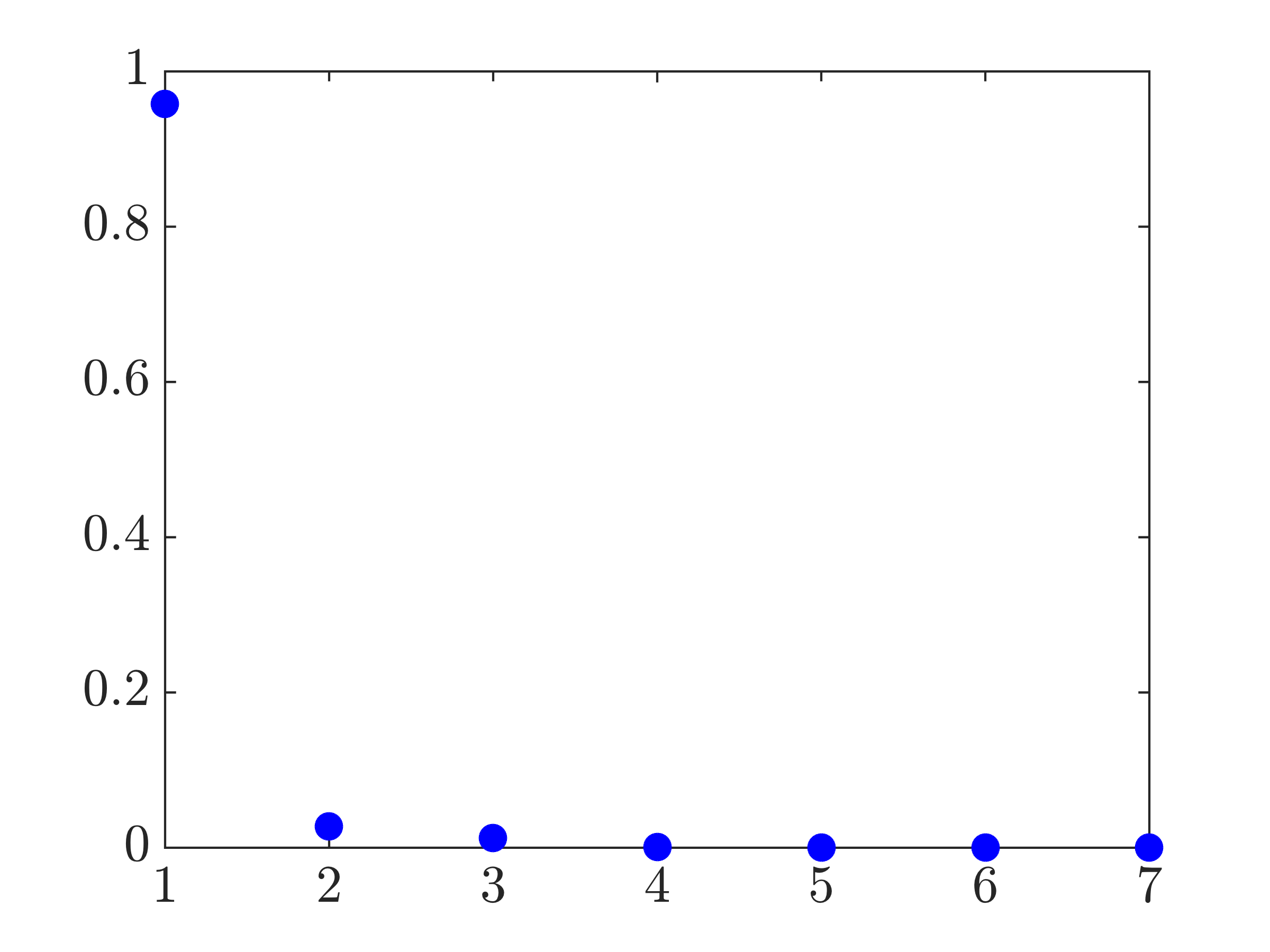}
		\\[-.1cm]
		~\,{\normalsize $j$}
	\end{tabular}
	\end{tabular}
\end{center}
\caption{Contribution of the response directions $\tilde{\xi}_j$ to the total energy in turbulent channel flow with $R_\tau=2003$ and $\bk=(1,6)$. The modified dynamics~\eqref{eq.feedback_dyn} are driven by harmonic excitation with temporal frequency (a) $\omega=21.4$ and (b) $\omega=26.5$.}
\label{fig.singularvalues_Twv}
\end{figure}

Figure~\ref{fig.singularvalues_Twv} shows the contribution of each output direction $\tilde{\xi}_j$ in~\eqref{eq.inout_analysis} to the energy amplification, $\sigma_j^2/(\sum_{i} \sigma_i^2)$. Since $B(\bk)$ is a tall matrix with $7$ columns, the frequency response $T_{\bv \bw}(\bk,\omega)$ has only $7$ non-zero singular values. For $\omega=21.4$, the principal output direction, which corresponds to the largest singular value $\sigma_{\max}$, approximately contains $58\%$ of the total energy. The second largest singular value contributes an additional $34\%$. Thus, the two most amplified output directions account for $92\%$ of the total energy. For  $\omega=26.5$, the largest singular value contains more than $95\%$ of the total energy. This further supports the finding that the turbulent velocity spectra and the Reynolds stress co-spectrum can be approximated with a few resolvent modes~\citep{moashatromck13,moajovtroshamckPOF14}.

\begin{figure}
	\begin{tabular}{cccc}
		\hspace{.2cm}\subfigure[]{\label{fig.principlesvd_wtov_kx1_kz6_c21p4}}&&\hspace{-.3cm}\subfigure[]{\label{fig.Vorticity_wtov_kx1_kz6_c21p4}}
		\\[-.6cm]
		&
		\hspace{-.3cm}
		\begin{tikzpicture}
                    \draw (0, 0) node[inner sep=0] {\includegraphics[width=6.15cm]{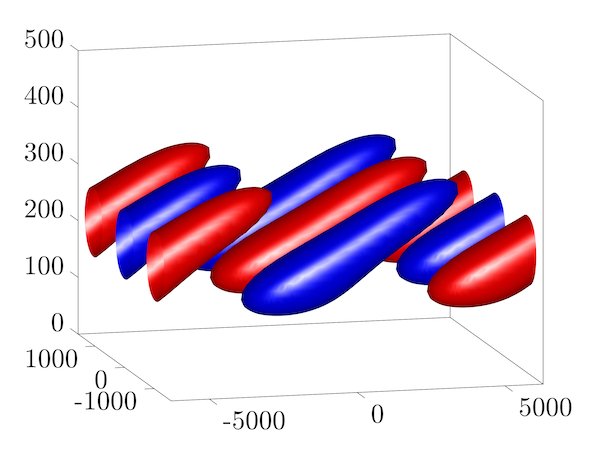}};
                    \draw (-3.2, .4) node {\normalsize $y^+$};
                     \draw (-3, -1.8) node {\normalsize $z^+$};
                      \draw (.8, -2.4) node {\normalsize $x^+$};
                      \draw [-latex] (-.2,1.3) -- (.8,1.3);
                      \draw (-0.4,1.3) node {\large $c$};
                \end{tikzpicture}
		&&
		\hspace{-.05cm}
		\begin{tikzpicture}
                    \draw (0, 0) node[inner sep=0] {\includegraphics[width=6.15cm]{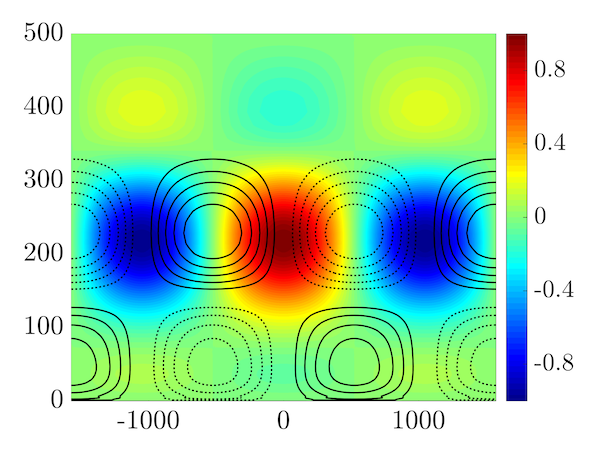}};
                      \draw (-0.1, -2.4) node {\normalsize $z^+$};
                \end{tikzpicture}
                \\[-.2cm]
                \hspace{.2cm}\subfigure[]{\label{fig.principlesvd_wtov_kx1_kz6_c26p5}}&&\hspace{-.3cm}\subfigure[]{\label{fig.Vorticity_wtov_kx1_kz6_c26p5}}
		\\[-.6cm]
		&
		\hspace{-.3cm}
		\begin{tikzpicture}
                    \draw (0, 0) node[inner sep=0] {\includegraphics[width=6.15cm]{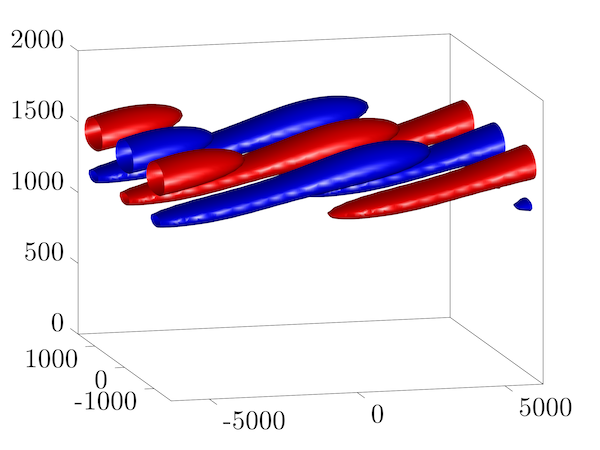}};
                    \draw (-3.2, .4) node {\normalsize $y^+$};
                     \draw (-3, -1.8) node {\normalsize $z^+$};
                      \draw (.8, -2.4) node {\normalsize $x^+$};
                      \draw [-latex] (-.2,-0.6) -- (.8,-0.6);
                      \draw (-0.4,-0.6) node {\large $c$};
                \end{tikzpicture}
		&&
		\hspace{-.05cm}
		\begin{tikzpicture}
                    \draw (0, 0) node[inner sep=0] {\includegraphics[width=6.15cm]{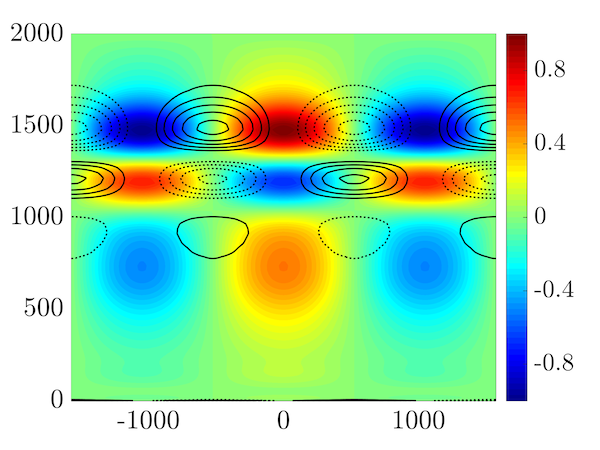}};
                      \draw (-0.1, -2.4) node {\normalsize $z^+$};
                \end{tikzpicture}
	\end{tabular}
	\caption{Spatial structure of the principal response directions of the frequency response $T_{\bv \bw}(\bk,\omega)$ in turbulent channel flow with $R_\tau=2003$, $\bk={(1,6)}$, at $t=0$ for (a,b) $\omega=21.4$ and (c,d) $\omega=26.5$. (a,c) Isosurfaces of the streamwise velocity; red and blue colors denote regions of high and low velocity at $60\%$ of their largest values. (b,d) Spatial structure of the streamwise velocity (color plots) and vorticity (contour lines).}
	\label{fig.Af_svd_vorticity_wtov_kx1_kz6}
\end{figure}

{
Relations~\eqref{eq.physical_uvw} can be used to visualize the spatial structure of each output direction $\tilde{\xi}_j$.} Figures~\ref{fig.Af_svd_vorticity_wtov_kx1_kz6}(a,b) show the spatial structure of the streamwise component of the principal output response in turbulent channel flow at $t=0$ and for $\omega=21.4$. These streamwise elongated structures are sandwiched between counter-rotating vortical motions in the cross-stream plane (cf.\ figure~\ref{fig.Vorticity_wtov_kx1_kz6_c21p4}) and they contain alternating regions of fast- and slow-moving fluid (which are slightly inclined to the wall). We see that they reside in the logarithmic region with their largest values roughly taking place in the middle of this region. Even though these structures do not capture the full complexity of turbulent flows, they are reminiscent of VLSMs that form at large Reynolds numbers. Figures~\ref{fig.Af_svd_vorticity_wtov_kx1_kz6}(c,d) show the spatial structure of the principal output response for the temporal frequency $\omega=26.5$. Compared to $\omega=21.4$, these streamwise elongated structures are longer and reside in the outer region of the channel with no protrusion to the logarithmic layer. 

A similar approach can be used to study the spatial structure of colored-in-time forcing to the linearized NS equations. This is accomplished by passing the output of filter~\eqref{eq.filter-tf} through the input matrix $B(\bk)$ and examining the resulting frequency response, $B(\bk) T_f(\bk,\omega)$. For $\omega=21.4$ and $26.5$, the principal output directions respectively contain $62\%$ and $88\%$ of the total energy of the forcing to the linearized NS equations.

\begin{figure}
	\begin{tabular}{cccc}
		\hspace{.2cm}\subfigure[]{\label{fig.principlesvd_wtof_kx1_kz6_c21p4}}&&\hspace{-.3cm}\subfigure[]{\label{fig.Vorticity_wtof_kx1_kz6_c21p4}}
		\\[-.6cm]
		&
		\hspace{-.3cm}
		\begin{tikzpicture}
                    \draw (0, 0) node[inner sep=0] {\includegraphics[width=6.15cm]{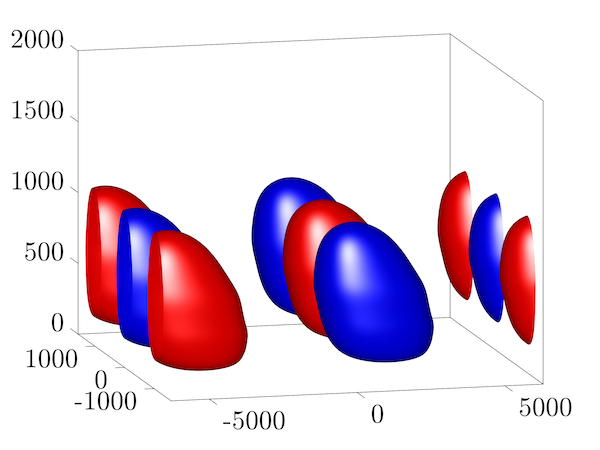}};
                    \draw (-3.2, .4) node {\normalsize $y^+$};
                     \draw (-3, -1.8) node {\normalsize $z^+$};
                      \draw (.8, -2.4) node {\normalsize $x^+$};
                      \draw [-latex] (-.2,1.3) -- (.8,1.3);
                      \draw (-0.4,1.3) node {\large $c$};
                \end{tikzpicture}
		&&
		\hspace{-.05cm}
		\begin{tikzpicture}
                    \draw (0, 0) node[inner sep=0] {\includegraphics[width=6.15cm]{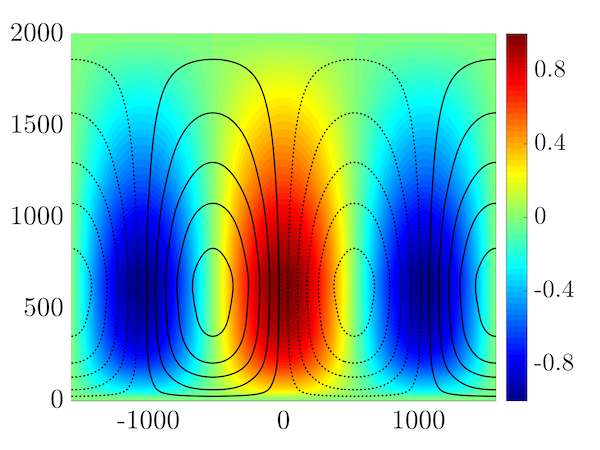}};
                      \draw (-0.1, -2.4) node {\normalsize $z^+$};
                \end{tikzpicture}
                \\[-.2cm]
                \hspace{.2cm}\subfigure[]{\label{fig.principlesvd_wtof_kx1_kz6_c26p5}}&&\hspace{-.3cm}\subfigure[]{\label{fig.Vorticity_wtof_kx1_kz6_c26p5}}
		\\[-.6cm]
		&
		\hspace{-.3cm}
		\begin{tikzpicture}
                    \draw (0, 0) node[inner sep=0] {\includegraphics[width=6.15cm]{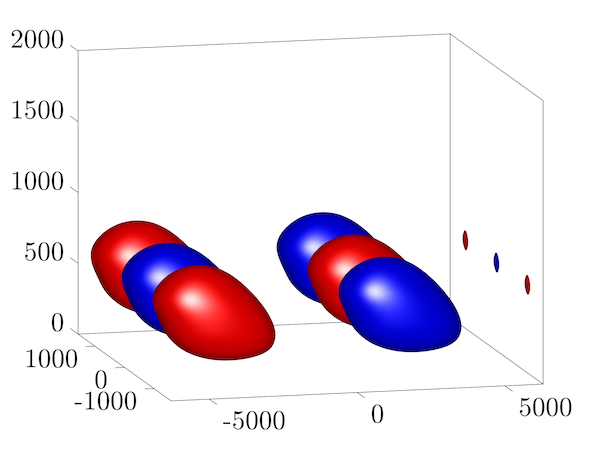}};
                    \draw (-3.2, .4) node {\normalsize $y^+$};
                     \draw (-3, -1.8) node {\normalsize $z^+$};
                      \draw (.8, -2.4) node {\normalsize $x^+$};
                      \draw [-latex] (-.2,1.3) -- (.8,1.3);
                      \draw (-0.4,1.3) node {\large $c$};
                \end{tikzpicture}
		&&
		\hspace{-.05cm}
		\begin{tikzpicture}
                    \draw (0, 0) node[inner sep=0] {\includegraphics[width=6.15cm]{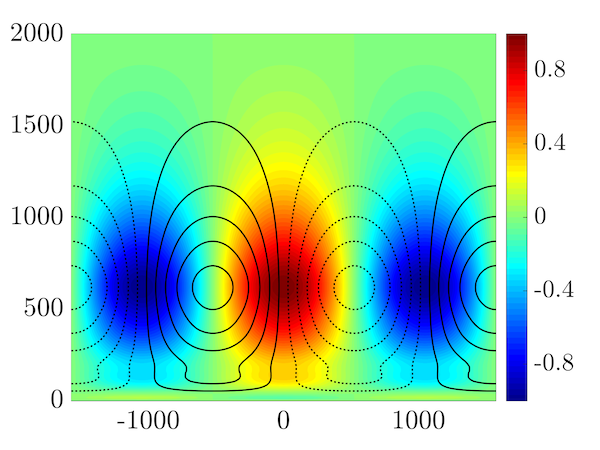}};
                      \draw (-0.1, -2.4) node {\normalsize $z^+$};
                \end{tikzpicture}
	\end{tabular}
	\caption{Spatial structure of the principal response of the operator $B(\bk)T_f(\bk,\omega)$ in turbulent channel flow with $R_\tau=2003$, $\bk=(1,6)$, at $t=0$ for (a,b) $\omega=21.4$ and (c,d) $\omega=26.5$. (a,c) Isosurfaces of the streamwise velocity; red and blue colors denote regions of high and low velocity at $60\%$ of their largest values. (b,d) Spatial structure of the streamwise component of the response (color plots) and vorticity (contour lines).}
	\label{fig.Af_svd_vorticity_wtof_kx1_kz6}
\end{figure}

Figures~\ref{fig.Af_svd_vorticity_wtof_kx1_kz6}(a,b) show the spatial structure of the principal wall-normal forcing component at $t=0$ and $\omega=21.4$. We see that the forcing to the Orr-Sommerfeld equation affects regions of the channel that begin in the logarithmic layer and extend to the middle of the channel. The color plot in figure~{\ref{fig.Vorticity_wtof_kx1_kz6_c21p4}} shows that the largest value of the normal forcing component is in the outer layer. This suggests that the turbulent flow structures that reside in the logarithmic layer are induced by a forcing which is not limited to the logarithmic layer. Similarly, for $\omega=26.5$, figures~\ref{fig.Af_svd_vorticity_wtof_kx1_kz6}(c,d) illustrate that the most energetic component of the wall-normal forcing begins in the logarithmic layer and extends to the outer region of the channel. Compared to $\omega=21.4$, these structures are shorter and do not influence the middle of the channel.

 	\vspace*{-1ex}
 \subsection{Temporal two-point correlations}
	\label{sec.temp_features}
	
\begin{figure}
\begin{center}
	\begin{tabular}{cccc}
	\hspace{-.3cm} \subfigure[]{\label{fig.autocorrelation_ruu}}
       & 
       &
       \hspace{-.6cm} \subfigure[]{\label{fig.autocorrelation_yp15}}
	&
	\\[-.35cm]
	\hspace{.05cm}
	\begin{tabular}{c}
		\vspace{.4cm}
		{\normalsize \rotatebox{90}{$y$}}
	\end{tabular}
	&
	\hspace{-.45cm}
	\begin{tabular}{c}
		\includegraphics[width=6.15cm]{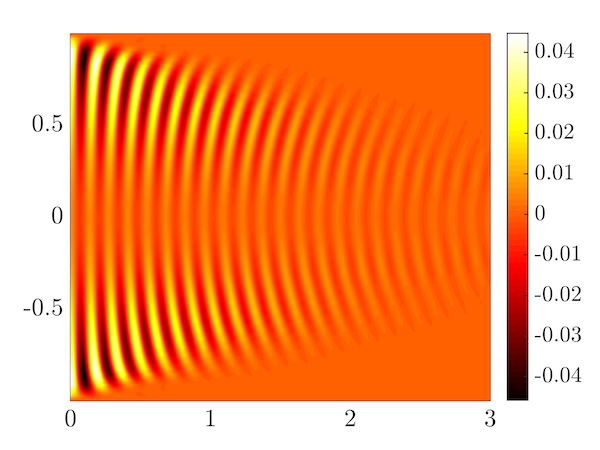}
		\\[-.1cm]
		\hspace{-.3cm}{\normalsize $\tau$}
	\end{tabular}
	&
	&
	\hspace{-.25cm}
	\begin{tabular}{c}
		\includegraphics[width=6.15cm]{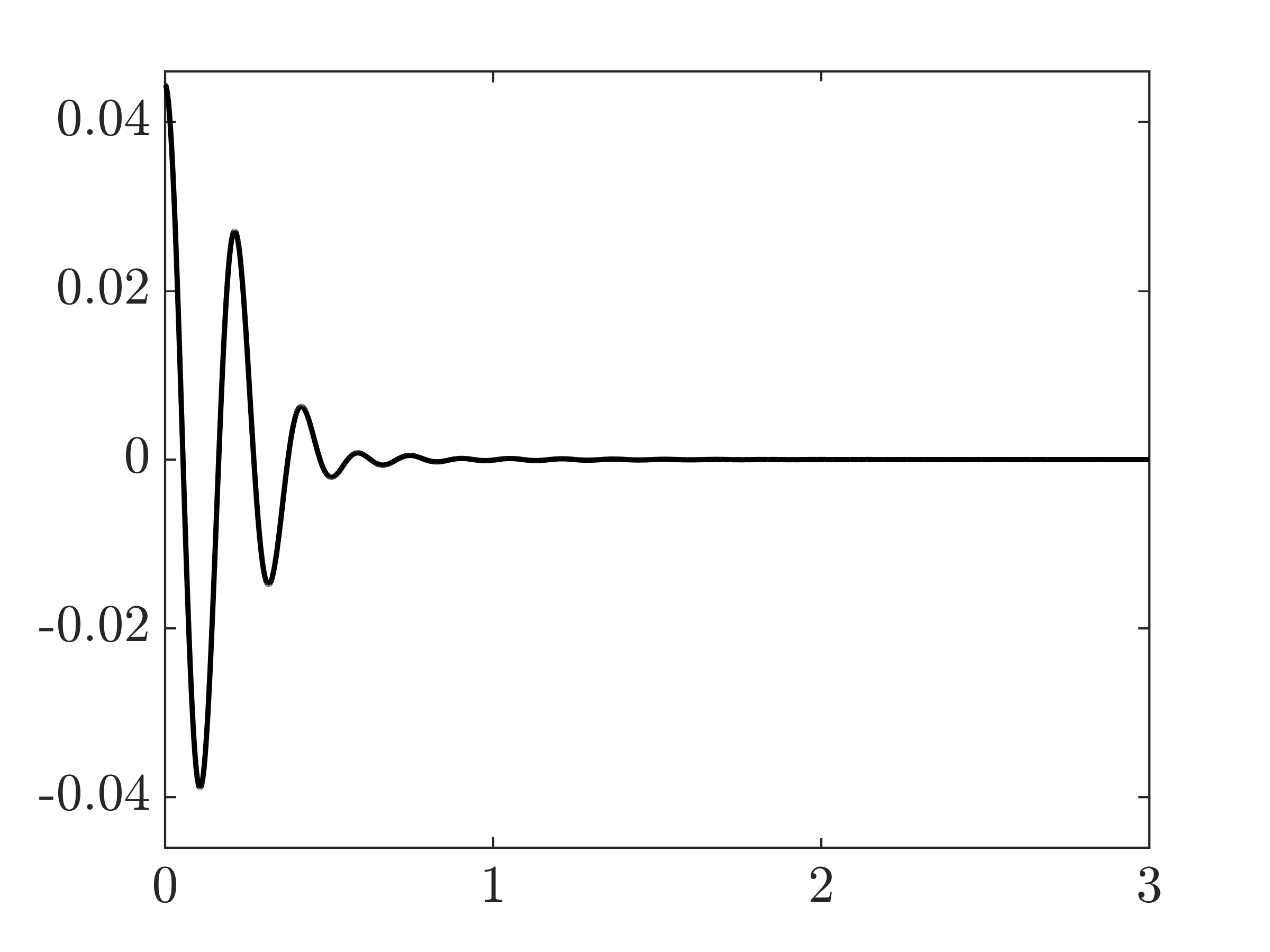}
		\\[-.1cm]
		~~{\normalsize $\tau$}
	\end{tabular}
	\end{tabular}
\end{center}
\caption{(a) The steady-state autocovariance $\Phi_{uu}(\bk,\tau)$ resulting from the modified dynamics~\eqref{eq.feedback_dyn} for turbulent channel flow with $R_\tau=186$ and $\bk=(2.5,7)$. (b) The same quantity plotted at $y^+=15$.}
\label{fig.autocorrelation_ruu_yp15}
\end{figure}

Figure~\ref{fig.autocorrelation_ruu} shows the dependence of the main diagonal of the steady-state autocovariance of streamwise velocity $\Phi_{uu}$ on the wall-normal coordinate $y$ and the time lag $\tau$ in flow with $R_\tau=186$ and $\bk=(2.5,7)$. Furthermore, figure~\ref{fig.autocorrelation_yp15} illustrates changes with $\tau$ at $y^+=15$. We observe attenuated oscillatory $\tau$-dependence where the period of oscillations increases as the wall is approached. For any wall-normal location $y$, the fundamental frequency $f_0$ of the corresponding diagonal entry of $\Phi_{uu}(\bk,\tau)$ can be used to estimate the streamwise convection velocity,
\be
	\label{eq.conv_velocity}
	c_u(\bk,y)
	\; = \;
	\dfrac{2\pi f_0}{k_x}.
	\ee

\begin{figure}
\begin{center}
	\begin{tabular}{cccc}
	\hspace{-.3cm} \subfigure[]{\label{fig.ConvectionVelocity_R180_kx2p5_kz7_a}}
       &&
       \hspace{-.6cm} \subfigure[]{\label{fig.ConvectionVelocity_R180_kx2p5_kz7_b}}
	&
	\\[-.5cm]
	\hspace{.05cm}
	\begin{tabular}{c}
		\vspace{.4cm}
		{\normalsize \rotatebox{90}{$c_u$}}
	\end{tabular}
	&
	\hspace{-.45cm}
	\begin{tabular}{c}
		\includegraphics[width=6.15cm]{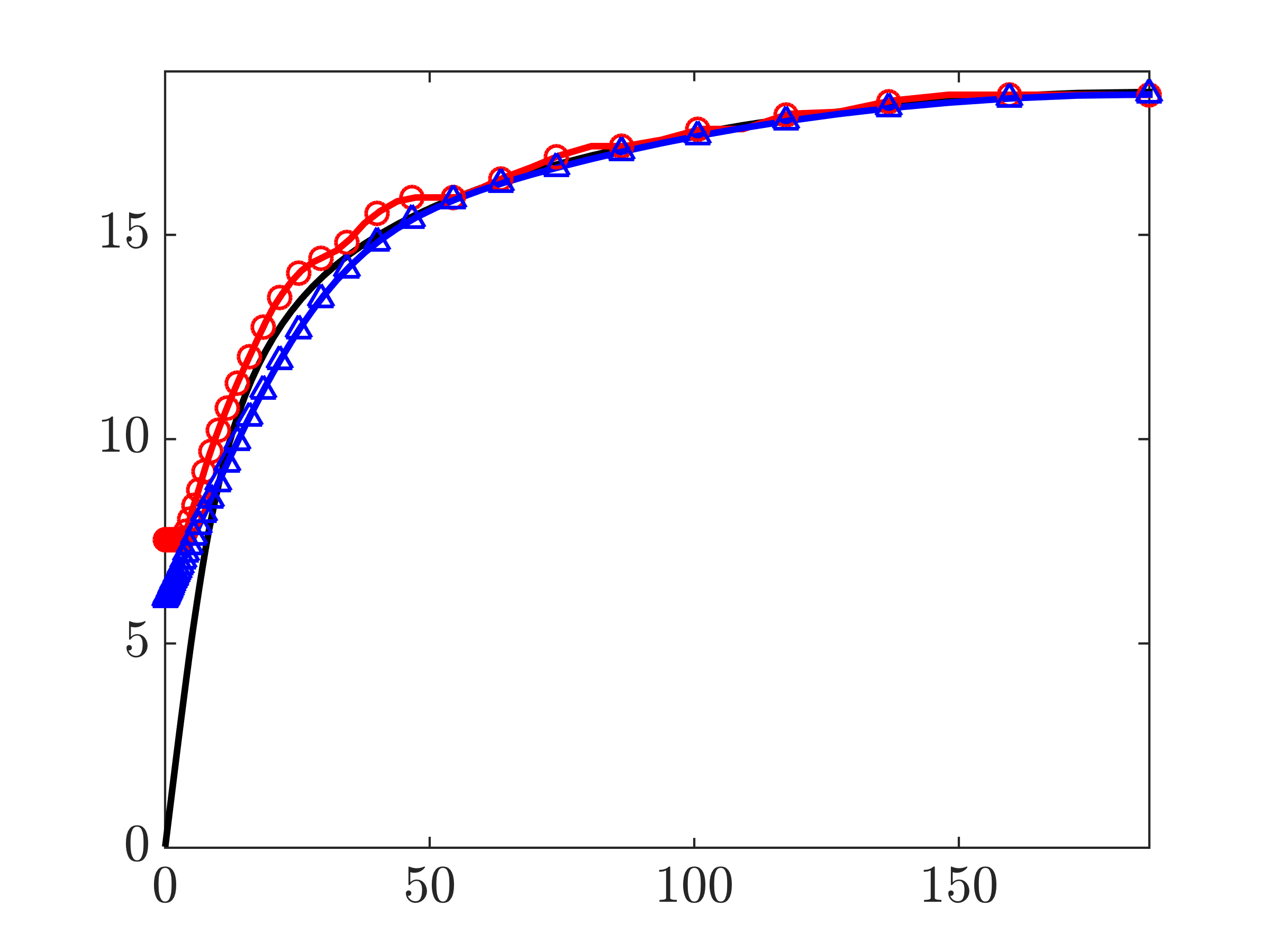}
		\\[-.1cm]
		{\normalsize $y^+$}
	\end{tabular}
	&
	\hspace{-.2cm}
	\begin{tabular}{c}
		\vspace{.4cm}
		{\normalsize \rotatebox{90}{$c_u$}}
	\end{tabular}
	&
	\hspace{-.46cm}
	\begin{tabular}{c}
		\includegraphics[width=6.15cm]{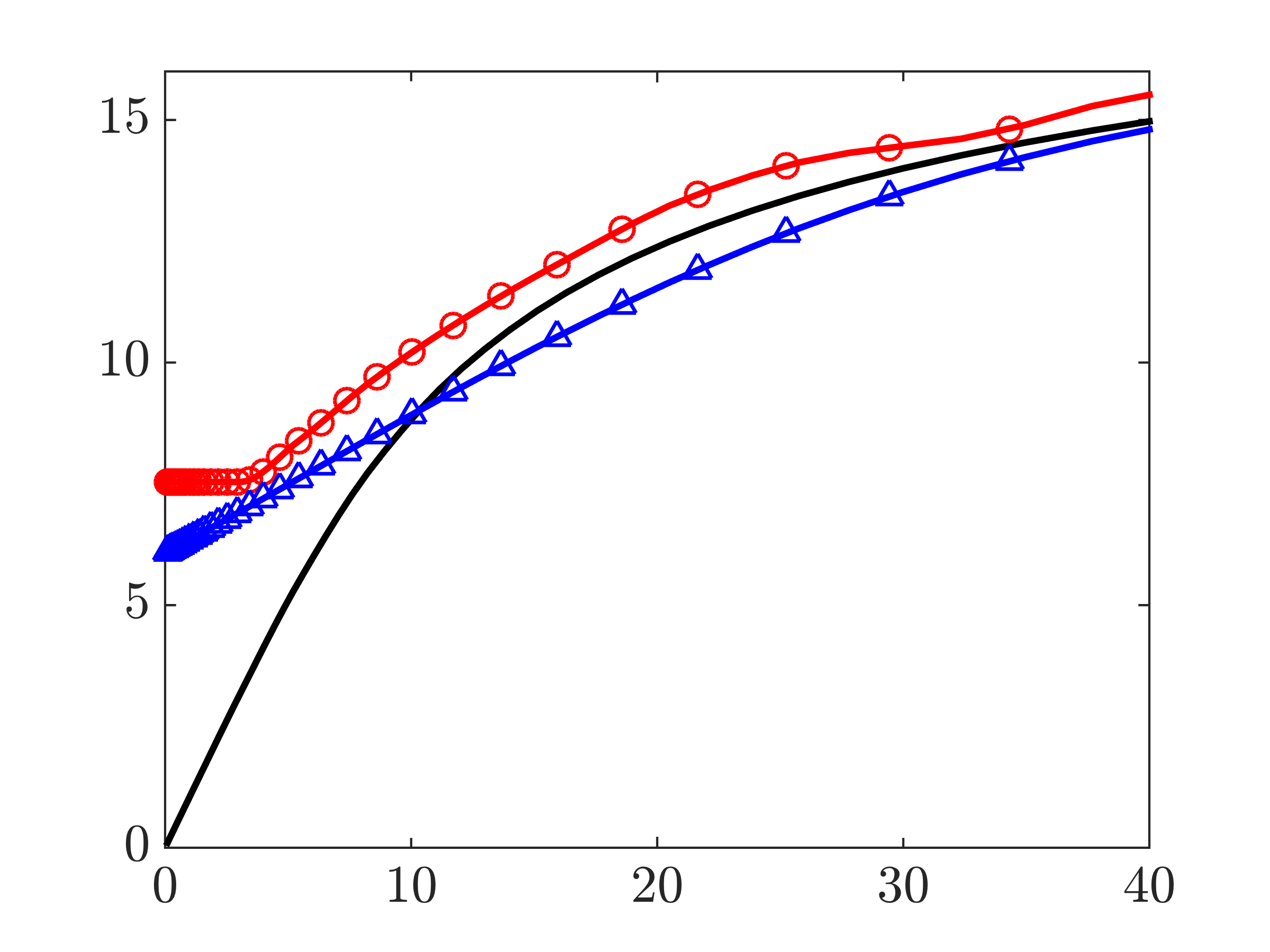}
		\\[-.1cm]
		~~{\normalsize $y^+$}
	\end{tabular}
	\end{tabular}
\end{center}
\caption{(a) Estimates of the streamwise convection velocity for $\bk=(2.5,7)$ computed using~\eqref{eq.conv_velocity} ($\Circle$) and~\eqref{eq.conv_velocity_empirical} ($\triangle$) as a function of wall distance $y^+$ for turbulent channel flow with $R_\tau=186$. The mean velocity profile is plotted for comparison ($-$). (b) Enlargement of the same plot for $y^+<40$.}
\label{fig.ConvectionVelocity_R180_kx2p5_kz7}
\end{figure}

Figure~\ref{fig.ConvectionVelocity_R180_kx2p5_kz7} compares this estimate of $c_u(\bk,y)$ to an estimate proposed by~\cite{deljim09}, 
	\be
	\label{eq.conv_velocity_empirical}
	c_u(\bk,y)
	\;=\;
	\ds{\int^1_{-1}} 
	W(\eta,\bk,y)
	\,
	U(\eta) 
	\, 
	\mrd \eta,
\ee
where $U$ is the mean velocity and $W$ is a Gaussian convolution window that accounts for the wall-normal structure of eddies with wavelength $\lambda_x=2\pi/k_x$ and $\lambda_z=2\pi/k_z$. The convolution window is tuned so that the resulting approximation agrees well with measurements of the convection velocity over a range of two-dimensional wavelengths, wall-normal distances, and Reynolds numbers~\citep{deljim09}. The convection velocity resulting from temporal correlations of the linear dynamics~\eqref{eq.feedback_dyn} is within $12\%$ of $c_u(\bk,y)$ computed from~\eqref{eq.conv_velocity_empirical}. Even though optimization problem~\eqref{eq.CP} constrains our model to only match one-point steady-state correlations, the modified dynamics~\eqref{eq.feedback_dyn} reproduce the essential features of the convection velocity of the most energetic modes. In particular, the deviation from Taylor's frozen turbulence hypothesis as the wall is approached is captured. This is a consequence of retaining the physics of the NS equations in our modeling and optimization framework. 

	\vspace*{-2ex}
\section{Discussion}
    \label{sec.discussion}

In this section, we provide insight and discuss broader implications of the framework.

\subsection{The role of the colored-in-time forcing}
	\label{sec.role-color}
	
	As noted earlier, the colored-in-time forcing introduces a low-rank modification to the dynamics of the linearized NS equations around turbulent mean velocity.
	This should be compared and contrasted to alternative modifications proposed in the literature.
	For instance, one such modification is obtained by augmenting molecular viscosity with turbulent eddy viscosity~\citep{reyhus72-3,deljimjfm06,cospujdep09,pujgarcosdep09,hwacosJFM10a,hwacosJFM10b}. Another modification adds a source of constant~\citep{farioa93c,delfar95,delfar96,del04} or dynamical~\citep{kra59,kra71} dissipation. The colored-in-time forcing similarly alters the dynamics but rather than postulating relations between Reynolds stresses and mean velocity gradients, it generates perturbations in a {\em data-driven\/} manner.	
	
\begin{figure}
	\begin{center}
%
%
%
%
%
%
\input{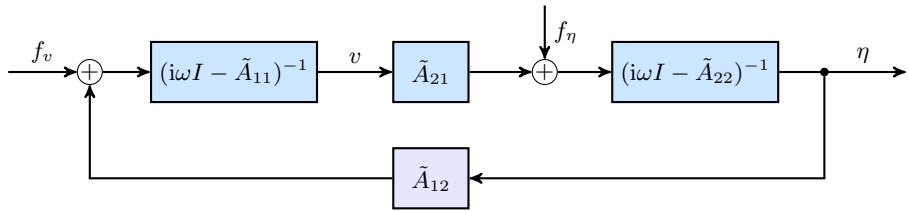}
%
%
\noindent
\begin{tikzpicture}[scale=1, auto, >=stealth']
  
    \small

    
     \node[block, minimum height = .8cm, top color=RoyalBlue!20, bottom color=RoyalBlue!20] (sys1) {$(\mri \omega I - \tilde A_{11})^{-1}$};
     
     \node[block, minimum height = .8cm, top color=RoyalBlue!20, bottom color=RoyalBlue!20] (sys2) at ($(sys1.east) + (1.5cm,0cm)$) {$\tilde A_{21}$};
     
     \node[block, minimum height = .8cm, top color=RoyalBlue!20, bottom color=RoyalBlue!20] (sys3) at ($(sys2.east) + (3cm,0cm)$) {$(\mri \omega I - \tilde A_{22})^{-1}$};
     
     \node[block, minimum height = .8cm, top color=blue!10, bottom color=blue!10] (sys4) at ($(sys2.south) - (0cm,1cm)$) {$\tilde A_{12}$};
     
     \node[] (input-node) at ($(sys1.west) - (2cm,0)$) {}; 
     
     \node[] (output-node) at ($(sys3.east) + (1.8cm,0)$) {};
     
     \node[] (mid-node1) at ($(sys2.east) + (1cm,1cm)$) {}; 
          
     \node[sum] (esum1) at ($(sys1.west) - (.8cm,0)$) {$+$};
     
     \node[sum] (esum2) at ($(sys2.east) + (1cm,0)$) {$+$};
      
%
%
%
%
%
%
%
     \node[branch] (R) at ($(sys3.east) + (.6cm,0.0cm)$){};
     
	

	
    \draw [connector] (input-node) -- node [midway, above] {$f_v$} (esum1.west);
    
     \draw [connector] (sys1.east) -- node [midway, above] {$v$} (sys2.west);
    
    \draw [connector] (sys2.east) -- (esum2.west);
    
    \draw [connector] (esum2.east) -- (sys3.west);
    
    \draw [connector] (mid-node1) -- node [midway, right] {$f_\eta$} (esum2.north);
    	
    \draw [line] (sys3.east) -- (R);
    \draw [connector] (R.west) -- node [midway, above] {$\eta$} (output-node);
    
    \draw [connector] (R.south) |- (sys4.east);
    
    \draw [connector] (sys4.west) -| (esum1.south);
    
    \draw [connector] (esum1.east) -- (sys1.west);
    
    
\end{tikzpicture}
	\end{center}
	\caption{Partitioning the state in~\eqref{eq.feedback_dyn} as $\bpsi=(v,\eta)$, and conformably the forcing $B\bw$ as $(f_v,f_\eta)$, the term $BC_f$ in~\eqref{eq.feedback_dyn} modifies the Orr-Sommerfeld, Squire, and coupling operators into $\tilde A_{11}$, $\tilde A_{22}$,
	and $\tilde A_{21}$, respectively. It also introduces an additional feedback term $\tilde A_{12}$.
	}
	\label{fig.modified-dynamics}
\end{figure}

	More specifically, in the linearized NS model, the generator $A(\bk)$ is lower block triangular. This means that wall-normal vorticity does not influence the dynamics of wall-normal velocity~\citep{schhen01}. In the context of channel flow, standard eddy-viscosity and dissipation models do not alter this structural feature. In contrast, the low-rank term $B(\bk)C_f(\bk)$ not only modifies the structure of the Orr-Sommerfeld, Squire, and coupling operators but it also introduces an additional feedback term $\tilde{A}_{12}$ as illustrated in figure~\ref{fig.modified-dynamics}. Thereby, besides an interpretation of colored-in-time forcing as a data-driven generalized eddy-viscosity refinement, the new framework points to potentially missing dynamical interactions in the linearized model. The nature and physical basis for such interactions calls for additional in-depth examination.

\subsection{Closure in mean flow equations}

\begin{figure}
	\begin{center}
%
%
%
%
%
%
\input{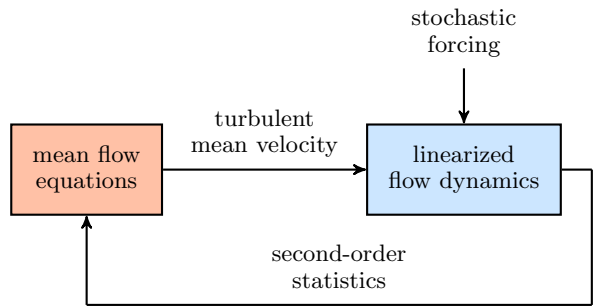}
%
%
\noindent
\begin{tikzpicture}[scale=1, auto, >=stealth']
  
    \small

    
     \node[block, minimum height = 1.2cm, top color=Orange!40, bottom color=Orange!40] (sys1) {$\ba{c} \mbox{mean flow} \\ \mbox{equations}\ea$};
     
     \node[block, minimum height = 1.2cm, top color=RoyalBlue!20, bottom color=RoyalBlue!20] (sys2) at ($(sys1.east) + (4cm,0cm)$) {$\ba{c} \mbox{linearized} \\ \mbox{flow dynamics}\ea$};
          
     \node[] (input-node) at ($(sys2.north) + (0,1.2cm)$) {$\ba{c} \mbox{stochastic} \\ \mbox{forcing}\ea$}; 
     
     
     \node[] (mid-node1) at ($(sys2.center) + (1.7cm,0cm)$) {};
     
     \node[] (mid-node2) at ($(sys2.center) + (1.7cm,-1.8cm)$) {};
     
     \node[] (mid-node3) at ($(sys1.center) - (0cm,1.8cm)$) {};
     
		
    \draw [connector] (sys1.east) -- node [midway, above] {$\ba{c} \mbox{turbulent} \\ \mbox{mean velocity}\ea$} (sys2.west);
    
    \draw [line] (sys2.east) -- (mid-node1.center);
    
    \draw [line] (mid-node1.center) -- (mid-node2.center);
    
     \draw [line] (mid-node2.center) -- node [midway, above] {$\ba{c} \mbox{second-order} \\ \mbox{statistics}\ea$} (mid-node3.center);

    \draw [connector] (mid-node3.center) -- (sys1.south);
    
    
    \draw [connector] (input-node) -- (sys2.north);
    
%
%
%
%
                                          
\end{tikzpicture}
	\end{center}
	\caption{For the linearized dynamics of fluctuations around turbulent mean velocity, the appropriate forcing is sought to reproduce partially available velocity correlations and complete the statistical signature of the turbulent flow field. Completed second-order statistics can then be brought into the mean flow equations in order to give turbulent mean velocity and provide equilibrium configuration.}
	\label{fig.equilibrium-loop}
\end{figure}

It is well known that the nonlinear nature of the NS equations makes the $n$th velocity moment depend on the $(n+1)$th. Colored-in-time forcing provides an alternative mechanism for developing a new class of {\em data-driven\/} turbulence closure models. More specifically, as shown in figure~\ref{fig.equilibrium-loop}, the turbulent mean velocity enters as a coefficient into the linearized flow equations. In turn these equations are used to compute second-order statistics which feed back into the mean flow equations.

A contribution of this paper is to identify power spectra of forcing to the linearized NS equations that yield velocity fluctuation statistics that are consistent with the DNS data in statistical steady state. The output of our model can be used to drive the mean flow equations in time-dependent simulations.
Thus, a correction to the mean velocity profile can be sought which perturbs the linearized NS dynamics.
This completes the loop by incorporating a two-way interaction between the mean flow and second-order statistics of the fluctuating velocity field.
It is an important topic to identify conditions under which the feedback connection of mean flow equations with stochastically-forced linearized equations, shown in figure~\ref{fig.equilibrium-loop}, converges.

Our methodology is conceptually related to recent work where streamwise-constant NS equations are combined with linearized flow equations driven by white-in-time forcing~\citep{farioa12,tholiejovfarioagayPOF14, conloznikfarioajim14}. It was demonstrated that self-sustained turbulence can indeed be maintained with such a model, although, correct statistics are not necessarily obtained. In this context, our approach offers a systematic framework for embedding data into physics-based models in order to capture correct turbulent statistics.

\subsection{Kinematic simulation of turbulent flow and the turbulent inflow generation}

Kinematic simulations of fully developed turbulence have been extensively used to generate synthetic flow fields. These typically involve the superposition of randomized Fourier modes that obey prescribed one- and two-point correlations~\citep{funhunmalper92,ellmaj96,malvas99,funvas98,camgodnicvas04,clavas11}.
Likewise, generating statistically consistent turbulent inflow conditions for numerical simulations of transitional/turbulent flows as well as flow control has been a topic of great interest~\citep{keapiobalkal04, hoenakfuk11}.
A common theme in these studies is that, in contrast to direct simulations, prescribed spatial and temporal correlations are used to generate statistically consistent flow fields.

Our approach is in line with this general theme in that it provides a data-driven method to generate statistically consistent velocity fluctuations using stochastically-driven linearized NS equations. 
The output of our modeling framework is a velocity field which can be calculated via {\em inexpensive stochastic linear simulations\/}.
While we only use one-point correlations as problem data, we have demonstrated that two-point spatial and temporal features are reasonably recovered.
This can be attributed to two elements of our framework. First, the underlying physics are intrinsic in the problem formulation and, second, the sought modifications of the linearized NS equations around turbulent mean velocity are of {\em low rank\/}.

\subsection{Model-based flow control}
  
Despite initial successes of model-based feedback~\citep{josspekim97,corspe98,bewliu98,leecorkimspe01,hogbewhen03,hogbewhen03b,kimbew07} and sensor-free~\citep{fratalbracos06,jovPOF08,moajovJFM10,liemoajovJFM10,moajovJFM12} control at low-Reynolds-numbers in wall-bounded flows many important challenges remain. One source of the problem is that, typically, sensing and actuation of the flow field is restricted to the surface of the domain. Thus, this limited actuation needs to rely on estimation of the flow field based on available noisy measurements of wall-shear stresses and pressure. 
\cite{hoechebewhen05,chehoebewhen06} recognized the importance of modeling the statistics of flow disturbances to obtain well-behaved estimation gains.
However, these initial studies, rely on assumptions on flow disturbances, e.g., whiteness-in-time, which often fail to hold in turbulent flows.  
In the present work, by departing from the white-in-time restriction, we demonstrate that turbulent flow statistics can be matched by linearized NS equations. Thus, the new methodology fits nicely into a Kalman filter estimation framework for turbulent flows and has the potential to open the door for a successful feedback control design at higher Reynolds numbers.

\subsection{Extension to complex geometries}

Channel and pipe flows allow Fourier transform techniques to exploit translational invariance in the homogeneous directions and, thereby, simplify computational aspects of the problem. In such cases, the governing equations for fluctuations around mean velocity can be decoupled across spatial wavenumbers. As a result, the optimization step in our theory deals with one pair of wall-parallel wavenumbers at a time; cf.~\eqref{eq.CP}. Non-parallel flows are spatially-developing and flows in more complex geometries may not be homogeneous in even a single spatial direction. One such example is that of boundary layer flows where experiments~\citep{hutmar07,mathutmar09,monhutngmarcho09,guametmck11,marmonhulsmi13} and simulations~\citep*{wumoi09,schorl10,siljimmos13,siljimmos14} have provided insight into coherent flow structures and statistics. 

Non-parallel flows and flows in complex geometries require dealing with a much higher number of degrees of freedom. Although our framework is pertinent to refining physics-based models of low complexity using data-driven methods, our current algorithms require $O(n^3)$ computations for an $n$-state discretized evolution model and fall short of dealing with the added computational overhead. At this point, development of more efficient optimization algorithms appears challenging. Thus, a possible direction is to examine approximations of the governing equations. Examples of such approximations, that have been used for the control and stability analysis of boundary layer flow, can be found in~\cite*{reesararn96,berherspa92,her97,hoghen02}. In general, model reduction techniques~\citep{sir87-1,berhollum93,row05,mez05,mez13,lum07,rowmezbagschhen09,sch10,cheturow12,jovschnicPOF14} can \mbox{also be used for this purpose.}

	
	\begin{figure}
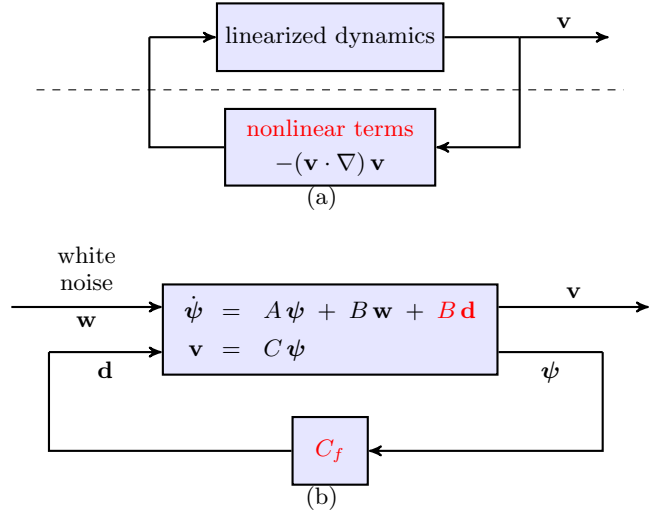

	\begin{center}
	\begin{tabular}{c}
		\subfigure[]{
%
%
%
%
%
\input{figures/Tikz_common_styles}
\noindent
\begin{tikzpicture}[scale=1, auto, >=stealth']
  
    \small
    
     \node[block, minimum height = .9cm, top color=blue!10, bottom color=blue!10] (sys1) {$\mbox{linearized dynamics}$};
     
     \node[block, minimum height = .9cm, top color=blue!10, bottom color=blue!10] (sys2) at ($(sys1.south) - (0cm,1cm)$) {$\ba{c} \mbox{\tc{red}{nonlinear terms}} \\[.1cm] -(\bv \cdot \nabla)\,\bv \ea$};
     
     \node[] (input-node) at ($(sys1.west) - (1cm,0)$) {}; 
     
     \node[] (output-node) at ($(sys1.east) + (2.3cm,0)$) {};
     
     \node[] (mid-node1) at ($(sys1.east) + (1cm,1cm)$) {}; 
                     
     
      \node[] (R2) at ($(sys1.west) - (2.5cm,0.7cm)$){};
      
      \node[] (R3) at ($(sys1.east) + (2.5cm,-0.7cm)$){};
     
                    	
    \draw [line] (sys1.east) -- ($(sys1.east) + (1cm,0cm)$);
    
    \draw [line] (sys2.west) -|  (input-node.east);
    
    \draw [connector] (input-node.east) --  (sys1.west);
    
    \draw [connector] ($(sys1.east) + (1cm,0cm)$) -- node [midway, above] {$\bv$} (output-node);
    
    \draw [connector] ($(sys1.east) + (1cm,0cm)$) |- (sys2.east);
    
    \draw [dashed] (R2) -- (R3);
                                       
\end{tikzpicture}
		         \label{fig.filter}
		         }
		        \\
		\subfigure[]{
%
%
%
%
%
\input{figures/Tikz_common_styles}
\noindent
\begin{tikzpicture}[scale=1, auto, >=stealth']
  
    \small
    
     \node[block, minimum height = 1.2cm, top color=blue!10, bottom color=blue!10] (sys1) {$\ba{rcl}\dot{\bpsi} & \!\! = \!\! & A\,\bpsi \;+\; B\, \bw \;+\; \tc{red}{B\, \bd} \\[.15cm] \bv & \!\! = \!\! & C\, \bpsi \ea$};
     
     \node[block, minimum height = .9cm, top color=blue!10, bottom color=blue!10] (sys2) at ($(sys1.south) - (0cm,1cm)$) {$\tc{red}{C_f}$};
                          
     \node[] (R1) at ($(sys1.east) + (1.5cm,-0.3cm)$){};
     
      \node[] (R2) at ($(sys1.west) - (2.5cm,0.7cm)$){};
      
      \node[] (R3) at ($(sys1.east) + (2.5cm,-0.7cm)$){};
     
    
    \draw [connector] ($(sys1.west) + (-2cm,.3cm)$) -- node [midway, above] {$\ba{c} \mbox{white} \\ \mbox{noise} \ea$} node [midway, below] {$\bw$} ($(sys1.west) + (0cm,.3cm)$);
                    	
    
    \draw [line] (sys2.west) -|  ($(sys1.west) + (-1.5cm,-.3cm)$);
    
    \draw [connector] ($(sys1.west) + (-1.5cm,-.3cm)$) -- node [midway, below] {$\bd$} ($(sys1.west) + (0cm,-.3cm)$);
    
    \draw [connector] ($(sys1.east) + (0cm,.3cm)$) -- node [midway, above] {$\bv$} ($(sys1.east) + (2cm,.3cm)$);
    
    \draw [line] ($(sys1.east) + (0cm,-.3cm)$) -- node[midway, below] {\bpsi} (R1.west);
    
    \draw [connector] (R1.west) |- (sys2.east);
    
                                       
\end{tikzpicture}
		         \label{fig.sys-modified}
		         }
	\end{tabular}
	\end{center}
	\caption{{
(a) Nonlinear NS equations as a feedback interconnection of the linearized dynamics with the nonlinear term; (b) Stochastically-driven linearized NS equations with low-rank state-feedback modification. The two representations can be made equivalent, at the level of second-order statistics, by proper selection of $B$ and $C_f$.}}
	\label{fig.nonlinear-vs-sf}
\end{figure}

\section{Concluding remarks}
    \label{sec.remarks}
    
	The focus of the paper is on how to account for statistical signatures of turbulent flows using low-complexity {linear} stochastic models. The complexity is quantified by the number of degrees of freedom in the NS equations that are directly influenced by stochastic forcing.
Models for colored-in-time forcing are obtained using a maximum entropy formulation together with a regularization that serves as a {penalty for model complexity.} 
We show that colored-in-time excitation of the NS equations can also be interpreted as a low-rank modification to the generator of the linearized dynamics. {Schematically, the correspondence between the nonlinear and stochastically-driven linearized NS equations is shown in figure \ref{fig.nonlinear-vs-sf}.} The modified dynamics {are designed to be equivalent, at the level of second-order statistics,} with DNS of turbulent channel flow.

{Our motivation has been to} develop a framework to complete unavailable statistics in a way that is consistent with the {linearized} dynamics around turbulent mean velocity. {The resulting dynamical model can be used for time-dependent} linear stochastic simulations and analyzed using tools from linear systems theory. We have verified the ability to match statistics of turbulent channel flow using such simulations. We have also analyzed the spatio-temporal responses to stochastic and deterministic excitation sources. In particular, by examining the power spectral density of velocity fluctuations, we have shown that the dynamical modification attenuates the amplification over all temporal frequencies. A similar effect has been observed in eddy-viscosity-enhanced linearization of the NS equations.
Although our models are based on one-point correlations in statistical steady-state, we have computed two-point temporal correlations to demonstrate that the essential features of the convection velocities of individual modes are reproduced.

Full scale physics-based models are often prohibitively complex. 
A value in our method is that it provides a data-driven refinement of models that originate from first principles. The method captures complex dynamics in a way that is tractable for analysis, optimization and control design. We have focused on a canonical flow configuration to demonstrate the ability to generate statistically consistent velocity fluctuations. The framework opens up the possibility to guide experimental data collection in an economic manner that, at the same time, allows faithful representation of structural and statistical flow features.
 
\section*{Acknowledgments}

Part of this work was performed during the 2014 CTR Summer Program with financial support from Stanford University and NASA Ames Research Center. We thank Prof.\ P.\ Moin for his interest in our work and for providing us with the opportunity to participate in the CTR Summer Program; Prof.\ J.\ Jimenez and Dr.\ J.\ A.\ Sillero for useful discussions regarding DNS-based turbulent statistics; Prof.\ J.\ W.\ Nichols and Dr.\ R.\ Moarref for their feedback on earlier versions of this manuscript{; and Profs.\ P.\ J.\ Schmid, S.\ I.\ Chernyshenko, and Y.\ Hwang for insightful discussions about our method.} Financial support from the National Science Foundation under Award CMMI 1363266, the Air Force Office of Scientific Research under Award FA9550-16-1-0009, the University of Minnesota Informatics Institute Transdisciplinary Faculty Fellowship, and the University of Minnesota Doctoral Dissertation Fellowship is gratefully acknowledged. The University of Minnesota Supercomputing Institute is acknowledged for \mbox{providing computing resources.}

\appendix

\section{Change of coordinates}
	\label{sec.coc}

The kinetic energy of velocity fluctuations in the linearized NS equations~\eqref{eq.lnse} is determined by,
\be
	E
	\;=\,
	\left< \bvarphi, \bvarphi \right>_e
	\,=\;
	\dfrac{1}{2} \,
	\ds{\int^1_{-1} \bvarphi^* \, \bQ \, \bvarphi \, \mrd y}
	\;\DefinedAsLtoR\,
	\left< \bvarphi, \bQ\, \bvarphi \right>,
\ee
where $\left<\cdot, \cdot\right>$ is the standard $L_2$ inner product and $\bQ$ is the operator that determines energy on the state-space $\mathbb{H}_\mathrm{OS} \times L_2 [-1, 1]$~\citep{redhen93,jovbamJFM05}. After wall-normal discretization, the energy norm is determined by
$
	E
	=
	\bvarphi^*\, Q\, \bvarphi,
$
where $Q$ is the finite-dimensional representation of the operator $\bQ$.

Since the matrix $Q$ is positive-definite, the state of the linearized NS equations~\eqref{eq.lnse} can be transformed into a set of coordinates in which the energy is determined by the standard Euclidean norm, i.e., $E=\bpsi^* \bpsi$ with $\bpsi \DefinedAs Q^{1/2} \bvarphi$. Equation~\eqref{eq.lnse1} results from the application of this change of coordinates on the discretized state-space matrices $\bar{A}$, $\bar{B}$, and $\bar{C}$
\be
	\label{eq.ABC-newcoordinate}
	A 
	\;=\; 
	Q^{1/2}\, \bar{A} \; Q^{-1/2},
	~~~~~~
	B
	\;=\;
	Q^{1/2}\, \bar{B} \, I_W^{-1/2},
	~~~~~~
	C
	\;=\;
	I_W^{1/2}\, \bar{C}\, Q^{-1/2},
\ee
and the discretized input $\bar{\bd}$ and velocity $\bar{\bv}$ vectors
\be
	\label{eq.inout-output-vectors}
	\bd \;=\; I_W^{1/2}\, \bar{\bd},
	~~~~~~
	\bv \;=\; I_W^{1/2}\, \bar{\bv}.
\ee
Here, $I_W$ is a diagonal matrix of integration weights on the set of Chebyshev collocation points. The form of the input and output matrices in~\eqref{eq.inout-output-vectors} follows from the definition of their respective energy norms which are given by the standard $L_2 [-1,1]$ inner product.

\section{Interpretation of the matrix $H$ solving \eqref{eq.lyap_BH}}
	\label{sec.H-interp}

From~\eqref{eq.psi-phi-chi} we see that the state $\bphi(\bk,t)$ of the linear filter~\eqref{eq.filter} is
\be
	\bphi(\bk,t)
	\;=\;
	\bpsi(\bk,t) 
	\,+\,
	\bchi(\bk,t),
\ee
where $\bchi(\bk,t)$ represents uncontrollable asymptotically stable modes of the cascade connection of the linearized NS equations and the filter given in \eqref{eq.cascade}.
Thus, $\bchi(\bk,t)\to 0$ as $t\to\infty$ and, consequently,
\be
	\lim_{t \, \to \, \infty}\left< \bpsi(\bk,t)\,\bphi^*(\bk,t)\right> 
	\,=\;
	\lim_{t \, \to \, \infty}\left< \bpsi(\bk,t)\,\bpsi^*(\bk,t)\right> 
	\,=\;
	X(\bk).
\ee
On the other hand, the cross-correlation between the colored-in-time forcing and the state of the linearized NS equations~\eqref{eq.lnse1} becomes
\be
	\ba{rcl}
	\ds{
	\lim_{t \, \to \, \infty}\left< \bpsi(\bk,t)\,\bd^*(\bk,t)\right>
	}
	&\!\!=\!\!&
	\ds{
	\lim_{t \, \to \, \infty}\left< \bpsi(\bk,t) \left(\bphi^*(\bk,t)\,C_f^*(\bk) \,+\, \bw^*(\bk,t)\right)\right>
	}
	\\[.2cm]
	&\!\!=\!\!&
	X(\bk)\,C_f^*(\bk),
	\ea
\ee
where we have used the fact that the state $\bpsi(\bk,t)$ and the white-in-time input $\bw(\bk,t)$ are not correlated. From the definition of $C_f(\bk)$ in~\eqref{eq.filter-A-C} we now have
\be
	\label{eq.H-relation}
	H(\bk)
	\;=\;
	\lim_{t \, \to \, \infty}\left< \bpsi(\bk,t)\,\bd^*(\bk,t)\right>
	\,+\,
	\dfrac{1}{2} B(\bk) \,\Omega(\bk).
\ee
Therefore, the solution $H(\bk)$ of \eqref{eq.lyap_BH} can be seen to be directly related to the cross-correlation between the forcing $\bd(\bk,t)$ and the state $\bpsi(\bk,t)$; see also \cite{geo02a}.

\section{The role of the regularization parameter $\gamma$}
	\label{sec.gamma}

When the true covariance matrices are not known, the regularization parameter $\gamma$ is typically chosen on an empirical basis or by cross-validation. In fact, the selection of the optimal value of $\gamma$ is an open theoretical challenge. If the DNS-generated two-point correlation matrix $\Phi_{\mathrm{dns}}(\bk)$ is known, we can use the following error criterion:
\be
	\label{eq.relative_error}
	\dfrac{\norm{\Phi(\bk) \,-\, \Phi_{\mathrm{dns}}(\bk)}_F}{\norm{\Phi_{\mathrm{dns}}(\bk)}_F} \times 100,
\ee
to assess {the} quality of approximation. 

\begin{figure}
\begin{center}
	\begin{tabular}{cccc}
	\subfigure[]{\label{fig.err}}
       & 
       &
       \hspace{-.4cm} 
       \subfigure[]{\label{fig.rankZ}}
	&
	\\[-.35cm]
	&
	\hspace{.1cm}
	\begin{tabular}{c}
		\includegraphics[width=6.15cm]{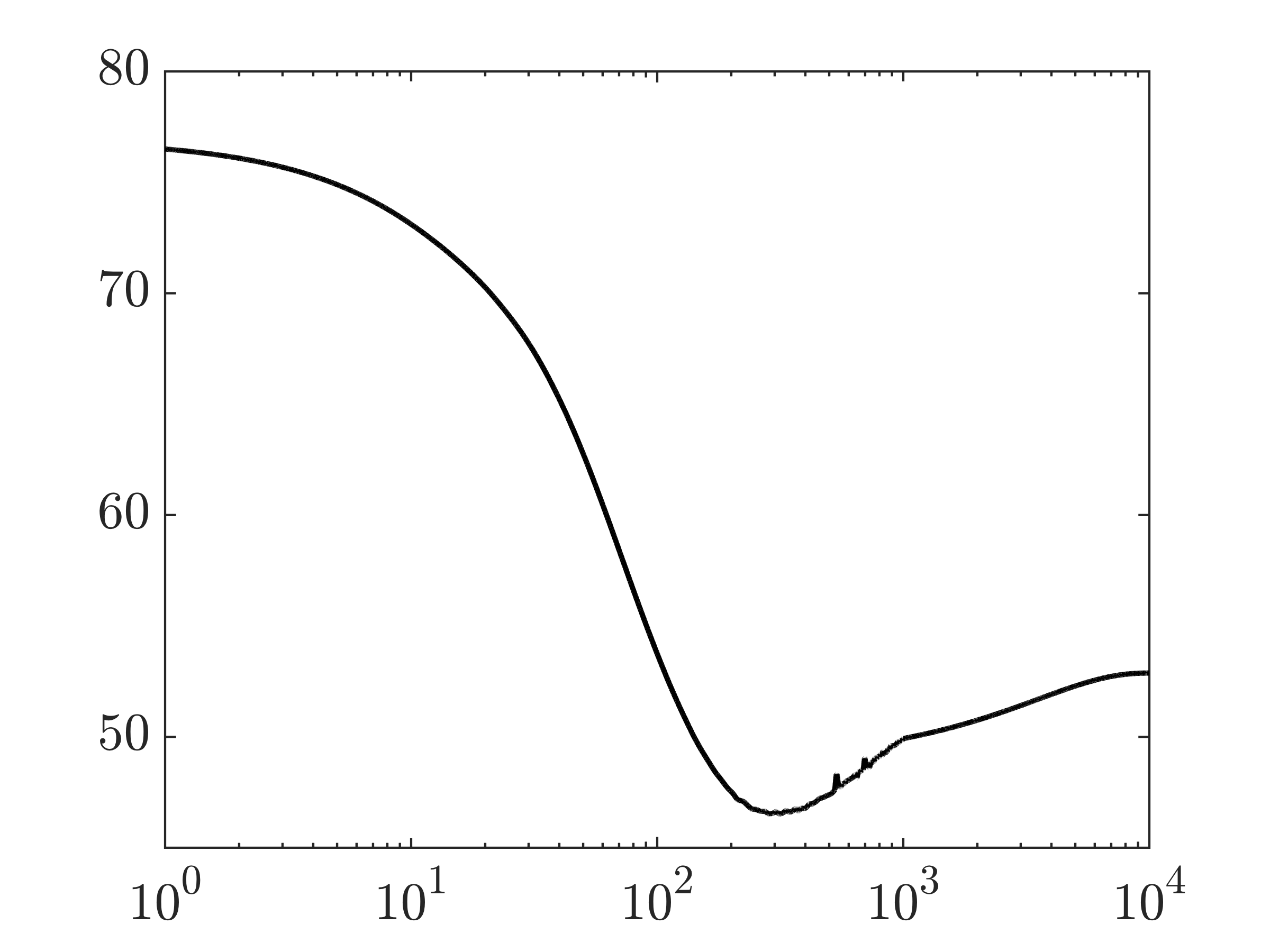}
		\\[-.1cm]
		\hspace{-.3cm}{\normalsize $\gamma$}
	\end{tabular}
	&
	&
	\hspace{.1cm}
	\begin{tabular}{c}
		\includegraphics[width=6.15cm]{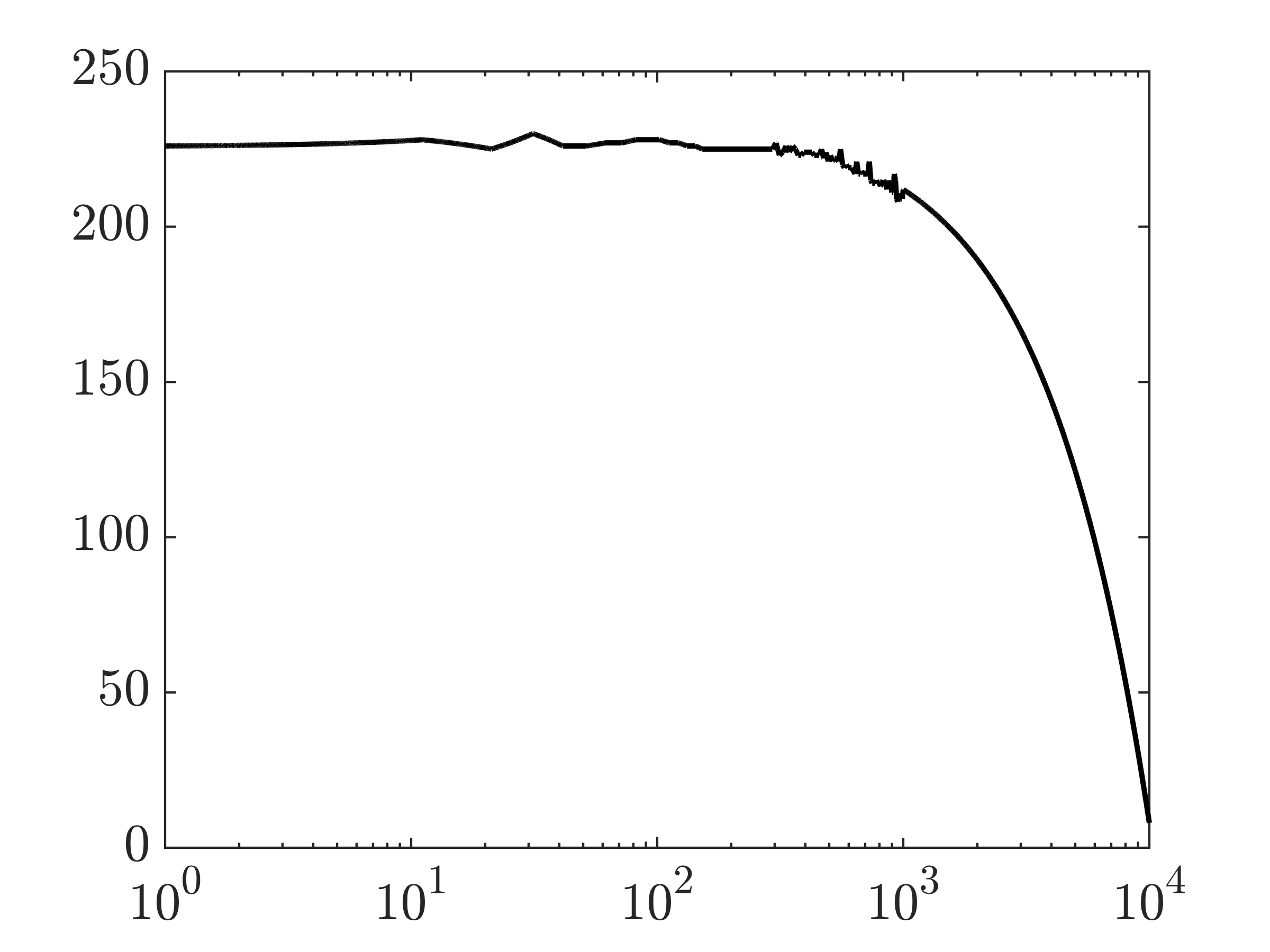}
		\\[-.1cm]
		~~{\normalsize $\gamma$}
	\end{tabular}
	\end{tabular}
\end{center}
\caption{The $\gamma$-dependence of (a) the relative error~\eqref{eq.relative_error}; and (b) the rank of the matrix $Z$ resulting from~\eqref{eq.CP} for turbulent channel flow with $R_\tau=186$ and $\bk=(2.5,7)$.}
\label{fig.gamma-path}
\end{figure}

For turbulent channel flow with $R_\tau=186$ and $\bk = (2.5,7)$, figure~\ref{fig.err} shows the $\gamma$ dependence of the above measure. Smallest error is achieved for $\gamma \approx 300$. On the other hand, figure~\ref{fig.rankZ} shows the $\gamma$ dependence of the rank of the matrix $Z$. Clearly, much larger values of $\gamma$ are needed to achieve $Z$ of lower rank. 


\end{document}